\documentclass{article}

\usepackage{fullpage}

\usepackage{xurl}
\usepackage{hyperref}
\usepackage{verified-badges}
\renewcommand{\leanproof}[1]{\begingroup\hypersetup{pdfborder={0 0 0}}\verifiedproof[lean]{#1}\endgroup}
\renewcommand{\leanformalized}[1]{\begingroup\hypersetup{pdfborder={0 0 0}}\formalizedstmt[lean-formalized]{#1}\endgroup}
\newcommand{\leanrepository}{https://github.com/BioDisCo/distributed_synthesis_lean}
\newcommand{\leansourceroot}{DistributedSynthesis}
\newcommand{\leanrevision}{ba94ebc0d82b04d70600f1a03e5ee2d24164ab8b}
\newcommand{\leansourceurl}[1]{\leanrepository/blob/\leanrevision/\leansourceroot/#1.lean}
\newcommand{\leanurl}[1]{\leansourceurl{Formalizations/Paper/#1}}
\newcommand{\leanline}[2]{\leansourceurl{#1}\#L#2}

\usepackage{acro}
\DeclareAcronym{LTL}{short = LTL, long = linear temporal logic}
\DeclareAcronym{LL}{short = LL, long = lossy link}
\DeclareAcronym{MSO}{short = MSO, long = monadic second-order logic}
\usepackage{graphicx}
\usepackage{tikz}
\usetikzlibrary{arrows.meta,decorations.pathreplacing,shapes.geometric}
\usepackage[ruled, linesnumbered]{algorithm2e}

\usepackage{amsmath,amsthm,amssymb}
\usepackage{mathtools}

\usepackage{aliascnt}
\usepackage[capitalize]{cleveref}

\newcommand{\newsharedtheorem}[2]{%
  \newaliascnt{#1}{theorem}%
  \newtheorem{#1}[#1]{#2}%
  \aliascntresetthe{#1}%
}

\theoremstyle{plain}
\newtheorem{theorem}{Theorem}[section]
\newsharedtheorem{lemma}{Lemma}
\newsharedtheorem{corollary}{Corollary}
\newsharedtheorem{proposition}{Proposition}
\newsharedtheorem{claim}{Claim}
\newsharedtheorem{conjecture}{Conjecture}

\theoremstyle{definition}
\newsharedtheorem{definition}{Definition}
\newsharedtheorem{example}{Example}

\theoremstyle{remark}
\newsharedtheorem{remark}{Remark}
\newsharedtheorem{observation}{Observation}
\newsharedtheorem{notation}{Notation}

\newcommand{\NN}{\mathbb N}
\newcommand{\ZZ}{\mathbb Z}

\newcommand{\llink}{(1 \leftarrow 2)}     
\newcommand{\rlink}{(1 \rightarrow 2)}    
\newcommand{\dlink}{(1 \leftrightarrow 2)}    
\newcommand{\liveext}[1]{{#1}^{(D)}_\text{live}}

\DeclareMathOperator{\Graph}{Graph}
\DeclareMathOperator{\inp}{in}
\DeclareMathOperator{\outp}{out}
\DeclareMathOperator{\link}{graph}
\DeclareMathOperator{\Pref}{Pref}
\DeclareMathOperator{\len}{len}
\DeclareMathOperator{\Views}{Views}
\DeclareMathOperator{\exec}{exec}
\DeclareMathOperator{\ind}{Pos}

\DeclareMathOperator{\ltlF}{\mathsf{F}}    
\DeclareMathOperator{\ltlG}{\mathsf{G}}    
\DeclareMathOperator{\ltlU}{\,\mathsf{U}\,}    
\DeclareMathOperator{\ltlX}{\mathsf{X}}    
\DeclareMathOperator{\ltlW}{\,\mathsf{W}\,}    

\newcommand{\countsymb}{\mathsf{c}} 
\newcommand{\graph}{g}  
\newcommand{\single}{\mathrm{counter}}     
\newcommand{\init}{\mathrm{init}}              
\newcommand{\zero}{\ensuremath{\mathit{zero}}}    
\newcommand{\rightplan}{\ensuremath{\mathit{plan}}}  
\newcommand{\carry}{\ensuremath{\mathit{carry}}}          
\newcommand{\pos}{\ensuremath{\mathit{pos}}}              
\newcommand{\realpos}{\ensuremath{\mathit{pos}}}              %
\newcommand{\view}{\ensuremath{\mathit{view}}}              
\newcommand{\mul}{\ensuremath{\mathit{mul}}}              
\newcommand{\pathsum}{\ensuremath{\mathit{sum}}}              
\newcommand{\posord}{\mathrm{PosOrd}}   
\newcommand{\pathagr}{\mathrm{PathAgr}} 
\newcommand{\localsum}{\mathrm{LocalSum}} 
\newcommand{\zeroind}{\mathrm{ZeroInd}} 
\newcommand{\leftmost}{\mathrm{LeftMost}} 
\newcommand{\rightmost}{\mathrm{RightMost}} 
\newcommand{\zeroclaim}{\mathrm{ZeroCheck}} 
\newcommand{\initcheck}{\mathrm{InitCheck}}   
\newcommand{\statetrans}{\mathrm{StateTrans}}   
\newcommand{\weighttrans}{\mathrm{WeightTrans}}   
\newcommand{\proptrans}{\mathrm{PropTrans}}   
\newcommand{\oneisleft}{\mathrm{OneIsLeft}} 
\newcommand{\childpos}{\mathrm{ChildPos}}
\newcommand{\proj}[2]{\ensuremath{{#1}_{#2}}}
\newcommand{\counter}{\ensuremath{\mathit{cnt}}}    

\author{Benedikt Bollig$^{1}$
\and Patricia Bouyer-Decitre$^{1}$
\and Matthias Függer$^{1}$
\and Dhrubajyoti Ghosh$^{1}$
\and Thomas Nowak$^{1,2}$
\and Nathalie Sznajder$^{3}$}

\title{Distributed \acs*{LTL} Synthesis under Arbitrary Message Adversaries}
\date{
    $^{1}$ Université Paris-Saclay, CNRS, ENS Paris-Saclay, LMF\\
    $^{2}$ Institut Universitaire de France\\
    $^{3}$ Sorbonne Universit\'e, CNRS, LIP6
}

\begin{document}
\pagenumbering{gobble}
\maketitle

\begin{abstract}
  Reactive synthesis is the problem of generating a program that produces outputs
    satisfying a logical specification for every input stream.
  We study distributed synthesis for two processes communicating in rounds
    over unreliable links, with at least one message delivered in every round.
  Despite this delivery guarantee, we show that synthesis from linear temporal
    logic specifications is undecidable, even when messages carry the processes'
    complete causal histories.

  More generally, we give a complete characterization of the communication
    restrictions that make synthesis decidable.
  Restrictions are described by a message adversary~$A$, a set of allowed
    infinite graph histories: sequences of communication graphs, each specifying
    which messages are delivered in a round.
  Iterated chromatic subdivision associates these histories with points in
    the unit interval, yielding an image $\pi(A) \subseteq [0,1]$.
  For every $\omega$-regular message adversary~$A$, synthesis is decidable if and
    only if~$\pi(A)$ is nowhere dense.

  In the decidable cases, our procedure takes a deterministic parity automaton
    for~$A$ and a specification, decides realizability, and constructs a realizing
    distributed algorithm whenever one exists.
  Further, our topological arguments shed new light on the fundamental
    requirements of synthesis.
  They show that, for $\omega$-regular message adversaries,
    synthesis is decidable if and only if consensus is repeatedly solvable
    in bounded time.
  This links two previously unrelated problems from formal verification and
    distributed computing through the processes' uncertainty about communication.
\end{abstract}

\newpage
\pagenumbering{arabic}

\section{Introduction}

\acresetall
Reactive synthesis asks for an algorithm that, given a logical specification,
  constructs a program satisfying it for every input stream,
  or determines that no such program exists.
It addresses a fundamental question in reactive-system design:
  how to ensure correctness of a program by construction.

The problem was introduced by Church~\cite{Church57,Church62}.
In its standard formulation, the input and output alphabets are finite,
  and the specification is given in \ac{MSO} over infinite words.
The output program to be constructed is a finite-state transducer that reads
  one input symbol and produces one output symbol per round.
Büchi and Landweber~\cite{BuchiLandweber69} proved decidability
  by viewing synthesis as an infinite game between the environment,
  which chooses inputs, and a controller, which chooses outputs.
A winning strategy for the controller defines a program satisfying the specification.
Shortly after, Rabin~\cite{Rabin72} provided an alternative proof using
  tree automata, exploiting the tree structure naturally formed by
  the finite prefixes of a program's possible executions.

Church's synthesis problem has received sustained attention
  and has been extended in many directions~\cite{Thomas08, BCJ18, Fijalkow25}. %
  An extension of particular relevance to distributed systems is synthesis under \emph{imperfect information}, 
  where a single process observes only part of the input~\cite{kv97,arnold&vincent&walukiewicz:2003:games}.
  Pnueli and Rosner~\cite{PnueliRosner90} posed the question for the distributed setting
  in which several processes, each observing only part of the input, must jointly realize a
  global specification.
They proved that distributed synthesis is undecidable even for
  specifications in \ac{LTL}~\cite{Pnueli77}
  and an architecture consisting of just two processes,
  each receiving a different part of the input and unable to communicate
  with the other.
By contrast, synthesis is decidable for pipeline architectures, in which
  processes can send messages only to their right neighbor along a line,
  and only the first process receives external input from the environment.
The proof successively eliminates processes from the pipeline,
  using tree automata to reduce the problem to synthesis for a single process.
Kupferman and Vardi~\cite{kv01} extended decidability to further architectures
  and Finkbeiner and Schewe~\cite{fs05} gave a complete characterization:
  the problem is decidable if and only if the architecture
  has no information fork.
  Distributed synthesis has also been studied for knowledge-based specifications~\cite{mw05}.
A recurring condition behind decidability in synchronous distributed synthesis
  is \emph{hierarchical information}:
  calling a process~$p$ more informed than process~$q$
  if it can deduce the information of~$q$ from its own,
  information is \emph{hierarchical} if all processes can be linearly ordered
  with respect to informedness.
Undecidability thus stems from processes holding incomparable information
  that they cannot reconcile: sharing information between processes
  is essential to solving the problem.
This observation led Gastin, Sznajder, and Zeitoun~\cite{GSZ09FMSD}
  to consider specifications that refer only to external inputs and outputs,
  and not to the communication channels between processes:
  processes are then free to use these channels to share information,
  which made it possible to extend the class of decidable architectures.

Asynchronous models push this idea further with \emph{causal memory}:
  whenever processes synchronize, they exchange their entire causal histories,
  so that each process knows everything that causally precedes its current state.
  Synthesis is then decidable for restricted architectures, such as
  series-parallel systems~\cite{gastinlermanzeitoun:04} or tree
  architectures~\cite{GGMW13,MuschollWalukiewicz14},
  but even this maximal form of information sharing does not make the problem
  decidable in general~\cite{Gimbert22}.

In synchronous models, another way to relax the rigidity of hierarchical information
  is to let the hierarchy itself evolve. Indeed, many distributed systems lack such a
   fixed hierarchy since communication is
  sporadic and process and communication faults may occur.
Moreover, even if it exists, it may be unknown to the processes.

Berwanger, Mathew, and van den Bogaard~\cite{BMV15hierarchical} studied
  synthesis in finite games where processes retain their entire observation
  history and the information hierarchy may change during play.
Observations in each round range over fixed finite alphabets, so the model
  does not directly capture messages carrying arbitrarily long causal histories.
Specifications are given by finite automata and depend only on signals
  observable to all processes.
They proved decidability in this setting and showed that each process
  can infer its position in the hierarchy from its own observation history.

The message adversary model~\cite{CBS09,AG15} unifies several widely used
  communication models from the theory of distributed computing, including process crashes and
  message losses by specifying which messages are received in each round by which process.
Reception of a message is expressed as an edge in a directed communication graph,
  which may change in each round.
The adversary is the set of allowed infinite graph histories.

In distributed computing, such submodels are used to study which problems
  can be solved under given communication assumptions.
\ac{LTL} can express commonly used message adversaries and decision tasks as well as long-lived tasks such as
  repeated consensus~\cite{Lamport98} and clock synchronization~\cite{srikanth87}.

B\'erard, Bollig, Bouyer, F\"ugger, and Sznajder~\cite{BBBFS22}
  studied synthesis for two processes and an unstable communication between them,
  however, additionally assuming acknowledgments that make each round's
  communication graph immediately known to both processes.
They allowed messages to carry their complete causal histories,
  allowing for the synthesis of full information protocols
  that are not restricted to bounded messages.
They proved that, for $\omega$-regular message adversaries, synthesis is decidable
  if and only if the number of consecutive rounds in which no message is
  delivered is uniformly bounded.
The acknowledgment assumption changes the solvability of even basic tasks,
  though.
For example, two processes can solve consensus in one round if at least
  one message is delivered and both processes know the communication graph.
The decidability of synthesis without this assumption thus remained open.

We resolve this question and characterize the message adversaries
  for which synthesis is decidable in the general case for message adversaries
  as they are studied in the theory of distributed computing.
We develop a new topological approach to establish this characterization.

Results marked with \leanproof{\leanurl{Characterization}} have machine-checked
  proofs in Lean~4 using Mathlib (v4.30.0).
Each proof badge links to a dedicated Lean file containing the corresponding
  statement.
Badges of the form \leanformalized{\leanurl{Model}} on definitions and in the
  Model and Lossy-Link Synthesis Problem subsections link directly to their
  formal counterparts.
The formalization only supplements the natural-language proofs in this paper and is not meant to replace them.
A short overview is given in Appendix~\ref{sec:lean-formalization}.

\subsection{Our Results}

In this paper, we study the synthesis problem of distributed algorithms
  for two processes under arbitrary message adversaries.
We assume that message adversaries are $\omega$-regular
  and given by a parity automaton.
In particular, any \ac{LTL}-definable message adversary is covered by our results.
This class includes all oblivious message adversaries~\cite{CG15},
  which have been shown to capture the computational power of synchronous
  and asynchronous models of computation, with processes communicating
  via shared registers~\cite{AG15} and message passing~\cite{CBS09},
  both without and with process faults~\cite{ACGN26}.
The class also includes models of dynamic communication networks~\cite{Cas12:time},
  including ones with eventual stabilization guarantees~\cite{WSS19}.

Authentication tells a recipient where a message came from, but need not tell
  its sender whether it arrived.
This distinction also appears in cryptographic channel abstractions, where
  authenticated communication may leave message delivery under adversarial
  control~\cite[Section~3.1]{KMTZ13}.
Consider two honest parties, Alice and Bob, communicating over such unreliable links.
Even if they exchange their complete causal histories, Alice may remain uncertain
  whether Bob received her latest message.

More precisely, we consider submodels of the \ac{LL} model~\cite{SWK09}.
In this model, any infinite graph history over the
  set $\Gamma = \left\{\llink, \rlink, \dlink \right\}$ can occur.
The parity automaton defining this message adversary is trivial.
In the \ac{LL} model, the consensus problem is not solvable~\cite{SW89}, but approximate consensus 
  is~\cite{CBFN15:icalp}.
Our first result is the undecidability of distributed synthesis in the \ac{LL} model:

\begin{theorem}[\leanproof{\leanurl{Undecidability}}]
    \label{thm:undecidability}
    The \textnormal{\textsc{LL-Synthesis}} problem is undecidable.
\end{theorem}

To characterize decidability for arbitrary message adversaries, we use the map
  $\pi\colon\Gamma^\omega\to[0,1]$ from infinite graph histories into the unit interval,
  defined by the limit of the iterated chromatic subdivision of the unit interval, which describes the set of
  possible local views of both processes~\cite{HKR-book,godard_back_2020}.
Decidability of the synthesis problem under a message adversary depends on a topological property of its
  image in the unit interval:

\begin{theorem}[\leanproof{\leanurl{Characterization}}]
    \label{thm:undecidability:adversary}
    Let~$A$ be an $\omega$-regular message adversary.
    The $\textnormal{\textsc{LL-Synthesis}}(A)$ problem is decidable if and only if~$\pi(A)$ is nowhere dense.
\end{theorem}

A subset of~$[0,1]$ is nowhere dense if its closure contains no interval of positive length.
This is a generalization of \Cref{thm:undecidability} since the full \ac{LL} message adversary~$A = \Gamma^\omega$ maps to the full unit interval $\pi(A) = [0,1]$, which is not nowhere dense.

The decidability direction is constructive and uniform in the message adversary:
  given a deterministic parity automaton for~$A$ with nowhere-dense image~$\pi(A)$ and an \ac{LTL} formula~$\phi$,
  our procedure decides whether~$\phi$ is realizable under~$A$ and constructs a realizing distributed algorithm
  whenever one exists.
Key to our decidability result is a connection underlying our characterization
  between the decidability of synthesis and the repeated solvability of consensus
  in bounded time.
While hierarchical information allows one to solve consensus, the reverse is not true.
Our decidability result does not require hierarchical information in each round
  and holds for adversaries where this is not given.
In fact we show that repeated solvability of consensus in bounded time is precisely what
  is required to solve synthesis.
This sheds new light on the nature of reactive synthesis.

\subsection{Our Techniques}

The proofs presented in this work are based on several new techniques that we summarize in the following.
Central to both directions of \Cref{thm:undecidability:adversary} is the use of the view complex,
  the graph whose vertices are the possible local views after round~$r$ and whose edges are the possible
  executions up to that round.
Distributed algorithms are naturally expressed as labeling each vertex with an output in this formalism.

\paragraph{Undecidability via encoding a 2-counter machine in a set of executions}
To show undecidability, we reduce from the non-halting problem for
  2-counter machines~\cite{minsky1967}.
In contrast to the undecidability result of Pnueli and Rosner~\cite{PnueliRosner90},
  our technique does not use independent inputs, but uses the very uncertainty about the system,
  in our case the communication graphs, to store unbounded integers.
Counters turn out to be particularly well suited to be encoded in such uncertainties.
The difficulty of this proof approach, however, lies in the fact that process output alphabets are finite,
  and thus no single execution round can store an unbounded counter value.
We therefore distribute the value over all executions of a round:
  every edge of the round-$r$ path carries one binary weight per counter,
  and a counter with value~$c$ at step~$r-1$ of the run is represented by a
  fraction~$3^{-c}$ of edges with weight~$1$.
An increment, a no-op, and a decrement multiply the number of weight-$1$ edges by~$1$, $3$, and~$9$,
  respectively.
All vertices of a round additionally output the current control state of the machine and a
  claim on whether each counter is zero (claim, since this is impossible to test locally, too),
  so that the sequence of labeled paths encodes the run.

\paragraph{Local indistinguishability enables \acs*{LTL} formulas to relate neighboring executions}
An \ac{LTL} formula is evaluated on one execution at a time,
  so in each round it can only compare the outputs at the two endpoints of a single edge.
However, to encode unbounded counter sizes, we need to argue about an unbounded number of executions
  in our formulas.
For this we use the fact that processes inherently are forced to behave identically on identical
  views: the output of a process is a function of its view, and adjacent edges in a view complex share a view.
Constraints on individual executions thus propagate along the path and become global constraints on a round.
For example, if both processes must always output the same control state and zero claims,
  then all vertices of a round do.

\paragraph{Safety formulas only depend on the closure of an adversary}
The formula of the undecidability reduction is a safety formula.
A violation of such a formula shows after finitely many rounds,
  and every finite prefix of a history in the closure~$\overline{A}$ is also a prefix of a history in~$A$.
Thus an algorithm realizes a safety formula under~$A$ if and only if it does so
  under~$\overline{A}$: liveness guarantees of an adversary, such as eventual stabilization,
  do not help for safety specifications.
If~$\pi(A)$ is not nowhere dense, then there is a finite graph history~$h$
  all of whose extensions are prefixes of histories in~$A$, which means that~%
  $h\cdot\Gamma^\omega \subseteq \overline{A}$.
Guarding the formula by
  ``if the graph history starts with~$h$, then the formula holds from round~$\lvert h\rvert+1$ on''
  reduces synthesis under the full adversary to synthesis under~$A$, yielding the undecidability
  direction of \Cref{thm:undecidability:adversary}.

\paragraph{Counters require growing uncertainty about the graphs}
For decidability, we organize connected subgraphs of complexes into a tree
  and bound their size while preserving the existence of a realizing algorithm.
Without inputs and under the full adversary, it is a path with~$3^r$ edges, one for each
  graph history in~$\Gamma^r$.
If~$\pi(A)$ is not nowhere dense, for which we show undecidability,
  the complex under~$A$ contains arbitrarily long subpaths,
  which can store the unbounded counters.
If~$\pi(A)$ is nowhere dense, the path breaks up into connected components of uniformly bounded size,
  and our decision procedure exploits this fragmentation.
In fact we show that this breaking-into-pieces is directly related to the consensus problem,
  showing a new relation between consensus and synthesis that was previously unknown.

\paragraph{Relating nowhere denseness to repeated consensus solvability}
Consensus is solvable under~$A$ if and only if
  $\pi(A)\neq[0,1]$~\cite{coutouly_topology_2023}.
A bounded-time algorithm cannot tell~$A$ from its closure, and compactness
  turns termination under a closed adversary into a uniform round bound.
Thus consensus is solvable in bounded time if and only if~$\pi(A)$ is not dense.
Removing a finite prefix amounts to rescaling its interval to~$[0,1]$,
  so consensus is solvable in bounded time after every live prefix
  if and only if~$\pi(A)$ is nowhere dense.
For an $\omega$-regular adversary, finitely many distinct left quotients
  yield a uniform bound~$D$, computable by exhaustive search.
The same bound allows agreement on a broadcaster after each prefix.
We use this agreement to show that all histories in a component at
  round~$r\geq D$ share the view of one process~$p$ at round~$r-D$,
  and the other process has heard from~$p$ since then.
In this sense, the processes recover a form of hierarchical information,
  with a delay of~$D$ rounds and a hierarchy that can change over time.

\paragraph{Synthesis via trees of components}
We represent an algorithm by a tree whose nodes store connected subgraphs of the
  view complex at rounds~$nD$, with automaton states on their edges.
At each node, we specify what each process outputs in the next~$D$ rounds for
  every possible local view.
After these~$D$ rounds, each connected component under~$A$ becomes a child node.
Different components remain disjoint under extension, so their output choices
  cannot conflict on the same local view.
A branch follows a sequence of components.
Based on this, we define a run as a sequence of edges along a branch.
The tree accepts if every run satisfies the parity condition for the guarded
  objective: the graph history lies outside~$A$, or the execution satisfies~$\phi$.
We then show that accepting component trees correspond exactly to realizing algorithms.
However, this approach faces a challenge: inputs can make their components arbitrarily
  large, \emph{e.g.}, under~$A=\{\rlink^\omega\}$ and a fixed input sequence of process~1,
  the round-$r$ component is a star with one edge for every input sequence of process~2.

\paragraph{Decidability via reductions}
To address this, we replace each component~$K$ by a subgraph of bounded size,
  using a reduction~$r\colon K\to K$ that preserves the state labels.
The next block then starts from~$r(K)$.
Using the common earlier view, we reduce each component to a bounded number of
  representative edges while preserving local indistinguishability and state labels.
The main difficulty is preserving acceptance.
Repeated replacements can create runs that do not correspond to executions
  of the original algorithm.
We choose representatives of minimum ordinal rank to preserve parity,
  following the progress-measure argument of Berwanger, Mathew, and
  Ramanujam~\cite{berwanger2016retraction}.
The resulting components have uniformly bounded size, so only finitely
  many labeled components, reductions, and block transitions remain
  up to isomorphism.
We express local consistency and acceptance of every compatible run in
  \ac{MSO}.
Rabin's theorem~\cite{rabin1969decidability} then decides whether an accepting
  tree exists and, if so, provides a finite description of one.
From this tree, we construct a realizing distributed algorithm.

\section{Model and Problem}

\subsection[Model]{Model\ \leanformalized{\leanurl{Model}}}

Fix the set of \textit{processes} to be~$\mathcal{P} := \left\{ 1, 2 \right\}$.
Every process~$p \in \mathcal{P}$ is assigned finite nonempty sets~$X_p$ and~$Y_p$ of possible \textit{inputs} and \textit{outputs} respectively.
We consider systems where computation and communication proceed in \textit{rounds}.
In round~$r \ge  1$, process~$p \in \mathcal{P}$ obtains input~$x_p[r] \in X_p$ from the \textit{environment} (different from the processes), sends and receives messages, does a compute step, and finally outputs~$y_p[r] \in Y_p$.
We allow processes to send all their knowledge, including their own inputs and messages received so far, without a bound on message size.
The value of~$y_p[r]$ depends on the knowledge acquired by~$p$ through round~$r$ and is formalized in the next section.

The set of possible \textit{communication graphs} is $\Gamma := \left\{\rlink, \dlink, \llink \right\}$.
A \textit{graph history}~$h \in \Gamma^* \cup \Gamma^\omega$ is a finite or infinite sequence of communication graphs.
Its $r$-th graph, denoted~$h[r]$, is the communication graph of round~$r$.
If~$h[r] = \rlink$, then only process 2 hears from process 1 in round~$r$; if~$h[r] = \llink$, then only process 1 hears from process 2; and if~$h[r] = \dlink$, then both processes hear from each other.

A \textit{message adversary}~$A \subseteq \Gamma^\omega$ describes the set of possible infinite graph histories that can occur in a system.
We call the message adversary~$A = \Gamma^\omega$ the \emph{full} message adversary.

Given a process~$p\in\mathcal{P}$, we write~$\bar{p} := 3 - p$ for the other process.

\subsection{Histories and Indistinguishability Relations}

Define the set of \textit{input signals} to be $\Sigma := X_1 \times X_2 \times \Gamma$.
An input signal $(x_1, x_2, \graph)$ occurring in round~$r$ means that process~$p$ receives input~$x_p$ from the environment and that the communication graph of round~$r$ is~$\graph$.
An \textit{input signal history}~$h \in \Sigma^* \cup \Sigma^\omega$ is a finite or infinite sequence of input signals.

For a finite history~$u$ and a finite or infinite history~$w$ over the same alphabet,
  we write~$u\cdot w$ or simply~$uw$ for their concatenation.
We write~$|h|$ for the length of a finite history~$h$.
For a finite or infinite word~$h$ and an integer~$r\geq1$, we write~$h[r]$ for its $r$-th letter and $h[r\dots]$ for the suffix beginning with that letter.
For an integer~$r\geq0$, we write $h[\dots r]$ for its prefix of length~$r$, with $h[\dots 0]=\varepsilon$.
When~$h$ is finite, these notations require~$r\leq |h|$.
For integers~$0\leq a\leq b$, we write $\Sigma^{[a,b]} := \bigcup_{t=a}^b \Sigma^t$.
For a language~$L$ of infinite words, let~$\Pref(L)$ denote its set of finite prefixes and~$\Pref_r(L)$ its set of prefixes of length~$r$.

We assume that messages contain the sender's current input and its entire previous \textit{view}.
A view records all information acquired through local inputs and received messages.
Once communication is described in this way, a distributed algorithm is determined by its choice of output for each possible local view.
In particular, histories that give a process the same view must lead that process to produce the same output, even when those histories require different behavior from a process with access to the global execution.

Given a history~$h \in \Sigma^*$, the view of process~$p$ at the end of~$h$, denoted by~$v_p(h)$, is a labeled tuple defined recursively.
Symbols~$\mathsf{I}_1, \mathsf{I}_2$ refer to the initial views of processes 1 and 2, which are simply their own process identifiers.
Symbols~$\mathsf{N}_p, \mathsf{C}_p$ are taken to mean no-communication and communication respectively and are used to label the tuples.
For~$h = \varepsilon$, define~$v_p(h) := \mathsf{I}_p$.
Otherwise suppose that $h = h'\cdot (x_1, x_2, \graph)$, \emph{i.e.}, $h$ has~$h'$ as a prefix and $(x_1, x_2, \graph)$ as its last letter.
Suppose that~$v_{1}(h') = u$ and~$v_{2}(h') = v$.
Table~\ref{tab:ViewDefn} defines~$v_p(h)$.

\begin{table}
    \centering
    \begin{tabular}{c|c c}
        $v_p(h)$ & $p = 1$ & $p = 2$ \\
        \hline
        $\graph = \rlink$ & $\mathsf{N}_1(x_1, u)$ & $\mathsf{C}_2(x_2, v, x_1, u)$ \\
        $\graph = \dlink$ & $\mathsf{C}_1(x_1, u, x_2, v)$ & $\mathsf{C}_2(x_2, v, x_1, u)$\\
        $\graph = \llink$ & $\mathsf{C}_1(x_1, u, x_2, v)$ & $\mathsf{N}_2(x_2, v)$ \\
    \end{tabular}
    \caption{Recursive view updates for $h=h'\cdot(x_1,x_2,\graph)$, with current inputs~$x_1,x_2$, communication graph~$\graph$, and previous views $u:=v_1(h')$, $v:=v_2(h')$.}
    \label{tab:ViewDefn}
\end{table}

Let $\Views_p := \left\{ v_p(h) \mid h \in \Sigma^* \right\}$ be the set of possible views of process~$p$.
Define histories~$h_1$ and~$h_2$ to be indistinguishable to process~$p$, or~$h_1 \sim_p h_2$, if $v_p(h_1) = v_p(h_2)$.

If two histories are indistinguishable to a process in round~$r + 1$, it must be so in round~$r$ as well:
\begin{lemma}[\leanproof{\leanurl{Indistinguishability}}]
    \label{lem:IndistDownClosed}
    Suppose that $h = h_0\cdot (x_1, x_2, \graph)$ and $h' = h_0' \cdot (x_1', x_2', \graph')$.
   If~$h \sim_p h'$ then~$h_0 \sim_p h_0'$, and moreover~$p$ has an incoming arrow in both~$\graph$ and~$\graph'$ or it does not have an incoming arrow in either graph.
\end{lemma}
\begin{proof}
    Assume that~$p = 1$; the case~$p = 2$ follows similarly.
    The proof that~$h_0 \sim_1 h_0'$ follows from the fact that each of $\textsf{N}_1(x_1, u) = \textsf{N}_1(x_1', u')$ and $\textsf{C}_1(x_1, u, x_2, v) = \textsf{C}_1(x_1', u', x_2', v')$ implies that~$u = u'$.

    If $v_1(h) = v_1(h')$ are of the form~$\textsf{N}_1(\cdot, \cdot)$, then~$\graph$ and~$\graph'$ are both of the form~$\rlink$.
    Otherwise they are of the form $\textsf{C}_1(\cdot, \cdot, \cdot, \cdot)$, in which case~$\graph$ and~$\graph'$ belong to~$\left\{ \dlink, \llink \right\}$, proving the latter claim.
\end{proof}

\subsection[Lossy-Link Synthesis Problem]{Lossy-Link Synthesis Problem\ \leanformalized{\leanurl{ProblemStatement}}}

We study the problem of synthesizing a distributed algorithm whose behavior is specified by an \ac{LTL} formula.
The formulas can talk about process inputs, process outputs, and communication graphs at each round.
The syntax and semantics are given in Appendix~\ref{app:ltl}.

In principle, the graph history in an execution is arbitrary from~$\Gamma^\omega$, modeling a highly dynamic network without any restrictions on stability, eventual convergence, \emph{etc.}
The specification is allowed to speak about the communication graphs along a history, however, with the possibility to restrict the behavior of the dynamic network and require process behavior to depend on the network dynamics.

A \textit{distributed algorithm}~$f$ is a pair~$(f_1, f_2)$ of functions $f_1\colon \Views_1 \to Y_1$ and $f_2\colon \Views_2 \to Y_2$ that associate with each view an output.
Given~$h\in\Sigma^\omega$, define $h_r:=h[\dots r]$.
Let~$\Omega := Y_1 \times  Y_2$ be the set of \textit{output signals}.
The execution of~$f$ on~$h$, denoted $\exec(f, h)$, has alphabet~$\Sigma \times \Omega$ with~$r$\textsuperscript{th} letter
\[
    \left( h[r], \left(y_1[r], y_2[r] \right) \right)\enspace,
\] 
where $y_1[r] := f_1(v_1(h_r))$ and $y_2[r] := f_2(v_2(h_r))$.

Define $\Graph\colon \Sigma^\omega \to \Gamma^\omega$ to project input signal histories to their graph histories.
For~$h\in\Sigma^\omega$ and~$r\geq1$, we have $\Graph(h)[r]=\graph$ whenever $h[r]=(x_1,x_2,\graph)$.
We extend~$\Graph$ to finite input signal histories by the same round-wise projection, with $\Graph(\varepsilon)=\varepsilon$.
For an execution~$e\in(\Sigma\times\Omega)^\omega$, we likewise write~$\Graph(e)$ for its graph history, ignoring inputs and outputs.

For a message adversary~$A$ and a language~$L \subseteq (\Sigma \times \Omega)^\omega$, we say that~$f$ \emph{LL-realizes}~$L$ under~$A$ if $\exec(f,h)\in L$ for every~$h\in\Sigma^\omega$ with $\Graph(h)\in A$.
The language~$L$ is \emph{LL-realizable} under~$A$ if such an algorithm exists.
For an \ac{LTL} formula~$\phi$, these notions apply to $L(\phi):=\{e\in(\Sigma\times\Omega)^\omega\mid e\models\phi\}$.
Unless an adversary is specified, realizability refers to the full adversary~$\Gamma^\omega$; we henceforth omit the prefix ``LL-''.

The synthesis decision problem~$\textsc{LL-Synthesis}(A)$~\leanformalized{\leanline{Formalizations/Paper/ProblemStatement}{47}} asks, given an \ac{LTL} formula~$\phi$, whether~$\phi$ is realizable under~$A$.
We write~$\textsc{LL-Synthesis}$~\leanformalized{\leanline{Formalizations/Paper/ProblemStatement}{57}} for the case~$A=\Gamma^\omega$.

\subsection{View Complexes}
\label{sec:view-complexes}

View complexes represent the processes' indistinguishability relations as a graph.
Each vertex represents a possible view of one process (we then say that said process \emph{owns} that vertex), and an edge joins two views that can occur together.
Histories whose edges share a vertex are indistinguishable to the process owning that vertex.
A path consequently represents a sequence of histories linked by individual indistinguishability.
Since we have only two processes, the view complex is a graph rather than a higher-dimensional simplicial complex~\cite{HKR-book}.

Let~$A$ be a message adversary and let~$r\geq0$ be a round number.
Let $\Views_p^{(A)}(r) := \{ v_p(h) \mid h \in \Sigma^r,\ \Graph(h)\in\Pref(A) \}$ be the set of possible views of process~$p$ under~$A$ after~$r$ rounds.
The view complex under~$A$ after round~$r$ is defined as the graph with the vertex set $\Views_1^{(A)}(r) \cup \Views_2^{(A)}(r)$ and the edges $\{ v_1(h) , v_2(h) \}$ where~$h\in\Sigma^r$ and $\Graph(h)\in\Pref(A)$.
The uncertainty of process~$p$ about the state of the other process~$\bar{p}$ is formalized as the set of views adjacent to its own.

Write $e(h):=\{v_1(h),v_2(h)\}$ for the edge associated with a history~$h$.
The pair of views determines the history: by \Cref{tab:ViewDefn}, it determines
  the last communication graph, both inputs, and both preceding views.
Repeating this reconstruction determines every letter of~$h$.
Thus distinct histories give distinct edges.

\begin{definition}[View complex with output\ \leanformalized{\leanline{Formalizations/Paper/Model}{95}}]\label{def:view-complex-output}
    The view complex after round~$r$ is \emph{labeled with output} if each vertex of process~$p$ is labeled with a value from~$Y_p$.
    A distributed algorithm~$f$ assigns the output label~$f_p(u)$ to every vertex~$u$ of process~$p$.
\end{definition}

Conversely, define~$f_p(u)$ as the label of the process-$p$ vertex~$u$.
This is well-defined across rounds because a view determines its process and round.
Assign arbitrary outputs to views that cannot occur under~$A$.

\subsection{Representing Graph Histories in the Unit Interval}
\label{sec:graph-history-interval}

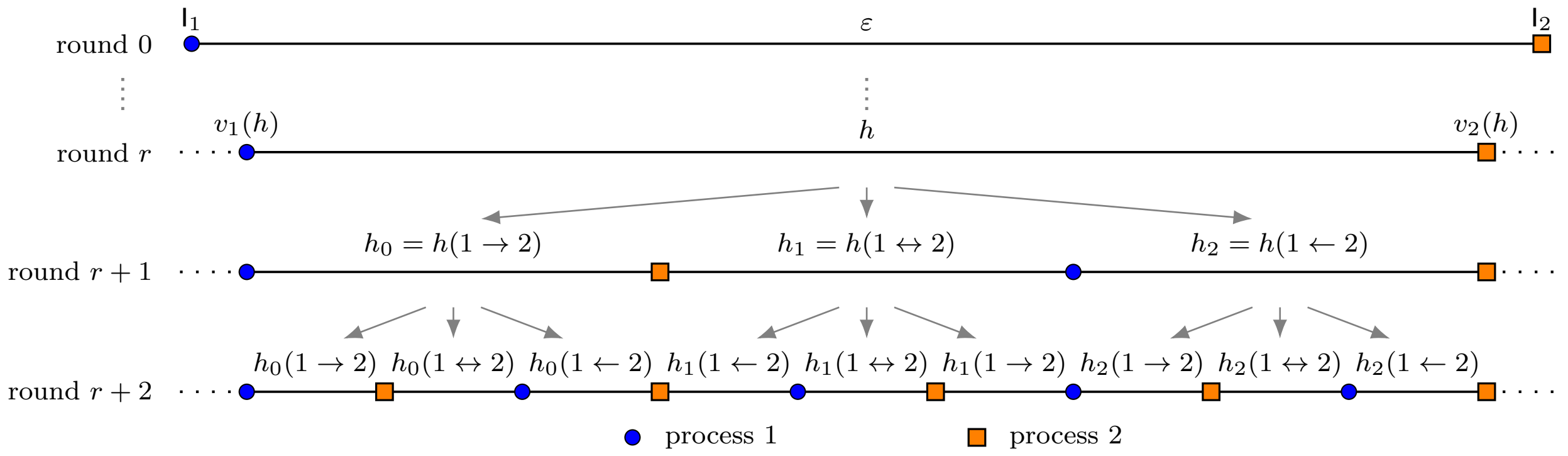
\begin{figure}
    \centering
\begin{tikzpicture}[
  x=1.45cm,y=.6cm,
  font=\footnotesize,
  process one/.style={circle,draw=black,fill=blue,
    minimum size=4.5pt,inner sep=0pt},
  process two/.style={rectangle,draw=black,fill=orange,
    minimum size=5pt,inner sep=0pt,line width=.6pt},
  edge label/.style={anchor=south,inner sep=1pt},
  position label/.style={anchor=north,inner sep=1pt},
  refinement/.style={-{Latex[length=2mm]},draw=black!50,line width=.5pt},
  group brace/.style={decorate,
    decoration={brace,mirror,amplitude=4pt},line width=.5pt}
]
  \draw[line width=.7pt] (-0.4,6.1) -- (9.4,6.1);
  \draw[line width=.7pt] (0,4.2) -- (9,4.2);
  \draw[line width=.7pt] (0,2.1) -- (9,2.1);
  \draw[line width=.7pt] (0,0) -- (9,0);
  \node[anchor=east] at (-.6,6.1) {round~$0$};
  \node[anchor=east] at (-.6,4.2) {round~$r$};
  \node[anchor=east] at (-.6,2.1) {round~$r + 1$};
  \node[anchor=east] at (-.6,0) {round~$r + 2$};

  \draw[loosely dotted, thick] (9, 0) -- (9.5, 0);
  \draw[loosely dotted, thick] (9, 2.1) -- (9.5, 2.1);
  \draw[loosely dotted, thick] (9, 4.2) -- (9.5, 4.2);

  \draw[loosely dotted, thick] (0, 0) -- (-0.5, 0);
  \draw[loosely dotted, thick] (0, 2.1) -- (-0.5, 2.1);
  \draw[loosely dotted, thick] (0, 4.2) -- (-0.5, 4.2);

  \node[process one] at (-0.4, 6.1) {};
  \node[process two] at (9.4, 6.1) {};
  \node[edge label] at (-0.4,6.27) {$\mathsf{I}_1$};
  \node[edge label] at (9.4,6.27) {$\mathsf{I}_2$};
  \node[edge label] at (4.5,6.27) {$\varepsilon$};
  \draw[draw=black!50,line width=.9pt, dotted] (4.5, 5.5) -- (4.5, 4.9);
  \draw[draw=black!50,line width=.9pt, dotted] (-.9, 5.5) -- (-.9, 4.9);

  \node[process one] at (0,4.2) {};
  \node[process two] at (9,4.2) {};
  \node[edge label] at (0,4.37) {$v_1(h)$};
  \node[edge label] at (9,4.37) {$v_2(h)$};
  \node[edge label] at (4.5,4.37) {$h$};
  \draw[refinement] (4.5,3.58) -- (4.5,3.03);
  \draw[refinement] (4.3,3.58) -- (1.7,3.03);
  \draw[refinement] (4.7,3.58) -- (7.3,3.03);

  \foreach \j in {0,1,2,3} {
    \ifodd\j
      \node[process two] at (3*\j,2.1) {};
    \else
      \node[process one] at (3*\j,2.1) {};
    \fi
  }
  \node[edge label] at (1.5,2.27) {$h_0 = h\rlink$};
  \node[edge label] at (4.5,2.27) {$h_1 = h\dlink$};
  \node[edge label] at (7.5,2.27) {$h_2 = h\llink$};

  \foreach \x in {1.5,4.5,7.5} {
    \draw[refinement] (\x,1.48) -- (\x,.93);
    \draw[refinement] (\x-.2,1.48) -- (\x-0.8,.93);
    \draw[refinement] (\x+.2,1.48) -- (\x+0.8,.93);
  }

  \foreach \j in {0,...,9} {
    \ifodd\j
      \node[process two] at (\j,0) {};
    \else
      \node[process one] at (\j,0) {};
    \fi
  }
  \foreach \x/\g in {
    .5/h_0\rlink,1.5/h_0\dlink,2.5/h_0\llink,
    3.5/h_1\llink,4.5/h_1\dlink,5.5/h_1\rlink,
    6.5/h_2\rlink,7.5/h_2\dlink,8.5/h_2\llink} {
    \node[edge label] at (\x,.17) {$\g$};
  }

  \node[process one] at (2.8,-.8) {};
  \node[anchor=west] at (2.95,-.8) {process~$1$};
  \node[process two] at (5.3,-.8) {};
  \node[anchor=west] at (5.45,-.8) {process~$2$};
\end{tikzpicture}
    \caption{Successive subdivisions of the inputless view complex. Each history has three extensions; their order reverses when the endpoint owners are exchanged.}
    \label{fig:ViewPathRefinement}
\end{figure}

To represent communication patterns geometrically, we erase input values from views.
Under the full adversary, the round-$r$ complex is a path with~$3^r$ edges.
Embed the initial edge as~$[0,1]$, with process~$1$ on the left; each round trisects every edge, as in \Cref{fig:ViewPathRefinement}.
This is iterated chromatic subdivision~\cite{HKR-book}, with the index and limit map of Godard and Perdereau~\cite{godard_back_2020}; see also the geometrization framework of Coutouly and Godard~\cite{coutouly_topology_2023}.
The construction is detailed in \Cref{sec:inputless-view-properties}.

For~$h\in\Gamma^r$, let $j(h)\in\{0,\dots,3^r-1\}$ be the index of its edge, numbered from left to right starting at~$0$; in particular, $j(\varepsilon)=0$.
Its interval is
\begin{equation}
    J(h):=\left[\frac{j(h)}{3^{\lvert h\rvert}},\frac{j(h)+1}{3^{\lvert h\rvert}}\right]\enspace.
    \label{eqn:J}
\end{equation}
For~$h\in\Gamma^\omega$, the nested intervals $J(h[\dots r])$ have lengths~$3^{-r}$.
Their unique common point defines~$\pi(h)$.

\begin{lemma}[\leanproof{\leanurl{CellInclusion}}]\label{J:inclusion:iff:prefix}
    For~$h,k\in\Gamma^*$, $h$ is a prefix of~$k$ if and only if $J(k)\subseteq J(h)$.
\end{lemma}

We equip~$\Gamma^\omega$ with the product topology for discrete~$\Gamma$; the cylinders~$h\cdot\Gamma^\omega$, for~$h\in\Gamma^*$, form a basis of clopen (open and closed) sets.
For~$B\subseteq X$, write~$\overline B$ and~$B^\circ$ for its closure and interior in~$X$.
The set~$B$ is \emph{dense} in~$X$ if~$\overline B=X$, and \emph{nowhere dense} if~$(\overline B)^\circ=\emptyset$.
In~$\Gamma^\omega$, closed sets are safety properties and dense sets are liveness properties~\cite[p.~184]{alpern_defining_1985}.

\begin{lemma}[\leanproof{\leanurl{GeometricMap}}]\label{pi:cont}
    The map $\pi\colon\Gamma^\omega\to[0,1]$ is continuous and surjective.
\end{lemma}

It is not injective: both~$\rlink\cdot\llink^\omega$ and~$\dlink\cdot\llink^\omega$ map to~$1/3$.

\section{Proof Overview}

We begin with the connection between repeated consensus and topology, then
  outline the undecidability reduction and the decision procedure.
\Cref{sec:inputless-view-properties,sec:consensus-details} develop the geometry
  of view complexes and the consensus characterization.
The detailed synthesis arguments appear in
  \Cref{sec:undecidability-lemmas,sec:tree-decidability}.

\subsection{Repeated Consensus}
\label{sec:consensus}

Fix a finite proposal set~$V$ with~$\lvert V\rvert\geq2$, and take~$X_1=X_2=V$ and $Y_1=Y_2=V\cup\{\bot\}$, where~$\bot$ denotes no decision.
A consensus algorithm under~$A$ eventually makes both processes decide irrevocably on the same one of their round-$1$ proposals in every execution with graph history in~$A$; later inputs remain arbitrary.

\begin{lemma}[{\cite[Theorem~26]{coutouly_topology_2023}}\ \leanproof{\leanurl{Consensus}}]\label{cons:char}
    Consensus is solvable under message adversary~$A$ if and only if $\pi(A) \neq [0,1]$.
\end{lemma}

Consensus is solvable in \emph{bounded time} if both processes decide by a common round bound~$R$ in every execution.
For $h\in\Pref(A)$, define $h^{-1}A:=\{k\in\Gamma^\omega\mid h\cdot k\in A\}$.
Consensus is \emph{repeatedly solvable in bounded time} if it is solvable in bounded time under every~$h^{-1}A$, for $h\in\Pref(A)$, allowing the algorithm and bound to depend on~$h$.
If one bound~$R$ works for all prefixes, it is repeatedly solvable in \emph{uniformly bounded time}.

\begin{lemma}[\leanproof{\leanurl{RepeatedConsensus}}]\label{lem:repeated_bounded_time_consensus}
    Consensus is repeatedly solvable in bounded time under message adversary~$A$ if and only if~$\pi(A)$ is nowhere dense.
    In this case, if~$A$ is $\omega$-regular, consensus is repeatedly solvable in uniformly bounded time.
\end{lemma}

A process is a \emph{broadcaster} for a finite graph history if at least one of its messages reaches the other process.
Agreement on a broadcaster lets both processes decide on its proposal; conversely, consensus yields agreement on a broadcaster within the same round bound~\cite[Theorem~7.12]{NSW24}.

\begin{lemma}[\leanproof{\leanurl{Broadcaster}}]\label{lem:cons:means:broadcaster}
    Let~$A$ be a message adversary under which consensus is solvable within~$R$ rounds.
    Then there is a distributed algorithm that makes both processes decide at round~$R$ on a common process~$p$, which is a broadcaster in the first~$R$ rounds.
\end{lemma}

\subsection{Undecidability Result}
\label{sec:undecidability}

We outline the counter-machine reduction proving \Cref{thm:undecidability}.

A \emph{(deterministic) $n$-counter machine} is a tuple $M = (Q, \Delta, q_\init, F)$
with finite state set~$Q$, initial state~$q_\init$, set of halting states~$F \subseteq Q$,
and counters $\countsymb_1, \dots, \countsymb_n$. The transition function~$\Delta$ assigns to each~%
$q \in Q \setminus F$ either an \emph{increment}
\texttt{increment$(\countsymb_i)$; goto~$q'$}, or a \emph{decrement}
\texttt{if~$\countsymb_i = 0$ then goto~$q'$ else decrement$(\countsymb_i)$; goto~$q''$}.
Configurations $(q, x_1, \dots, x_n) \in Q \times \NN^n$ evolve accordingly, and~$M$
\emph{halts} if its run from $(q_\init, 0, \dots, 0)$ reaches a state in~$F$.
Halting is undecidable for 2-counter machines~\cite[p.~255--258]{minsky1967}.

We show that the non-halting problem for 2-counter machines can be reduced to the $\textsc{LL-Synthesis}$ problem.
We will then show that it can be further reduced to the $\textsc{LL-Synthesis}(A)$ problem where~$A$ is omega-regular and~$\pi(A)$ is not nowhere-dense, proving one direction of~\Cref{thm:undecidability:adversary}.

Throughout this reduction, the input sets~$X_1$ and~$X_2$ are singletons, so we identify~$\Sigma$ with~$\Gamma$ by omitting the fixed input values.
Specifications are therefore interpreted over~$(\Gamma\times\Omega)^\omega$, where $\Omega=Y_1\times Y_2$.
The reduction constructs output sets~$Y_1$ and~$Y_2$ of finite size.%

To ease the presentation, we explain the reduction from a $1$-counter machine. 
The adaptation to a 2-counter machine is straightforward.
Let $M_{\single} = (Q, \Delta, q_\init, F, \left\{\countsymb_1 \right\})$ be a 1-counter machine.
Since the output sets~$Y_1, Y_2$ have to be finite, having an unbounded counter means that an \ac{LTL} formula trying to simulate~$M_{\single}$ cannot talk about the exact counter values of the machine configurations.
The insight here is to use the view complexes of an algorithm realizing $\phi(M_\single)$ and distribute the counter value at step~$r$ over the view complex at round~$r+1$.

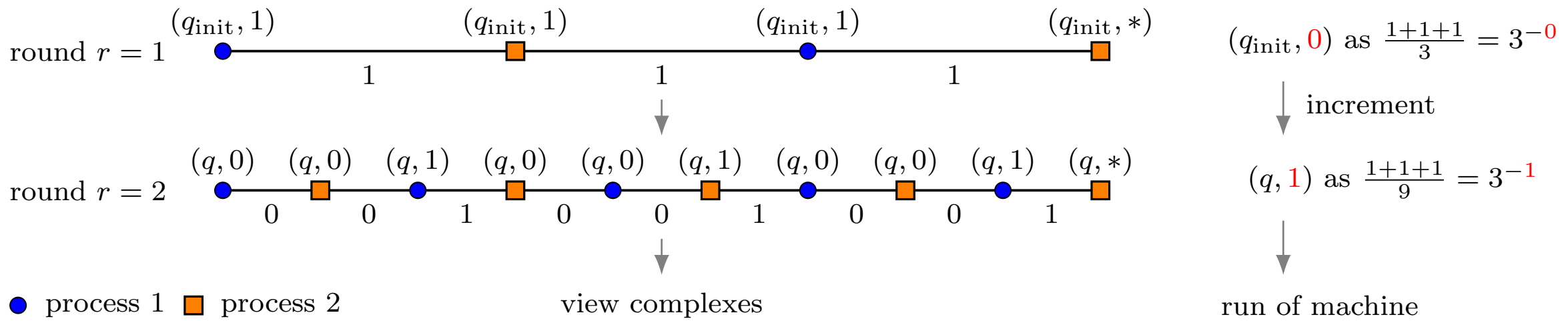
\begin{figure}
    \centering
\begin{tikzpicture}[
  x=1.45cm,y=.6cm,
  font=\footnotesize,
  process one/.style={circle,draw=black,fill=blue,
    minimum size=4.5pt,inner sep=0pt},
  process two/.style={rectangle,draw=black,fill=orange,
    minimum size=5pt,inner sep=0pt,line width=.6pt},
  edge label/.style={anchor=south,inner sep=1pt},
  position label/.style={anchor=north,inner sep=1pt},
  refinement/.style={-{Latex[length=2mm]},draw=black!50,line width=.5pt},
  group brace/.style={decorate,
    decoration={brace,mirror,amplitude=4pt},line width=.5pt}
]
  \draw[line width=.7pt] (0,2.3) -- (6,2.3);
  \draw[line width=.7pt] (0,0) -- (6,0);
  \node[anchor=east] at (-.3,2.3) {round~$r=1$};
  \node[anchor=east] at (-.3,0) {round~$r=2$};


  \foreach \j in {0,1,2,3} {
    \ifodd\j
      \node[process two] at (2*\j,2.3) {};
    \else
      \node[process one] at (2*\j,2.3) {};
    \fi
  }
  \node[position label] at (1,2.13) {$1$};
  \node[position label] at (3,2.13) {$1$};
  \node[position label] at (5,2.13) {$1$};

  \node[edge label] at (0,2.47) {$(q_\init, 1)$};
  \node[edge label] at (2,2.47) {$(q_\init, 1)$};
  \node[edge label] at (4,2.47) {$(q_\init, 1)$};
  \node[edge label] at (6,2.47) {$(q_\init, *)$};


  \foreach \j in {0,...,9} {
    \ifodd\j
      \node[process two] at (\j*2/3,0) {};
    \else
      \node[process one] at (\j*2/3,0) {};
    \fi
  }
  \foreach \x/\g in {
      0/0, 1/0, 2/1,
      3/0, 4/0, 5/1,
      6/0, 7/0, 8/1
    } {
    \node[position label] at (1/3+\x*2/3,-.17) {$\g$};
  }

  \foreach \x/\g in {
    0/{(q,0)}, 1/{(q, 0)}, 2/{(q, 1)},
    3/{(q,0)}, 4/{(q, 0)}, 5/{(q, 1)},
    6/{(q,0)}, 7/{(q, 0)}, 8/{(q, 1)},
    9/{(q,*)}
  }{
    \node[edge label] at (\x*2/3,.17) {$\g$};
  }

  \node[process one] at (-1.4,-1.9) {};
  \node[anchor=west] at (-1.30,-1.9) {process~$1$};
  \node[process two] at (-.2,-1.9) {};
  \node[anchor=west] at (-.10,-1.9) {process~$2$};

  \node[edge label] at (8,2.1) {$(q_\init, \textcolor{red}{0})$ as $\frac{1 + 1 + 1}{3} = 3^{-\textcolor{red}{0}}$};
  \node[edge label] at (8, -0.2) {$(q, \textcolor{red}{1})$ as $\frac{1 + 1 + 1}{9} = 3^{-\textcolor{red}{1}}$};

  \draw[refinement]  (3, 1.5) -- (3, 0.9);
  \draw[refinement]  (3, -0.8) -- (3, -1.4);

  \draw[refinement]  (7.25, 1.8) -- (7.25, 0.9);
  \node[edge label] at (7.85, 1.2) {increment};
  \draw[refinement]  (7.25, -0.5) -- (7.25, -1.4);

  \node at (3, -1.9) {view complexes};
  \node at (7.5, -1.9) {run of machine};

\end{tikzpicture}
    \caption{The basic idea for encoding configurations with one counter.
       Actual process outputs will contain more parameters.    }
    \label{fig:example_one_counter}
\end{figure}

The encoding of the machine configurations in the view paths is implemented as follows.
Consider the sequence of view complexes of an algorithm~$f$ realizing $\phi(M_\single)$.
Every vertex is required to output an individual binary value for its right edge. %
The following invariant has to be satisfied when encoding a counter value into the line of a view complex:
\emph{The fraction of edges at round~$r$ with non-zero weights is~$3^{-c}$, where~$c$ is the counter value at step~$r-1$ in the run of~$M_{\single}$.}
Moreover, every vertex at round~$r$ should output the state of~$M_{\single}$ at round~$r-1$.
One thus obtains a sequence of view complexes with output labels that simulate a run of~$M_{\single}$. See Figure~\ref{fig:example_one_counter} for an example.

Other issues that $\phi(M_\single)$ needs to address are: $(i)$ ensuring that the view complex simulation correctly 
detects whether counter~$\countsymb_1$ is zero or not, $(ii)$  the exact manner of encoding the counter value at each round so that it can be locally checked.

We now formally specify the output sets and the formula $\phi(M_{\single})$.
Every vertex outputs a tuple of the form 
    $(q, \zero, \rightplan, \carry, \pos)$,
consisting of a control state~$q \in Q$, a \emph{zero indicator} variable $\zero \in \left\{ \top, \bot \right\}$ claiming
whether~$\countsymb_1$ is zero or not, %
a tuple~$\rightplan$ for its right edge, variables $\carry \in \left\{ 0, \dots, 8 \right\}$ and $\pos \in \left\{ 0, 1, 2 \right\}$.

A \emph{plan} is a tuple which specifies the weights of the current (right) edge and its three child edges from left to right, denoted respectively~$w\in\{0,1\}$ and $a_0, a_1, a_2\in\{0,1\}$.

The~$\pos$ variable is meant to be equal to the vertex's position in its view path modulo 3.
The modulo operator is needed in order to keep the output set finite.

The~$\carry$ variable is to help a vertex determine how many child edges in a plan should have non-zero weight, in the following way.
The number of edges in the view complex at round~$r+1$ is three times the number of edges in the view complex at round~$r$.
When the counter is unchanged (respectively incremented), the number of weight-1 edges must be multiplied by 3 (respectively
stay constant), and this could be checked edge by edge.
Implementing a decrement is more complicated, since it requires multiplying the total number of edge weights with value 1 by 9. This cannot
be checked only by comparing the value of an edge weight and those of its children, and will be enforced through the carries. Assume an edge weight is 0 and
its children have weight 0, 1, 1. If we focus only on the current edge, nine times its weight is still 0. At the next level though, the total number
of edges with value 1 will be 2. There is an excess of two, compared to the expected value, which will be \emph{carried} to the neighboring edge. For each
plan, if the current weight is 1, the expected total value of edge weights at the next level is 9. If the carry computed by its left neighbor is 7, then it is enough for its
own child edges to have weights 0, 1, 1. This meets the requirement of multiplying the number of weight-1 edges by 9. If the child edges have
instead weights 1, 1, 1, then there is an excess of 1, that will be \emph{carried over} to the right neighbor. So the 
carry is the cumulative excess, scanning the path from left to right, of children of weight 1 produced so far, over those required so far. It must stay between 0 and 8, and  must be equal to 0 at both ends, so that the computation does not
start with undue carry, and no carry is left over at the end of the line. 
To be uniform, the same mechanism is used for the other operations, with factors 1 and 3 instead of 9.

\paragraph[Realizing phi\_single means satisfying the conditions]{Realizing $\phi(M_\single)$ means satisfying conditions on the view complex}
Any algorithm~$f$ realizing $\phi(M_\single)$ needs to satisfy locally checkable conditions (C0)--(C5) on its
view complex, which we explain below.

 We will use the following shorthands for the \ac{LTL} formulas: instead of writing~$\outp_p=y$, since the outputs are tuples, we will write~%
$q_p=q$, $\zero_p=z$, and so on, for each component of the output. Moreover, since~$\rightplan$ is also a tuple, we will write
$\rightplan_p.w= x$, $\rightplan_p.a_i=a$, for~$i\in\{0,1,2\}$.
  In the view complex at round~$r$,
we refer to the different components of the output of the~$i^\textit{th}$ vertex (numbered from 0 to~$3^r$, from left to right) by~$q(r,i)$, $\zero(r,i)$, $\carry(r,i)$, and~%
$\pos(r,i)$. Moreover, 
the different components of~$\rightplan$ are referred to by $\rightplan(r,i).w$, $\rightplan(r,i).a_0$,
$\rightplan(r,i).a_1$, $\rightplan(r,i).a_2$.

Each condition below
can be expressed by \ac{LTL} formulas on executions: they are able to relate a specific edge of the view complex 
(\emph{i.e.}, a finite history) with
the outputs of its two endpoints (the corresponding views of the processes). However, the specification must determine which process's view is the left endpoint of the current edge. To this end, each process also outputs the position of its view along the path, modulo 3 to keep the output set finite. The left endpoint is then the vertex whose position precedes the other’s by one, modulo 3.
Three labels suffice because the two possible position differences, $1$ and~$-1$, are distinct modulo~3.

\begin{center}
    \fbox{
        \begin{minipage}{0.97\linewidth}
            (C0) Position Order: \textit{For every round~$r$, $\pos(r,i)=i\bmod 3$.}
             \end{minipage}
    }

\end{center}

We will enforce this by an \ac{LTL} formula thanks to the following inductive reasoning: after a history~$h$, if process 1's view has integer position~$u$
and process 2's view has integer position~$v$, then we can determine their positions at the next round~$h\graph$: observe that going from
the view of process 1 at position~$3u$ to the view of process 2 at position~$3v$, we always encounter the following
sequence of~$\graph$: $\rlink$, then~$\dlink$, then~$\llink$. Since~$v-u=1$ or~$v-u=-1$, depending on whether process
1's view or process 2's view is on the left, the different positions encountered are then~$3u+(v-u)$, $3u+2(v-u)$,
and $3u+3(v-u)=3v$. Since process 1's and process 2's views alternate in the line, this gives the following
inductive computation:

Let $\ZZ_3 := \left\{ 0, 1, 2 \right\}$, and write~$a \bmod 3 \in \ZZ_3$ for the least nonnegative remainder of an integer~$a$ modulo~3.
For~$\graph \in \Gamma$, define $\childpos_\graph\colon \ZZ_3^2 \to \ZZ_3^2$ by
\begin{equation}
    \childpos_\graph(u, v) :=
    \begin{cases}
        (3u \bmod 3, (2u + v) \bmod 3) &\text{if }\graph = \rlink \\
        ((u + 2v) \bmod 3, (2u + v) \bmod 3) &\text{if }\graph = \dlink \\
        ((u + 2v) \bmod 3, 3v \bmod 3) & \text{if }\graph = \llink\enspace.
    \end{cases}
\end{equation}

For~$u, v \in \ZZ_3$ and~$\graph \in \Gamma$, define
\begin{equation}
    \phi(u,v, \graph)
    :=
    (\pos_1, \pos_2) = (u, v)
    \rightarrow
    \ltlX
    \left[
        \link = \graph
        \rightarrow
        (\pos_1,\pos_2) = \childpos_\graph(u,v)
    \right]
\end{equation}

\begin{equation}
    \label{eqn:posord}
    \posord :=
    \bigwedge_{\graph\in\Gamma} \bigl(\link=\graph \rightarrow (\pos_1, \pos_2) = \childpos_\graph(0, 1)\bigr)
    \wedge
    \ltlG 
    \left( \bigwedge_{u\in\ZZ_3}
        \bigwedge_{v\in\ZZ_3}
        \bigwedge_{\graph\in\Gamma}
    \phi(u,v, \graph) \right)
\end{equation}%

Formula~$\posord$ ensures that if the pair of positions of processes 1 and 2 in edge~$h$ is~$(u, v)$, then for every~$\graph \in \Gamma$, the pair of positions in edge~$h\graph$ is given by $\childpos_\graph(u, v)$.

Now that the processes are bound to output the position of their view modulo 3 on the line, it is easy to write an
\ac{LTL} formula to determine whether process 1's view is on the left endpoint of the current edge, or not:
$\oneisleft:= \bigvee_{\substack{(p,q)\in\ZZ_3^2\\ q \equiv p+1 \mod 3}} (\pos_1 = p \wedge \pos_2 = q).$

      \begin{center}
    \fbox{
        \begin{minipage}{0.97\linewidth}
            (C1) Path Agreement: {\itshape For every round~$r$, for all~$0\leq i,j\leq 3^r$, $q(r,i)=q(r,j)$ and $\zero(r,i)=\zero(r,j)$.}
        \end{minipage}
    }\end{center}

The state of the counter machine, as well as the status of the counter with respect to 0 at the~$i^\textit{th}$ step of the computation
is then claimed consistently by all the processes in round~$i+1$. This global constraint on each line of the
view complex is ensured thanks to the indistinguishability relation linking the different edges. It is enough to
require that processes~1 and~2 always agree on these components of their outputs.

The current state and current zero-status of the counter are enough to determine how the counter has to be changed,
 and thus, which multiplication factor has
to be applied to the weights of the view complex. Given~$q\in Q$ and~$z\in\{\top,\bot\}$ the zero-status of the counter, we let~$m(q,z)$ be this factor.
The next condition ensures that the ratio of the sum of edge weights to the total number of edges at each round changes appropriately depending on the operation.
\begin{center}
    \fbox{
        \begin{minipage}{0.97\linewidth}
            (C2) Local Sum Check: \textit{For every round~$r$, for all~$0\leq i\leq 3^r-1$, $
                \sum_{k= 0}^2 \rightplan(r,i).a_k + \carry(r, i) = m(q(r,i), \zero(r,i)) \cdot \rightplan(r, i).w + \carry(r, i+1)$.}
        \end{minipage}
    }
\end{center}
Additionally, we require in the next constraint that $\carry(r,0)=\carry(r,3^r)=0$, ensuring that  the computation does not start with undue
carry, and no carry is left over at the end of the line. Since for all~$0\leq i,j\leq 3^r$, $q(r,i)=q(r,j)=q$ and $\zero(r,i)=\zero(r,j)=z$, thanks to (C1), summing over all the vertices of the line will make carries cancel out and will ensure that
$\sum_{i=0}^{3^r-1} \sum_{k= 0}^2 \rightplan(r,i).a_k = m(q,z)\cdot \sum_{i=0}^{3^r-1} \rightplan(r, i).w$, as required.

Importantly, this can be checked locally on the view complexes of~$f$. It is easy to write an \ac{LTL} formula that,
after any history~$h\in\Gamma^r$, relates the~$\carry$ and~$\rightplan$ outputs of the two processes to ensure the
equation of (C2), using $\oneisleft$. 
As for (C1), this local equation propagates along the line because a process’s carry is the same for its left and right edges.

\begin{center}
    \fbox{
        \begin{minipage}{0.97\linewidth}
            (C3) Zero Claim Check: \textit{For every round~$r\geq 1$, $\carry(r, 0) = \carry(r, 3^r) = 0$. Moreover, if $\zero(r,0)=\top$, then for all~$j \le 3^r - 1$,
            $\rightplan(r, j).w = 1$, and if $\zero(r,0)=\bot$, then $\rightplan(r, 0).w = 0$.      
            }
        \end{minipage}
    }
\end{center}
This condition actually ensures two different things: one is the value of the carries for the extreme vertices, as explained above. The second is to check zero claims made by the processes. If checking that the counter is
0 simply requires that all the weights attributed along the line are 1, checking that it is not requires to
check that at least one of the weights is 0: but how could it be checked on a single run? Observe that for the
decrement mechanism to work, the weight of the first edge is necessarily 0: since
the first vertex receives no carry from the left (and a carry cannot become negative), its three child edges contribute at most 3; if its own
current edge is 1, it will not be possible to meet the requirement. This is why it is safe for the specification
to check the validity of a non-zero claim on the first edge. It remains to explain
how the specification can constrain the outputs of the extreme vertices of the view complex. For that, the $\pos$-component of the output is useless. Instead, we use the fact that the view positioned on the left of every round~$r$
always corresponds to the view of process 1 after the history~$\rlink^r$ (and the view of process 2 after the history~$\llink^r$ for the rightmost vertex). Hence the formula $ \leftmost := \left[ \carry_1 = 0 \wedge \left( \zero_1 = \bot \rightarrow \rightplan_1.w = 0 \right) \right] \ltlW (\link \ne \rlink)$. The other formulas for (C3) are given in
Section~\ref{sec:undecidability-lemmas}
.%

The following condition ensures correct encoding of the initial configuration of the counter machine. 
\begin{center}
    \fbox{
        \begin{minipage}{0.97\linewidth}
            (C4) Initial Check: \textit{For all~$0\leq j\leq 3$, $q(1,j)=q_{\init}$, and for~$0\leq j\leq 2$, $\rightplan(1,j).w=1$.}
        \end{minipage}
    }
\end{center}

Lastly, the algorithm should check that the state values output in each round are correct in accordance with the zero claims and that the weight of each edge is equal to the weight promised in the previous round. For that we
can define the function $\operatorname{next}(q,z)$ that gives the next state given by~$\Delta$, according to the
current state~$q$ and the $\zero$-status of the counter.

\begin{center}
    \fbox{
        \begin{minipage}{0.97\linewidth}
            (C5) Proper Transition: For every round~$r\geq 1$, \textit{$q(r+ 1,0) = \operatorname{next}(q(r,0), \zero(r,0))$ and, for all~$0\leq j\leq 3^{r}-1$, for all~$x\in\{0,1,2\}$,
                $\rightplan(r+1, 3j+x).w = \rightplan(r, j).a_x.$}
        \end{minipage}
    }
\end{center}
When the last edge of the history was~$\dlink$, the current weight has to be compared with the promise~$a_1$ made
at the previous round. When it was~$\llink$ or~$\rlink$ however, it has to be compared with either~$a_0$ or~$a_2$ from the
previous round depending on the orientation of the edge with respect to process 1 and process 2. This can be easily computed and ensured thanks to the formula $\oneisleft$.

Combining conditions (C0)--(C5) with $\ltlG(\bigwedge_{q \in F} q_1 \ne q)$ gives a formula $\phi(M_\single)$.
This formula is realizable if and only if~$M_\single$ does not halt.

We say that an \ac{LTL} formula~$\phi$ is a safety formula if its language~$L(\phi)$ is a safety property in~$(\Sigma\times\Omega)^\omega$. A safety property $P$ is such that, for all $t\in (\Sigma\times\Omega)^\omega\setminus P$, there exists $i\geq 0$ such that, for all $u\in (\Sigma\times\Omega)^\omega$, $t[\dots i]\cdot u\notin P$. Formulas in negation normal form built from atomic propositions using only
$\wedge$, $\vee$, $\ltlX$, $\ltlG$, and $\ltlW$ are safety formulas~\cite{Sistla94}.
Formula~$\phi(M_\single)$ is of this form: antecedents of implications
contain only atomic propositions and~$\ltlX$, which is self-dual, and no $\ltlF$ or $\ltlU$ operator occurs.

\paragraph{Not Nowhere Dense Adversaries}
Let~$A$ be an $\omega$-regular message adversary such that~$\pi(A)$ is not nowhere dense.
There exists a finite graph history~$h$ such that~$h^{-1}A$ is dense in~$\Gamma^\omega$.
Define
$    \operatorname{Prefix}_h := \bigwedge_{r=1}^{\lvert h\rvert} \ltlX^{r-1} (\link = h[r])$
and, for an \ac{LTL} formula~$\phi$, define 
   $ \widehat{\phi}
    :=
    \operatorname{Prefix}_h \rightarrow \ltlX^{\lvert h\rvert} \phi
    \enspace.$

For a safety formula~$\phi$, the guarded formula~$\widehat{\phi}$ is realizable under~$A$ if and only if~$\phi$ is realizable under~$\Gamma^\omega$.
Applying this to the safety formula~$\phi(M)$ constructed from a 2-counter machine~$M$ yields undecidability of $\textsc{LL-Synthesis}(A)$.
\Cref{sec:undecidability-lemmas} establishes both the counter-machine encoding and this transfer of realizability.

\subsection{Decidability Result}
\label{sec:decidability}

We prove the decidability direction of \Cref{thm:undecidability:adversary} by giving a procedure
  that also constructs a realizing distributed algorithm whenever one exists.

\begin{theorem}[\leanproof{\leanurl{Decidability}}]\label{thm:nowhere-dense-decision-procedure}
Given a deterministic parity automaton for a message adversary~$A$ with nowhere-dense
  image~$\pi(A)$ and an \ac{LTL} formula~$\phi$, one can decide whether a
  distributed algorithm realizing~$\phi$ under~$A$ exists and construct one if it does.
\end{theorem}

We represent distributed algorithms by trees whose nodes carry connected
  subgraphs of the view complex, with automaton states on edges and outputs
  on vertices of their extensions.
The uniform consensus bound from \Cref{lem:repeated_bounded_time_consensus}
  lets us bound the number of edges at each node while preserving acceptance.
Up to isomorphism, their node data range over a finite set.
Acceptance is expressible in \ac{MSO}, so tree-automaton nonemptiness yields
  a decision procedure.

\subsubsection{Notation and Consensus Bound Computation}

In the constructions below, $K$, $K'$, and their indexed variants denote
  subgraphs of a view complex at a fixed round, with no isolated vertices.
Write~$E(K)$ for the edge set of~$K$.
For an edge~$e$, write~$e_p$ for its process-$p$ endpoint.
Applying \Cref{tab:ViewDefn} to the endpoints of~$e$ and a signal~$a\in\Sigma$
  gives an edge~$ea$ at the next round.
Iterating this operation defines~$ew$ for~$w\in\Sigma^*$, with~$e\varepsilon=e$.
In particular, $e(h)w=e(hw)$.
For every integer~$t\geq0$, let~$K^{(t)}$ be the graph with edge set
\[
  E(K^{(t)}):=\{ew \mid e\in E(K),\ w\in\Sigma^t\} \enspace.
\]
Each edge in~$K^{(t)}$ has a unique such decomposition.
The graph~$K^{(t)}$ includes all continuations over~$\Sigma$, including those
  whose graph histories do not extend to~$A$.
For every~$p\in{\cal P}$, $e,e'\in E(K)$, and~$w,w'\in\Sigma^t$,
  repeated application of \Cref{lem:IndistDownClosed} gives
\[
  (ew)_p=(e'w')_p\implies e_p=e'_p \enspace.
\]
Consequently, extensions of distinct connected components cannot share a vertex.

In the following, fix a deterministic parity automaton $\mathcal A_\text{adv}$
  recognizing a message adversary~$A$ with nowhere-dense image~$\pi(A)$.
We assume that~$A\neq\emptyset$, since otherwise any distributed algorithm suffices.

From $\mathcal A_\text{adv}$, one can construct a
  deterministic finite automaton $\mathcal A_\text{live} = (S, s_\text{init}, \tau, F_\text{live})$
  with set of states~$S$, initial state $s_\text{init}$, transition function~$\tau\colon S \times \Gamma \to S$,
  and accepting states $F_\text{live}$, that recognizes~$\Pref(A)$.
Keep the states and transitions of $\mathcal A_\text{adv}$, and declare a state accepting in $\mathcal A_\text{live}$ if an accepting infinite continuation exists from that state in $\mathcal A_\text{adv}$.

It is further well known that one can construct an automaton that checks whether an execution satisfies~$\phi$ whenever its graph history belongs to~$A$:

\begin{lemma}[\leanproof{\leanurl{SpecificationAutomaton}}]\label{lem:A_phi}
From $\mathcal A_\text{adv}$ and an \ac{LTL} formula~$\phi$, one can construct a deterministic parity
  automaton $\mathcal A_\phi = (Q_\phi, q_\phi, \delta_\phi, \chi_\phi)$, with states~$Q_\phi$, initial state~$q_\phi$,
  transition function $\delta_\phi\colon Q_\phi \times (\Sigma \times \Omega) \to Q_\phi$, and priorities~$\chi_\phi\colon Q_\phi \to \NN$,
  that accepts the executions~$e \in (\Sigma \times \Omega)^\omega$ with $\Graph(e) \notin A$ or~$e \models \phi$.
\end{lemma}
\begin{proof}
From~$\phi$, construct the parity automaton~$\mathcal A'_{\phi}$ that accepts the
  executions~$(\Sigma \times \Omega)^\omega$ with~$e \models \phi$.
Complementing $\mathcal A_\text{adv}$ and taking the union with~$\mathcal A'_{\phi}$ yields~$\mathcal A_\phi$.
Both operations are computable~\cite{thomas_automata_1990}.
\end{proof}

For an input signal history~$h \in \Sigma^*$, we write~$s(h)$ for the state of $\mathcal A_\text{live}$ reached on~$\Graph(h)$.
We also write $s(e(h)):=s(h)$ and call an edge~$e$ \emph{live} if
  $s(e)\in F_\text{live}$.

A uniform bound on the rounds needed for consensus after every live prefix
  can be computed from~$\mathcal A_\text{adv}$.

\begin{lemma}[\leanproof{\leanurl{ConsensusBound}}]\label{lem:computable-consensus-bound}
From $\mathcal A_\text{adv}$, one can compute a~$D \geq 1$
  such that consensus is solvable within~$D$ rounds under~$h^{-1}A$
  for every $h \in \Pref(A)$.
\end{lemma}

In the following, let~$D \geq 1$ be a uniform consensus bound for adversary~$A$,
  computed using \Cref{lem:computable-consensus-bound}.

\subsubsection{Component Trees}

We group rounds into blocks of length~$D$, with outputs still produced in every round.
Using the automata $\mathcal A_\text{live}$ and~$\mathcal A_\phi$ constructed
  above, we obtain an automaton that reads these blocks and also records
  whether the communication prefix is live.

\begin{lemma}[\leanproof{\leanurl{BlockAutomaton}}]\label{lem:tree-combined-automaton}
One can construct a deterministic parity automaton
  $\mathcal B=(Q,q_\text{init},\delta,\chi)$ over~%
  $\Sigma^D\times\Omega^D$ that accepts exactly the executions,
  grouped into blocks, that satisfy~$\phi$ or whose graph history
  is outside~$A$.
Moreover, one can compute a subset $Q_\text{live}\subseteq Q$ such that,
  after any finite sequence of blocks, the automaton is in
  $Q_\text{live}$ if and only if the corresponding graph history
  extends to an infinite graph history in~$A$.
\end{lemma}

A \emph{position}~$(K,q)$ at round~$nD$ consists of a nonempty connected subgraph~%
  $K$ of the view complex under~$A$ and an edge labeling~$q\colon E(K)\to Q$.
For each $e=e(h)\in E(K)$, there must be an output sequence such that~%
  $\mathcal B$ reaches~$q(e)$ after reading the resulting execution
  in blocks of length~$D$.
Thus $q(e)\in Q_\text{live}$.
The initial position~$(K_0,q_0)$ consists of the single edge
  $e_\text{init}:=\{\mathsf I_1,\mathsf I_2\}$, labeled with $q_\text{init}$.

An \emph{output labeling from~$K$} assigns an output~%
  $\lambda_p(v)\in Y_p$ to every process-$p$ vertex~$v$
  of each extension graph~$K^{(t)}$, $1\leq t\leq D$.
For an edge~$e\in E(K)$ and an input signal history~$u\in\Sigma^{[1,D]}$, write
\[
  \lambda(e,u):=\bigl(\lambda_1((eu)_1),\lambda_2((eu)_2)\bigr) \enspace.
\]
Since each local view receives a single output label, each process's output
  depends only on its view.
For a complete block~$u\in\Sigma^D$, let~$q_\lambda(e,u)$ be the successor
  of~$q(e)$ in~$\mathcal B$ on~$u$ with the successive outputs
  $\lambda(e,u[\dots1]),\ldots,\lambda(e,u)$.
We call $\chi(q_\lambda(e,u))$ the \emph{block priority}.
Let~$\liveext{K}$ be the subgraph of~$K^{(D)}$ with edge set
\[
  E(\liveext{K}):=\{eu\mid e\in E(K),\ u\in\Sigma^D,\ q_\lambda(e,u)\in Q_\text{live}\} \enspace.
\]
Its edges are precisely the extensions allowed under~$A$.

\begin{definition}[\leanformalized{\leanline{Formalizations/ComponentTrees/Trees}{62}}]\label{def:tree-uncompressed}
An \emph{uncompressed component tree} is a rooted tree carrying a position~%
  $(K,q)$ and an output labeling~$\lambda$ from~$K$ at each node.
The root carries~$(K_0,q_0)$.
At a node carrying~$(K,q)$ and~$\lambda$, every connected component~$K'$ of~$\liveext{K}$
  gives exactly one child carrying~$(K',q')$, where
\[
  q'(eu):=q_\lambda(e,u)
\]
for all $eu\in E(K')$.
These are all its children. The same output labeling is used for every child.

A \emph{run} consists of an infinite branch from the root with successive
  positions~$(K_n,q_n)$, edges~$e_n\in E(K_n)$, and blocks~$u_n\in\Sigma^D$ such that
\begin{equation}\label{eq:tree-uncompressed-run}
  e_0=e_\text{init},\qquad e_{n+1}=e_nu_n\in E(K_{n+1}) \enspace.
\end{equation}
The tree is \emph{accepting} if the state sequence $(q_n(e_n))_{n\geq0}$
  of every run satisfies the parity condition.
\end{definition}

\begin{figure}
  \centering
  \includegraphics[width=\textwidth]{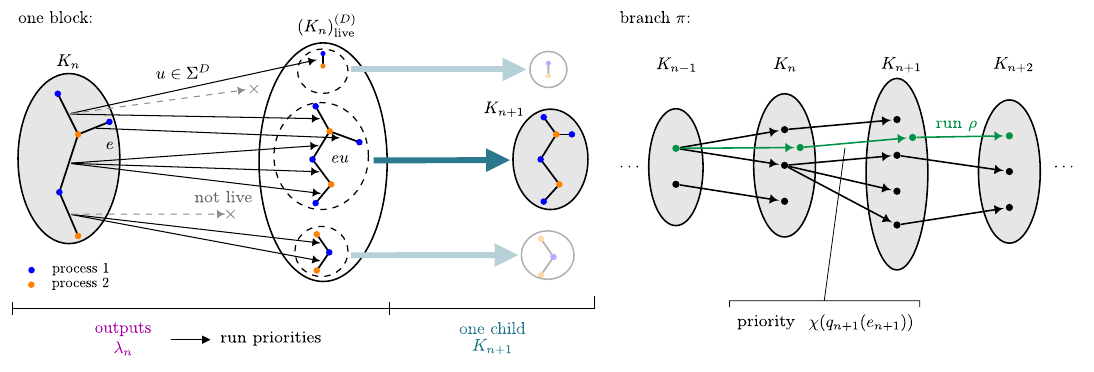}
  \caption{One block of an uncompressed component tree.
    On the left, colored vertices represent process views, black edges join
    views that can occur together, and crosses mark extensions that cannot
    be continued under~$A$.
    Each connected component of~$\liveext{(K_n)}$ gives a child; the middle child is highlighted.
    On the right, dots represent edges along one branch, and the green path is a run.
    The state at its next edge determines the block priority.}
  \label{fig:component-tree}
\end{figure}

Every edge of~$K$ has an extension allowed under~$A$, so every node has a child.
A branch specifies successive components, while a run follows successive edges
  through them. Several runs may therefore share a branch; see \Cref{fig:component-tree}.

A finitely branching labeled tree is \emph{effectively given} if, after numbering
  the children at each node, an algorithm computes the label and the number of
  children at any valid finite node address.
To \emph{effectively construct} such a tree means to construct such an algorithm.

We now relate accepting uncompressed component trees to realizing distributed algorithms.
\begin{lemma}[\leanproof{\leanurl{UncompressedTrees}}]\label{lem:tree-uncompressed-equivalence}
A distributed algorithm realizing~$\phi$ under~$A$ exists if and only if an
  accepting uncompressed component tree exists.
From an effectively given accepting uncompressed tree, one can compute the
  local output functions of a realizing algorithm.
\end{lemma}

An algorithm provides the vertex labels. Conversely, distinct components
  have disjoint vertex sets, and \Cref{lem:IndistDownClosed} implies that
  their extensions cannot share a vertex. The labels at different nodes
  therefore define consistent local output functions.

Components can have arbitrarily many edges, and their views grow with the round.
We use the uniform consensus bound to choose, at each node, a reduction~%
  $r\colon K\to K$ that preserves edge labels and has a bounded image.
The next block starts from~$r(K)$.
Suitable choices give accepting trees whose nodes have a uniformly bounded
  number of edges.
Up to isomorphism, only finitely many labeled graphs and block transitions remain.
We call trees of these finite descriptions \emph{symbolic component trees}.
Each child also records which reduced parent edge and block generated each
  of its edges.
\ac{MSO} checks that these labels fit together and that
  every compatible run accepts, yielding a decision procedure by Rabin's
  theorem~\cite{rabin1969decidability}.
\Cref{sec:tree-decidability} constructs these bounded representations and the resulting decision procedure.

\section{Properties of the Inputless View Complex}
\label{sec:inputless-view-properties}

Formally, for a finite graph history~$h\in\Gamma^*$, define the \emph{inputless view}~$v_p^0(h)$ by $v_p^0(\varepsilon):=\mathsf{I}_p$ and the recursion in \Cref{tab:ViewDefn} with all input arguments omitted.
Let~$E_p$ be the map that recursively erases input values from a view of process~$p$.
An induction on~$\lvert h'\rvert$ using the view updates gives, for every input signal history~$h'\in\Sigma^*$,
\begin{equation}
    E_p(v_p(h'))=v_p^0(\Graph(h'))
    \enspace.
\end{equation}
The \emph{inputless} round-$r$ view complex under~$A$ has the edges $\{v_1^0(h),v_2^0(h)\}$ for the graph histories~$h\in\Gamma^r$ that extend to an element of~$A$, and their endpoints as vertices.
Throughout this subsection, we use the full message adversary~$A=\Gamma^\omega$ and write~$v_p(h)$ for~$v_p^0(h)$.

\begin{figure}
    \centering
\begin{tikzpicture}[
  x=1.45cm,y=.6cm,
  font=\footnotesize,
  process p/.style={diamond,draw=black,fill=black,
    minimum size=5pt,inner sep=0pt},
  other process/.style={regular polygon, regular polygon sides=3,draw=black,fill=black,
    minimum size=5pt,inner sep=0pt,line width=.6pt},
  edge label/.style={anchor=south,inner sep=1pt},
  position label/.style={anchor=north,inner sep=1pt},
  refinement/.style={-{Latex[length=2mm]},draw=black!50,line width=.5pt},
  group brace/.style={decorate,
    decoration={brace,mirror,amplitude=4pt},line width=.5pt}
]
  \draw[line width=.7pt] (0,4.1) -- (6,4.1);
  \draw[line width=.7pt] (0,2.3) -- (6,2.3);
  \node[anchor=east] at (-.7,4.1) {round~$r$};
  \node[anchor=east] at (-.7,2.3) {round~$r + 1$};

  \node[process p] at (0,4.1) {};
  \node[other process] at (6,4.1) {};
  \node[edge label] at (0,4.27) {$u = v_p(h)$};
  \node[edge label] at (6,4.27) {$v = v_{\bar{p}}(h)$};
  \node[position label] at (3,3.93) {$h$};

  \foreach \j in {0,1,2,3} {
    \ifodd\j
      \node[other process] at (2*\j,2.3) {};
    \else
      \node[process p] at (2*\j,2.3) {};
    \fi
  }

  \node[edge label] at (0,2.47) {$\textsf{N}_p(u)$};
  \node[edge label] at (2,2.47) {$\textsf{C}_{\bar{p}}(v,u)$};
  \node[edge label] at (4,2.47) {$\textsf{C}_p(u,v)$};
  \node[edge label] at (6,2.47) {$\textsf{N}_{\bar{p}}(v)$};

  \node[position label] at (1,2.13) {$h(p \to \bar{p})$};
  \node[position label] at (3,2.13) {$h(p \leftrightarrow \bar{p})$};
  \node[position label] at (5,2.13) {$h(p \leftarrow \bar{p})$};

  \node[process p] at (7,3.65) {};
  \node[anchor=west] at (7.15,3.65) {process~$p$};
  \node[other process] at (7,2.85) {};
  \node[anchor=west] at (7.15,2.85) {process~$\bar{p}$};

\end{tikzpicture}
    \caption{View labels in a single subdivision. The left endpoint of the original edge is a view of process~$p$, its right endpoint one of process~$\bar{p}$. The three extensions share the two receiving views, while the outer endpoints carry the updated nonreceiving views.}
    \label{fig:SubdivLabels}
\end{figure}
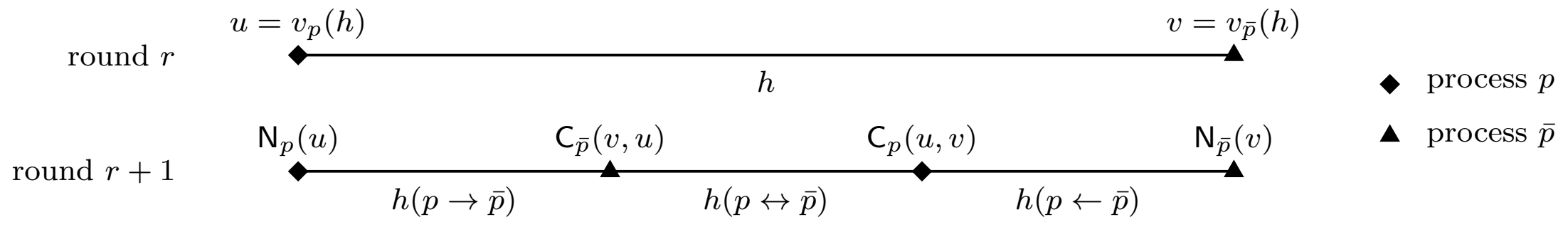

To establish the link of these subdivisions of~$[0,1]$ with the view complexes, we show that the round-$r$ view complex is a line of length~$3^r$ and that the histories that correspond to its edges line up with the above construction.
\Cref{fig:SubdivLabels} shows the main argument.
Suppose an original edge has endpoint views~$u:=v_p(h)$ on the left and~$v:=v_{\bar{p}}(h)$ on the right.
Its three extensions form a path with successive vertices
\begin{equation}
    \mathsf{N}_p(u),\quad \mathsf{C}_{\bar{p}}(v,u),\quad \mathsf{C}_p(u,v),\quad \mathsf{N}_{\bar{p}}(v)\enspace.
\end{equation}
The outer endpoints retain their positions, but their view labels change from~$u$ and~$v$ to~$\mathsf{N}_p(u)$ and~$\mathsf{N}_{\bar{p}}(v)$.
Neighboring extensions share a receiving process's view: that process cannot distinguish whether its own message was also delivered.
Moreover, subdivisions of neighboring original edges meet at the same updated view~$\mathsf{N}_p(u)$ whenever their common endpoint was~$u$ of process~$p$.
Consequently, the subdivisions fit together into a single path.

The endpoints of the line are those in which one of the processes has perfect information about the graph history.
A process that does not receive a message in a round knows that its message was received by the other process.
Thus, the view of a process that never received a message is an endpoint of the line.
There are exactly two such views; one for each process.

\begin{lemma}[\leanproof{\leanurl{SingleSubdivision}}]\label{lem:complex:is:line:single}
Let~$h\in\Gamma^r$.
Then the subgraph of the inputless round-$(r+1)$ view complex induced by the set of views $\{ v_p(h \graph) \mid p\in\mathcal{P} \wedge \graph \in \Gamma\}$ is a line of length~$3$.
\end{lemma}
\begin{proof}
    By the recusive definition of~$v_p(h\graph)$ in \Cref{tab:ViewDefn}, the induced subgraph has the four vertices~$\mathsf{N}_1(u)$, $\mathsf{C}_2(v,u)$, $\mathsf{C}_1(u,v)$, and~$\mathsf{N}_2(v)$ joined by the edges $\{\mathsf{N}_1(u) , \mathsf{C}_2(v,u)\}$, $\{\mathsf{C}_2(v,u) , \mathsf{C}_1(u,v)\}$, and $\{\mathsf{C}_1(u,v) , \mathsf{N}_2(v)\}$, where~$u := v_1(h)$ and~$v := v_2(h)$.
\end{proof}

\begin{lemma}[\leanproof{\leanurl{ViewPath}}]\label{lem:complex:is:line}
    The inputless round-$r$ view complex is a line of length~$3^r$.
\end{lemma}
\begin{proof}
    By induction on~$r$.
    The base case~$r = 0$ is trivial.
    Now suppose that the inputless round-$r$ view complex is a line of length~$3^r$.
    By \Cref{lem:complex:is:line:single}, the round-$(r+1)$ view complex contains at least~$3^r$ lines of length~$3$ each.

    These~$3^r$ lines are edge-disjoint, but not vertex-disjoint:
    If $\{ v_p(h) , v_{\bar{p}}(h) \}$ and $\{ v_{\bar{p}}(h') , v_p(h') \}$ are two distinct incident edges in the round-$r$ view complex, with $v_{\bar{p}}(h) = u = v_{\bar{p}}(h')$, then $v_{\bar{p}}(h\graph) = \mathsf{N}_{\bar{p}}(u) = v_{\bar{p}}(h'\cdot \graph)$ where~$\graph = (p \leftarrow \bar{p})$.
    By \Cref{lem:IndistDownClosed}, the~$3^r$ lines share no other vertices.
    Thus, the round-$(r+1)$ view complex contains a line of length~$3^{r+1}$.

    Since there are no more than~$3^{r+1}$ round-$(r+1)$ graph histories in~$\Gamma^{r+1}$, the complex does not contain any other edges.
    The round-$(r+1)$ view complex is thus a line of length~$3^{r+1}$.
\end{proof}

\begin{lemma}[\leanproof{\leanurl{ExtremeVertices}}]\label{lem:extreme-vertex}
    The view $v_p\big( (p \rightarrow \bar{p})^r \big)$ has a single adjacent view in the inputless round-$r$ view complex.
\end{lemma}
\begin{proof}
    By induction on~$r$.
    The base case~$r = 0$ is trivial since there are only two views in the view complex.
    Now suppose that $u = v_p\big( (p \rightarrow \bar{p})^r \big)$ has $v = v_{\bar{p}}\big( (p\rightarrow \bar{p})^r \big)$ as the only adjacent view in the inputless round-$r$ view complex.
    Then, by the recursive definition in \Cref{tab:ViewDefn} and \Cref{lem:IndistDownClosed}, the view $v_p\big( (p \rightarrow \bar{p})^{r+1} \big) = \mathsf{N}_p(u)$ has $v_{\bar{p}}\big( (p \rightarrow \bar{p})^{r+1} \big) = \mathsf{C}_{\bar{p}}(v,u)$ as the only adjacent view in the inputless round-$(r+1)$ view complex.
\end{proof}

The order of the three extensions depends on which process's view is the left endpoint.
If this process is~$p$, then the extensions appear from left to right as
\begin{equation}
    h(p\rightarrow\bar{p}),\qquad h(p\leftrightarrow\bar{p}),\qquad h(p\leftarrow\bar{p})\enspace.
\end{equation}
Thus, the bidirectional graph always corresponds to the middle subinterval.
The two unidirectional graphs exchange positions when the processes at the endpoints are reversed.
This happens, for example, when subdividing the middle edge~$h_1$ in \Cref{fig:ViewPathRefinement}.

By convention we will take the view~$v_1\big( \rlink^r \big)$ as the left endpoint of the inputless round-$r$ view complex.
Based on this orientation, we assign consecutive positions to the edges in the complex, starting with position~$0$ on the leftmost edge.
Since the edges in the complex are in a one-to-one correspondence with the length-$r$ graph histories, this amounts to assigning positions~$j(h)$ to the length-$r$ graph histories~$h \in \Gamma^r$.
By the above convention, the graph history~$\rlink^r$ has position $j\big( \rlink^r \big) = 0$ and the graph history~$\llink^r$ has position $j\big( \llink^r \big) = 3^r - 1$.

For the recurrence below, define the offsets
\[
    \delta(\rlink):=0,\qquad \delta(\dlink):=1,\qquad \delta(\llink):=2\enspace.
\]

\begin{lemma}[\leanproof{\leanurl{PathIndex}}]\label{lem:j}
For all~$h\in\Gamma^*$ and~$\graph\in\Gamma$, we have
\begin{equation}
    j(h\graph) = 3j(h) + 
    \begin{cases}
        \delta(\graph) & \text{if~$j(h)$ is even} \\
        2 - \delta(\graph) & \text{if~$j(h)$ is odd}\enspace.
    \end{cases}
    \label{eqn:j}
\end{equation}
\end{lemma}
\begin{proof}
    Let~$r := \lvert h\rvert$.
    As shown in the proof of \Cref{lem:complex:is:line}, each edge of the round-$r$ view complex is replaced by the line of length~$3$ from \Cref{lem:complex:is:line:single}, with each old vertex~$w$ of process~$p$ replaced by~$\mathsf{N}_p(w)$.
    Since $\mathsf{N}_1\big(v_1(\rlink^r)\big) = v_1(\rlink^{r+1})$, this replacement preserves the left endpoint and hence the order of the edges.
    Thus, the three extensions of~$h$ occupy positions~$3j(h)$, $3j(h)+1$, and~$3j(h)+2$.

    Every edge joins views of different processes, and the leftmost vertex belongs to process~$1$.
    Hence, the left endpoint of the edge corresponding to~$h$ belongs to process~$1$ if~$j(h)$ is even, and to process~$2$ if~$j(h)$ is odd.
    By the recursive definition in \Cref{tab:ViewDefn}, the three extensions are ordered from left to right as~$h\rlink$, $h\dlink$, $h\llink$ in the even case, and in the reverse order in the odd case.
    The offset of~$j(h\graph)$ from~$3j(h)$ is therefore~$\delta(\graph)$ in the even case and~$2-\delta(\graph)$ in the odd case.
\end{proof}

\begin{proof}[Proof of \Cref{J:inclusion:iff:prefix}]
    ($\Rightarrow$):
    By \Cref{lem:j}, the intervals~$J(h\graph)$, for~$\graph\in\Gamma$, are the three equal subdivisions of~$J(h)$.
    Repeated application shows that $J(k)\subseteq J(h)$ whenever~$h$ is a prefix of~$k$.

    ($\Leftarrow$):
    Suppose that $J(k)\subseteq J(h)$.
    Since these intervals have lengths~$3^{-\lvert k\rvert}$ and~$3^{-\lvert h\rvert}$, respectively, we have~$\lvert k\rvert\ge\lvert h\rvert$.
    Let~$u$ be the prefix of~$k$ of length~$\lvert h\rvert$.
    By the ($\Rightarrow$) direction, we have $J(k)\subseteq J(u)$ and thus $J(k)\subseteq J(h)\cap J(u)$.
    Thus, this intersection has positive length.
    By the definition of~$J$, distinct histories of the same length correspond to intervals with disjoint interiors.
    Hence~$u=h$, which proves that~$h$ is a prefix of~$k$.
\end{proof}

The product topology on~$\Gamma^\omega$ is metrizable via the ultrametric $d(h, h') := 2^{-\inf\{r \geq 1 \mid h[r] \neq h'[r]\}}$~\cite[Ch.~III, Prop.~3.2]{perrin_infinite_2004}.

\begin{proof}[Proof of \Cref{pi:cont}]
    We prove uniform continuity.
    Let~$\varepsilon>0$ and choose an integer~$r\geq 0$ such that~$3^{-r}<\varepsilon$.
    If~$h,h'\in\Gamma^\omega$ satisfy $d(h,h')<2^{-r}$, then they have a common prefix~$k$ of length~$r$.
    By \Cref{J:inclusion:iff:prefix} and the definition of~$\pi$, both~$\pi(h)$ and~$\pi(h')$ belong to~$J(k)$.
    Hence
    \begin{equation}
        \lvert\pi(h)-\pi(h')\rvert \leq 3^{-r}<\varepsilon \enspace,
    \end{equation}
    proving continuity.

    For surjectivity, let~$x\in[0,1]$.
    By \Cref{lem:j}, the three intervals~$J(h\graph)$, for~$\graph\in\Gamma$, cover~$J(h)$.
    Starting from $x\in J(h_0) = J(\varepsilon) = [0,1]$, we can therefore recursively choose~$h_{r+1}=h_r\graph$ such that $x\in J(h_{r+1})$.
    These compatible prefixes define an infinite history~$h\in\Gamma^\omega$.
    Both~$x$ and~$\pi(h)$ belong to~$J(h_r)$ for every~$r$, so their distance is at most~$3^{-r}$ for every~$r$.
    Thus~$\pi(h)=x$.
\end{proof}

\section{Nowhere-Dense Message Adversaries and Consensus}
\label{sec:consensus-details}

We characterize bounded-time consensus using the topology of graph histories, extend this characterization to repeated consensus, and relate consensus to agreement on a broadcaster.

\subsection{Consensus and Bounded Termination}

A \emph{decision problem} is defined by finite nonempty sets~$I$ of possible inputs and~$O$ of possible outputs, and a relation~$\Delta \subseteq I^2 \times O^2$ of allowed input--output pairs.
We take~$X_1=X_2=I$ and $Y_1=Y_2=O\cup\{\bot\}$, where~$\bot\notin O$ denotes that no decision has been made.
The inputs received in round~$1$ are the proposals for the decision problem.
Subsequent inputs remain arbitrary, but the validity of the decisions depends only on the proposals.
Let~$A\subseteq \Gamma^\omega$ be a message adversary.
A distributed algorithm~$f$ solves a decision problem under message adversary~$A$ if (1) all its non-$\bot$ outputs are irrevocable (\emph{i.e.}, don't change over time) and (2) in every execution with graph history in~$A$ and round-$1$ inputs~$(i_1,i_2)$, both processes eventually output non-$\bot$ values~$(o_1,o_2)$ such that $\left( (i_1,i_2) \,,\, (o_1,o_2) \right) \in \Delta$.

This notion of solvability agrees with the usual model in which inputs are supplied only initially.
An algorithm for that model can ignore subsequent inputs.
Conversely, an algorithm that solves the problem with arbitrary subsequent inputs can be simulated using each process's proposal as its input in every round.

The consensus problem with value set~$V$ is the decision problem with~$I=O=V$ and~$\Delta$ defined by requiring both outputs to be equal and to be one of the inputs.
To make the problem nontrivial, we choose any finite~$V$ with~$\lvert V \rvert \geq 2$.

The construction of~$\pi$ describes graph histories independently of the input values.
We use them to study the message adversary~$A\subseteq\Gamma^\omega$, while algorithms solving decision problems operate on views containing inputs.
\Cref{cons:char} relates consensus solvability in this model to the geometry of~$A$.

Ordinary consensus permits the decision time to depend on the execution.
The geometric characterization reflects this flexibility: it is enough that~$\pi(A)$ omit a single point, even if admissible histories have images arbitrarily close to that point.
Requiring a common decision bound imposes a stronger condition.

We say that a distributed algorithm solves a decision problem in bounded time under~$A$ if it solves it under~$A$ and there is a round number~$R$ such that all~$\bot$ outputs happen before round~$R$ in all executions considered for solvability.

Taking the closure of an adversary introduces no new finite graph histories.
Consequently, an algorithm that decides within a fixed number of rounds cannot use the distinction between an adversary and its closure.
Conversely, compactness turns termination under a closed adversary into a uniform decision bound.
This explains why bounded-time consensus is governed by the closure of~$\pi(A)$, and hence by the existence of an interval missing from its image.

\begin{lemma}[\leanproof{\leanurl{Closure}}]\label{pi:closure}
    For any message adversary~$A \subseteq \Gamma^\omega$, we have $\pi(\overline{A}) = \overline{\pi(A)}$.
\end{lemma}
\begin{proof}
    Since~$\Gamma^\omega$ is compact by Tychonoff's theorem~\cite[Ch.~I, \S9.5, Thm.~3]{bourbaki_general_topology} and~$[0,1]$ is Hausdorff, continuity of~$\pi$ (\Cref{pi:cont}) implies $\pi(\overline{A}) = \overline{\pi(A)}$ \cite[Ch.~I, \S9.4, Cor.~2 to Thm.~2 and Ch.~I, \S5.4, Prop.~9]{bourbaki_general_topology}.
\end{proof}

\begin{lemma}[\leanproof{\leanurl{BoundedConsensus}}]\label{bounded:cons:char}
    Consensus is solvable in bounded time under message adversary~$A$ if and only if~$\pi(A)$ is not dense in~$[0,1]$.
\end{lemma}
\begin{proof}
    ($\Leftarrow$):
    Since~$\pi(A)$ is not dense in~$[0,1]$, using \Cref{pi:closure}, we have $\pi(\overline{A}) = \overline{\pi(A)} \neq [0,1]$.
    By \Cref{cons:char}, consensus is solvable under the message adversary~$\overline{A}$ by some distributed algorithm~$f$.
    The space of admissible infinite input signal histories under~$\overline{A}$ is naturally identified with~$(I\times I)^\omega\times\overline{A}$, which is compact since~$I$ is finite and~$\overline{A}$ is closed.
    In every execution, both processes have decided after some finite prefix of the input signal history.
    Any admissible history with the same prefix yields the same outputs up to that round, so both processes have also decided by then.
    For each such prefix, the set of admissible histories extending it is open.
    These sets cover all admissible histories.
    A finite subcover gives a uniform decision bound~\cite[Lemma~1.1]{lubitch_closed_1995}.
    Thus~$f$ solves consensus in bounded time under~$\overline{A}$, and under~$A$ as well since~$A\subseteq\overline{A}$.

    ($\Rightarrow$):
    Assume by contradiction that~$f$ solves consensus in bounded time under~$A$ and that~$\pi(A)$ is dense in~$[0,1]$, \emph{i.e.}, that $\overline{\pi(A)} = [0,1]$.
    Let~$R$ be an upper bound for~$\bot$ outputs of~$f$ under~$A$.
    Consider the message adversary $B := \Pref_R(A) \cdot \Gamma^\omega$ whose graph histories are the length-$R$ prefixes of~$A$ followed by an arbitrary infinite graph history.
    Let~$f'$ follow~$f$ until each process's first non-$\bot$ output, which that process then repeats forever.
    Every length-$R$ input signal history under~$B$ extends to an infinite input signal history under~$A$ with the same inputs and graphs in the first~$R$ rounds.
    Hence~$f'$ makes valid, agreeing decisions by round~$R$ under~$B$, and solves consensus under~$B$.
    Since~$B$ is a safety property, hence closed~\cite[p.~184]{alpern_defining_1985}, and~$A\subseteq B$, we have~$\overline{A} \subseteq B$ and thus $[0,1] = \overline{\pi(A)} = \pi(\overline{A}) \subseteq \pi(B) \subseteq [0,1]$.
    But then $\pi(B) = [0,1]$, which contradicts the fact that~$f'$ solves consensus under~$B$ by \Cref{cons:char}.
\end{proof}

\subsection{Removing Finite Prefixes}

A decision bound from the beginning of an execution does not ensure that consensus remains solvable after an arbitrary prefix.
The prefix restricts the possible continuations, and the gap that originally permitted consensus may lie outside the part of the adversary that remains relevant.
We must therefore apply the characterization to each left quotient of the adversary.

For a finite graph history $h\in\Pref(A)$, define the left quotient $h^{-1}A := \{ k \in \Gamma^\omega \mid h \cdot k \in A \}$.

Geometrically, this requires restricting attention to histories extending~$h$ and expressing their images in the coordinates of~$J(h)$.
Removing the prefix rescales this interval to~$[0,1]$, reversing its orientation when~$j(h)$ is odd.
The next three lemmas establish this correspondence at the level of edge positions, finite intervals, and infinite histories.

\begin{lemma}[\leanproof{\leanurl{ConcatenationIndex}}]\label{concat:pos}
    $\displaystyle
    j(k) = 
    \begin{cases}
        j(h\cdot k) - 3^{\lvert k\rvert} j(h) & \text{if $j(h)$ is even}\\ 
        3^{\lvert k\rvert}(j(h)+1) - j(h\cdot k) - 1 & \text{if $j(h)$ is odd}
    \end{cases}$
\end{lemma}
\begin{proof}
    By induction on~$\lvert k\rvert$.
    The base case~$k=\varepsilon$ follows from~$j(\varepsilon)=0$.
    Now let~$k=\ell\graph$ and assume the claim holds for~$\ell$.
    Set~$r:=\lvert\ell\rvert$ and $a:=j(\ell\graph)-3j(\ell)$.

    If~$j(h)$ is even, the induction hypothesis gives $j(h\cdot\ell)=3^rj(h)+j(\ell)$.
    Hence~$j(h\cdot\ell)$ and~$j(\ell)$ have the same parity.
    By \Cref{lem:j}, appending~$\graph$ therefore gives the same offset~$a$ in both cases, so
    \begin{equation}
        \begin{split}
            j(h\cdot k)
            &=3j(h\cdot\ell)+a\\
            &=3^{r+1}j(h)+3j(\ell)+a\\
            &=3^{r+1}j(h)+j(k)
            \enspace.
        \end{split}
    \end{equation}

    If~$j(h)$ is odd, the induction hypothesis gives $j(h\cdot\ell)=3^r(j(h)+1)-j(\ell)-1$.
    Hence~$j(h\cdot\ell)$ and~$j(\ell)$ have opposite parity.
    By \Cref{lem:j}, the offset for appending~$\graph$ to~$h\cdot\ell$ is therefore~$2-a$, so
    \begin{equation}
        \begin{split}
            j(h\cdot k)
            &=3j(h\cdot\ell)+2-a\\
            &=3^{r+1}(j(h)+1)-3j(\ell)-1-a\\
            &=3^{r+1}(j(h)+1)-j(k)-1
            \enspace.
        \end{split}
    \end{equation}
    Rearranging these equalities proves the claim.
\end{proof}

\begin{lemma}[\leanproof{\leanurl{ConcatenationCell}}]\label{concat:interval}
    $\displaystyle
    J(k) = 
    \begin{cases}
        3^{\lvert h\rvert} J(h\cdot k) - j(h) & \text{if $j(h)$ is even}\\ 
        j(h) + 1 - 3^{\lvert h\rvert} J(h\cdot k) & \text{if $j(h)$ is odd}
    \end{cases}$
\end{lemma}
\begin{proof}
    Let~$r:=\lvert k\rvert$.
    By the definition of~$J$,
    \begin{equation}
        3^{\lvert h\rvert}J(h\cdot k)
        =
        \left[
            \frac{j(h\cdot k)}{3^r},
            \frac{j(h\cdot k)+1}{3^r}
        \right]
        \enspace.
    \end{equation}
    If~$j(h)$ is even, \Cref{concat:pos} shows that this interval is $j(h)+J(k)$.
    If~$j(h)$ is odd, the same lemma shows that it is
    \begin{equation}
        \left[
            j(h)+1-\frac{j(k)+1}{3^r},
            j(h)+1-\frac{j(k)}{3^r}
        \right]
        =
        j(h)+1-J(k)
        \enspace.
    \end{equation}
    Rearranging gives the claimed identities.
\end{proof}

\begin{lemma}[\leanproof{\leanurl{ConcatenationMap}}]\label{concat:pi}
    $\displaystyle
    \pi(h^{-1}A) = 
    \begin{cases}
        3^{\lvert h\rvert} \pi(A \cap h\cdot\Gamma^\omega) - j(h) & \text{if $j(h)$ is even}\\ 
        j(h) + 1 - 3^{\lvert h\rvert} \pi(A \cap h\cdot \Gamma^\omega) & \text{if $j(h)$ is odd}
    \end{cases}$
\end{lemma}
\begin{proof}
    Let~$k\in\Gamma^\omega$ and denote its length-$r$ prefix by~$k_r$.
    By the definition of~$\pi$, the intervals~$J(k_r)$ and~$J(h\cdot k_r)$ shrink to~$\pi(k)$ and~$\pi(h\cdot k)$, respectively.
    Applying \Cref{concat:interval} and taking limits gives
    \begin{equation}
        \pi(k)=
        \begin{cases}
            3^{\lvert h\rvert}\pi(h\cdot k)-j(h)
            & \text{if~$j(h)$ is even}\\
            j(h)+1-3^{\lvert h\rvert}\pi(h\cdot k)
            & \text{if~$j(h)$ is odd}\enspace.
        \end{cases}
    \end{equation}
    As~$k$ ranges over~$h^{-1}A$, the history~$h\cdot k$ ranges over~$A\cap h\cdot\Gamma^\omega$.
    Taking the sets of values on both sides proves the claim.
\end{proof}

\subsection{Repeated Consensus}

The preceding identities allow us to compare consensus after different prefixes.
We now require solvability under every left quotient and distinguish bounds that depend on the prefix from a common bound.

We say that a decision problem is \emph{repeatedly solvable} under a message adversary~$A$ if it is solvable starting after any finite prefix, \emph{i.e.}, if it is solvable under all~$h^{-1}A$ where $h \in \Pref(A)$.
Analogously, it is repeatedly solvable in bounded time if it is solvable in bounded time under all~$h^{-1}A$.
Finally, it is repeatedly solvable in \emph{uniformly} bounded time if there is a uniform bound~$R$ and distributed algorithms~$f_h$ such that, for all $h\in\Pref(A)$, the algorithm~$f_h$ solves the problem under~$h^{-1}A$ and all~$\bot$ outputs happen before time~$R$.

\begin{lemma}[\leanproof{\leanurl{EmptyInterior}}]\label{lem:empty-interior-repeated-consensus}
    If~$\pi(A)$ has empty interior, then consensus is repeatedly solvable under message adversary~$A$.
\end{lemma}
\begin{proof}
    Let~$\pi(A)$ have empty interior and let $h\in\Pref(A)$.
    We have  $\pi(A \cap h\cdot\Gamma^\omega) \subseteq \pi(A) \cap \pi(h\cdot\Gamma^\omega) = \pi(A) \cap J(h) \subsetneq J(h)$ since~$J(h)$ is an interval of positive length and~$\pi(A)$ has empty interior.
    But then, by \Cref{concat:pi}, we get $\pi(h^{-1}A) \subsetneq [0,1]$.
    \Cref{cons:char} then shows that consensus is solvable under~$h^{-1}A$.
\end{proof}

The converse does not hold.
Let~$A$ consist of the infinite graph histories~$k$ whose ternary digits $j(k_r)-3j(k_{r-1})$ are not eventually zero, where~$k_r$ is the length-$r$ prefix of~$k$.
Then $\pi(A)=(0,1]$, while~$\pi(h^{-1}A)$ is~$(0,1]$ if~$j(h)$ is even and~$[0,1)$ if~$j(h)$ is odd, for every finite graph history~$h$.
Thus consensus is repeatedly solvable under~$A$ by \Cref{cons:char}, although~$\pi(A)$ has nonempty interior.

Repeated bounded-time consensus requires a gap after every admissible prefix.
In the interval representation, these requirements amount to finding interval gaps at arbitrarily small scales throughout~$[0,1]$.
Nowhere density captures precisely this condition.
The next two lemmas allow us to pass between the restricted adversaries and their intervals.

\begin{lemma}[\leanproof{\leanurl{CellInterior}}]\label{pi:inverse:J:interior}
    For every finite graph history~$h\in\Gamma^*$, we have $\pi^{-1}(J(h)^\circ) \subseteq h\cdot\Gamma^\omega$.
\end{lemma}
\begin{proof}
    Let $k \in \pi^{-1}(J(h)^{\circ})$, \emph{i.e.}, $\pi(k) \in J(h)^\circ$.
    Then, by definition of~$\pi$, since~$J(h)^\circ$ is open, there exists a finite prefix~$\ell$ of~$k$ such that $J(\ell) \subseteq J(h)^\circ \subseteq J(h)$.
    But this, by \Cref{J:inclusion:iff:prefix}, implies that~$h$ is a prefix of~$\ell$.
    Hence~$h$ is also a prefix of~$k$, which means that~$k\in h\cdot\Gamma^\omega$.
\end{proof}

\begin{lemma}[\leanproof{\leanurl{ClopenCones}}]\label{cones:are:clopen}
    For every finite graph history~$h\in\Gamma^*$, the set~$h\cdot\Gamma^\omega$ of infinite graph histories extending~$h$ is clopen.
\end{lemma}
\begin{proof}
    The set~$h\cdot\Gamma^\omega$ is open since it is equal to the open ball of radius~$2^{-\lvert h\rvert}$ using the ultrametric~$d$.
    It is also closed since we can write it as
    \begin{equation}
        h\cdot\Gamma^\omega
        =
        \Gamma^\omega \setminus \bigcup_{\substack{k\in\Gamma^{\lvert h\rvert}\\k\neq h}} k\cdot\Gamma^\omega
        \enspace,
    \end{equation}
    that is, as the complement of an open set.
\end{proof}

The remaining distinction concerns the decision bounds: they may initially depend on the prefix, but an $\omega$-regular adversary has only finitely many distinct left quotients, allowing a common bound to be chosen.

\begin{proof}[Proof of \Cref{lem:repeated_bounded_time_consensus}]
    ($\Leftarrow$):
    Let~$\pi(A)$ be nowhere dense and let $h \in \Pref(A)$.
    Using \Cref{pi:closure,cones:are:clopen}, and the fact that~$\overline{A\cap B} = \overline{A} \cap B$ if~$B$ is clopen~\cite[Ch.~I, \S1.6, Prop.~5]{bourbaki_general_topology}, we have
    \begin{equation}
        \overline{\pi(A \cap h\cdot\Gamma^\omega)} = \pi(\overline{A \cap h\cdot\Gamma^\omega}) = \pi(\overline{A} \cap h\cdot\Gamma^\omega) \subseteq \pi(\overline{A}) \cap \pi(h\cdot\Gamma^\omega) = \pi(\overline{A}) \cap J(h) \subsetneq J(h)
    \end{equation}
    since~$J(h)$ is an interval of positive length and~$\overline{\pi(A)}$ has empty interior.
    But then, by \Cref{pi:closure,concat:pi}, we get $\pi(\overline{h^{-1}A}) = \overline{\pi(h^{-1}A)} \subsetneq [0,1]$.
    \Cref{bounded:cons:char} then shows that consensus is solvable in bounded time under~$h^{-1}A$.

    ($\Rightarrow$):
    Let consensus be repeatedly solvable in bounded time under~$A$ and assume by contradiction that~$I \subseteq [0,1]$ is a nonempty open interval included in~$\overline{\pi(A)}$.
    Since~$\pi$ is surjective (\Cref{pi:cont}) and $\len(J(h)) \to 0$ as~$\lvert h\rvert \to \infty$, there exists an $h\in\Pref(A)$ such that $J(h)\subseteq I \subseteq \overline{\pi(A)}$.
    By \Cref{pi:inverse:J:interior}, we have
    \begin{equation}
        J(h)^\circ
        =
        J(h)^\circ \cap \overline{\pi(A)}
        =
        J(h)^\circ \cap \pi(\overline{A})
        \subseteq
        \pi(\overline{A} \cap h\cdot\Gamma^\omega)
        =
        \pi(\overline{A \cap h\cdot\Gamma^\omega})
        =
        \overline{\pi(A \cap h\cdot\Gamma^\omega)}
        \enspace.
    \end{equation}
    Since the rightmost set in this inclusion chain is closed, we have $J(h) = \overline{J(h)^\circ} \subseteq \overline{\pi(A \cap h\cdot\Gamma^\omega)}$.
    But, by \Cref{pi:closure,bounded:cons:char}, we have $\pi(\overline{h^{-1}A}) \subsetneq [0,1]$,
    and thus $\overline{\pi(A \cap h\cdot\Gamma^\omega)} \subsetneq J(h)$ by \Cref{concat:pi}, a contradiction.

    Finally, let~$A$ be $\omega$-regular and~$\pi(A)$ be nowhere dense.
    Then there are only finitely many distinct left quotients~$h^{-1}A$ \cite[Sec.~2, p.~142]{thomas_automata_1990}.
    Since consensus is solvable in bounded time under each of them, the maximum of their finitely many time bounds is a uniform bound.
\end{proof}

\subsection{Agreement on a Broadcaster}

Call a process a \emph{broadcaster} for a finite graph history if at least one of its messages is delivered to the other process.
Consensus is equivalent to agreement on a broadcaster~\cite[Theorem~7.12]{NSW24}.
In fact, once both processes have selected the same broadcaster, they can decide on its proposal.
\Cref{lem:cons:means:broadcaster} establishes the converse without increasing the number of rounds.

\begin{proof}[Proof of \Cref{lem:cons:means:broadcaster}]
    Let~$f$ solve consensus under~$A$ within~$R$ rounds.
    Choose two distinct proposal values, identified with~$1$ and~$2$, and
    let~$p$ be the common decision.
    Suppose that~$\bar p$ has received no message from~$p$ during the first~$R$ rounds.
    Changing the simulated inputs of~$p$ to~$\bar p$ leaves the view of~$\bar p$ unchanged, so it still decides~$p$.
    But both proposals are now~$\bar p$, contradicting validity.
    Hence~$p$ is a broadcaster in the first~$R$ rounds.
\end{proof}

\section{Undecidability under Not-Nowhere-Dense Message Adversaries}
\label{sec:undecidability-lemmas}

Throughout this section, we use the notation and constructions of \Cref{sec:undecidability}.
In particular, input sets are singletons, and we identify input signal histories with graph histories.
\subsection{Undecidability under the Full Message Adversary}

We consider algorithms $f=(f_1,f_2)$ with $f_p\colon \Views_p\to Y_p$. We let $\proj{f_p(v_p(h))}{q}$, $\proj{f_p(v_p(h))}{\zero}$,
 $\proj{f_p(v_p(h))}{\rightplan}$, $\proj{f_p(v_p(h))}{\carry}$, and $\proj{f_p(v_p(h))}{\pos}$ be the projections of
 $f_p(v_p(h))$ onto the respective components of the output.

The first lemma relies on the indistinguishability relation that forms a line in the view complex and shows that local
agreement implies global agreement for all histories at a given round. 
\begin{lemma}[\leanproof{\leanurl{PathAgreement}}]\label{lem:agreement_on_path}
    Let~$f$ be a distributed algorithm, and let~$r\geq 1$.
    If processes~$1$ and~$2$ output the same value in round~$r$ of every execution, then all vertices of the inputless round-$r$ view complex have the same output label under~$f$.
\end{lemma}
\begin{proof}
    Every edge of the view complex represents a history that extends to an execution, so its endpoints have equal output labels by assumption.
    Since the inputless round-$r$ view complex is connected by \Cref{lem:complex:is:line}, all vertices have the same output label.
\end{proof}

We next verify that~$\posord$ encodes vertex positions modulo~3 and determines the orientation of each edge.
\begin{lemma}[\leanproof{\leanurl{PositionOrder}}]\label{lem:posord}
    Let~$f$ realize~$\posord$. For all~$r\geq 1$, in the view complex of~$f$ at round~$r$, $\pos(r,i) = i \bmod 3$ for all~$0\leq i\leq 3^r$.
    Moreover, for all~$h\in\Gamma^r$, the left endpoint of~$J(h)$ is process 1's view if and only if  $\proj{f_2(v_2(h))}{\pos}\equiv \proj{f_1(v_1(h))}{\pos} +1 \mod 3$.
\end{lemma}

\begin{proof}
    We prove the first statement using induction on~$r$.
    For all~$h\in\Gamma^\omega$, we have $\exec(f, h) \models \posord$.
    Thus, for each such~$h$,
    \[
        (\proj{f_1(v_1(h[\dots 1]))}{pos}, \proj{f_2(v_2(h[\dots 1]))}{pos}) = \childpos_{h[1]}(0, 1)\enspace.
    \] 
    As $\childpos_{\rlink}(0, 1) = (0, 1), \childpos_{\dlink}(0, 1) = (2, 1)$, and $\childpos_{\llink}(0, 1) = (2, 0)$, the statement holds for~$r = 1$.

    We also have for all~$h \in \Gamma^r$ and~$\graph \in \Gamma$ that
    \begin{equation}
        \label{eqn:posord_transition}
        \left( \proj{f_1(v_1(h))}{pos}, \proj{f_2(v_2(h))}{pos} \right) = (u, v)\implies\left( \proj{f_1(v_1(h\graph))}{pos}, \proj{f_2(v_2(h\graph))}{pos} \right) = \childpos_\graph(u, v)\enspace.
    \end{equation}
    Suppose the first statement is true at round~$r \ge 1$.
    Let~$m \le 3^r - 1$ and let~$h \in \Gamma^r$ such that~$j(h) = m$.
    By the induction hypothesis, $\left( \proj{f_1(v_1(h))}{pos}, \proj{f_2(v_2(h))}{pos} \right)$ is either $(m \bmod 3, (m + 1) \bmod 3)$ or $((m + 1) \bmod 3, m \bmod 3)$.
    In either case, doing casework on~$\graph \in \Gamma$ and using (\ref{eqn:posord_transition}) gives that the vertices at the left and right endpoints of~$J(h\graph)$ output~$j(h\graph) \bmod 3$ and $(j(h\graph) + 1) \bmod 3$ as their $\pos$-outputs respectively.
    Iterating~$m$ over $0, 1, \dots, 3^r- 1$ completes the inductive step.

    As a consequence of the first statement, if process 1's view is the left endpoint of~$J(h)$ then $\proj{f_2(v_2(h))}{pos} - \proj{f_1(v_1(h))}{pos} \equiv 1 \mod 3$ and if process 2's view is the left endpoint then $\proj{f_2(v_2(h))}{pos} - \proj{f_1(v_1(h))}{pos} \equiv -1 \mod 3$.
    As~$1$ and~$-1$ are distinct modulo 3, the condition in the second statement is necessary and sufficient.
\end{proof}

We provide an \ac{LTL} formula~$\pathagr$ for ensuring (C1).
\begin{align*}
    \pathagr &:= \ltlG \bigl[(q_1 = q_2) \wedge (\zero_1 = \zero_2)\bigr]%
\end{align*}

\begin{lemma}[\leanproof{\leanurl{ControlAgreement}}]\label{lem:pathagr}
    Let~$f$ realize~$\pathagr$.
    The view complexes of~$f$ satisfy (C1).
\end{lemma}

\begin{proof}
Let~$h\in\Gamma^r$ be an edge at round~$r$ in the view complex. Since for all~$h\in\Gamma^\omega$, $\exec(f,h)\models\pathagr$, we have
that $f_1(v_1(h))_q=f_2(v_2(h))_q$ and $f_1(v_1(h))_{\zero}=f_2(v_2(h))_{\zero}$. Lemma~\ref{lem:agreement_on_path} applied to the~$q$
and the~$\zero$ components of~$f$ gives that all the vertices at round~$r$ output
the same state~$q$ and truth value for~$\zero$.
\end{proof}

Given~$q\in Q$ and~$z\in\{\top,\bot\}$ we let~$m(q,z)$ be the multplying factor applied to the number of weight-1
edges, defined
as follows: 
\begin{center}
\begin{tabular}{@{}lll@{}}
\hline
Change of counter & $m(q,z)$\\
\hline
$c\mapsto c+1$ & $1$\\
$c\mapsto c$ & $3$\\
$c\mapsto c-1$ & $9$\\
\hline
\end{tabular}
\end{center}

We give an \ac{LTL} formula~$\localsum$ for (C2).
Observe that since every component ranges over a finite set, every arithmetic constraint over them can be expressed in \ac{LTL} as a disjunction, over all the possible
valuations, of conjunctions of atomic equalities of the form~$a=p$. Hence we use such constraints as atomic propositions.
For~$p \in \left\{ 1, 2 \right\}$, let~$\bar p$ be the other process.
For~$q \in Q$ and~$z \in \left\{ \top, \bot \right\}$, define the \ac{LTL} formulas
\begin{equation}
    \psi_p(q, z) := (q_p=q \wedge \zero_p = z)\rightarrow \left(\sum_{k = 0}^2 \rightplan_p.a_k + \carry_p = m(q, z) \cdot \rightplan_p.w + \carry_{\bar{p}}\right)\enspace.
\end{equation}

Define the \ac{LTL} formula~$\localsum$ as
\begin{equation}
    \localsum := \ltlG \left( \bigwedge_{q \in Q} \bigwedge_{z \in \left\{ \top, \bot \right\}} \left( \oneisleft \rightarrow \psi_1(q, z) \right) \wedge \left( \neg\oneisleft \rightarrow \psi_2(q, z) \right) \right)\enspace.
\end{equation}

\begin{lemma}[\leanproof{\leanurl{LocalSum}}]\label{lem:local-sum}
    Let~$f$ realize~$\localsum$ and~$\posord$.
    The view complexes of~$f$ satisfy (C2).
\end{lemma}

\begin{proof}
Let~$h\in\Gamma^\omega$, and~$h'\in \Gamma^r$ its prefix of length~$r$. Assume, without loss of generality, that the left endpoint of~$h'$ in the view complex
of~$f$ is process 1's view. Since~$f$ realizes~$\posord$, by Lemma~\ref{lem:posord}, $f_2(v_2(h'))_{\pos}\equiv f_1(v_1(h'))_{\pos}+1 \mod 3$.
Then $\exec(f,h)[r\dots]\models\oneisleft$ and $\exec(f,h)[r\dots]\models \psi_1(q,z)$ for all~$q\in Q$ and~%
$z\in \{\top,\bot\}$. Let $q:=f_1(v_1(h'))_q$ and $z:=f_1(v_1(h'))_{\zero}$. Then,
$\sum_{k=0}^2f_1(v_1(h'))_{\rightplan}.a_k+ f_1(v_1(h'))_{\carry} = m(q,z) \cdot f_1(v_1(h'))_{\rightplan.w}+
f_2(v_2(h'))_{\carry}$. If the left endpoint of~$h'$ is the~$i^\textit{th}$ vertex, and corresponds to~$v_1(h')$,
then the~$(i+1)^{\textit{th}}$ vertex
is its right endpoint, and it corresponds to~$v_2(h')$, by definition of the view complex. Then the view complex of~%
$f$ satisfies~$(C2)$.
\end{proof}

We define,
\begin{align}
    \zeroind &:= \ltlG \left[ (\oneisleft \wedge (\zero_1 = \top \rightarrow \rightplan_1.w = 1)) \vee(\neg \oneisleft\wedge (\zero_2=\top\rightarrow \rightplan_2.w=1))\right]\enspace, \\
    \leftmost &:= \left[ \carry_1 = 0 \wedge \left( \zero_1 = \bot \rightarrow \rightplan_1.w = 0 \right) \right] \ltlW (\link \ne \rlink)\enspace, \\
    \rightmost &:= (\carry_2 = 0) \ltlW [\link\ne \llink]\enspace.
\end{align}
and based on this,
\begin{equation}
    \zeroclaim := \zeroind \wedge \leftmost \wedge \rightmost\enspace.
\end{equation}
For $\zeroclaim$ we then show:

\begin{lemma}[\leanproof{\leanurl{ZeroCheck}}]\label{lem:zero-claim}
    Let~$f$ realize $\zeroclaim$, $\pathagr$, and~$\posord$.
    The view complexes of~$f$ satisfy (C3).
\end{lemma}

\begin{proof}
Let~$r\geq 1$. Let~$h:=\rlink^r$ and consider any continuation~$h'\in\Gamma^\omega$.
Since~$f$ realizes~$\leftmost$, $\exec(f,h')\models  \carry_1 = 0 \ltlW (\link \ne \rlink)$.
Since for all~$d\leq r$, $\exec(f,h')[d\dots]\models \link=\rlink$, we have $\exec(f,h')[d\dots]\models
\carry_1=0$ and hence $f_1(v_1(\rlink^r))_{\carry}=0$. Similarly, since it realizes $\rightmost$,
$f_2(v_2(\llink^r))_{\carry}=0$. Hence, $\carry(r,0)=\carry(r, 3^r)=0$, by Lemma~\ref{lem:extreme-vertex}.

Assume now that $\zero(r,0)=\top$
in the view complex of~$f$, \emph{i.e.}, that $f_1(v_1(\rlink^r))_{\zero}=\top$.
Let~$h\in \Gamma^r$ and assume that its left endpoint is~$p$'s view.
Since~$f$ realizes~$\pathagr$ and~$\posord$, by Lemma~\ref{lem:pathagr}, for all~$h\in\Gamma^r$, $f_1(v_1(h))_{\zero}=f_2(v_2(h))_{\zero}$. Applying Lemma~\ref{lem:agreement_on_path} to the~$\zero$ part of the output
shows that $f_p(v_p(h))_{\zero}=\top$. Since~$f$ realizes~$\zeroind$, $f_p(v_p(h))_{\rightplan}.w=1$.

Finally, assume that $\zero(r,0)=\bot$. Let~$h:=\rlink^r$ and consider any continuation~$h'\in\Gamma^\omega$.
Since $\zero(r,0)=\bot$ in the view complex of~$f$, $f_1(v_1(h))_{\zero}=\bot$.
Since~$f$ realizes~$\leftmost$, $\exec(f,h')\models \left( \zero_1 = \bot \rightarrow \rightplan_1.w = 0 \right) \ltlW (\link \ne \rlink)$.
Since for all~$d\leq r$, $\exec(f,h')[d\dots]\models \link=\rlink$, we have $\exec(f,h')[r\dots]\models
\left( \zero_1 = \bot \rightarrow \rightplan_1.w = 0 \right)$ and hence $f_1(v_1(\rlink^r))_{\rightplan}.w=0$.
\end{proof}

To check (C4), we define the \ac{LTL} formula
\begin{equation}
    \initcheck := (q_1 = q_\init) \wedge (\zero_1=\top)\enspace.
\end{equation}

\begin{lemma}[\leanproof{\leanurl{InitialCheck}}]\label{lem:init-check}
    Let~$f$ realize $\initcheck$, $\zeroclaim$, $\pathagr$, and~$\posord$.
    The view complexes of~$f$ satisfy (C4).
\end{lemma}

\begin{proof}
Let~$\graph\in\Gamma$. Since~$f$ realizes $\initcheck$, $f_1(v_1(\graph))_q=q_\init$, and since it realizes~$\pathagr$ and~%
$\posord$,
$f_2(v_2(\graph))_q=f_1(v_1(\graph))_q=q_\init$. Hence, for every~$\graph\in\Gamma$, its two endpoints' outputs are~%
$q_\init$ on their $q$-components in the view complex of~$f$, \emph{i.e.}, $q(1,j)=q_\init$ for~$0\leq j\leq 3$. Moreover,
$f_1(v_1(\graph))_{\zero}=\top$. Hence, $\zero(1,0)=\top$ in the view complex of~$f$. Since~$f$ realizes $\zeroclaim$,
$\pathagr$ and~$\posord$,
Lemma~\ref{lem:zero-claim} implies that $\rightplan(1,j).w=1$, for all~$0\leq j\leq 2$.
\end{proof}

For~$\ell_1 \in \left\{ \top, \bot \right\}$ that is equal to~$\top$ if and only if process 1 is on the left side of the edge,
and~$\graph \in \Gamma$ define the function~$D(\tau, \graph)$ as follows:

\begin{equation}
    \begin{array}{c|c|c|c}
        D(\ell_1, \graph)  & \graph = \rlink & \graph = \dlink & \graph = \llink \\
        \hline
        \ell_1= \top & 0               & 1               & 2               \\
        \ell_1 = \bot & 2               & 1               & 0               
    \end{array}
\end{equation}
We can then define formulas~$D_k(\ell_1, \graph)$, for~$k\in\{0,1,2\}$,  that are true if and only if $D(\ell_1, \graph)=k$.
We use the following shorthand formulas: let~$x$ denote one of the components of~$\rightplan$,
\begin{align*}
\rightplan_\ell.x(v)&:= (\oneisleft \wedge \rightplan_1.x=v) \vee (\neg\oneisleft \wedge \rightplan_2.x=v)\\
D_k(\graph)&:=  (\oneisleft \wedge D_k(\top,\graph))\vee (\neg\oneisleft \wedge D_k(\bot,\graph))
\end{align*}
Define
\begin{align}
    \statetrans &:= \bigwedge_{q \in Q} \bigwedge_{z \in \left\{ \top, \bot \right\}} \left[ (q_1= q \wedge \zero_1=z) \rightarrow \ltlX \left( q_1 = \operatorname{next}(q, z) \right) \right] \\
   \weighttrans &:= \bigwedge_{k \in \left\{ 0, 1, 2 \right\}} \bigwedge_{b \in \left\{ 0, 1 \right\}}\bigwedge_{\graph\in\Gamma}  \Big[ \big( \ltlX(\link = \graph) \wedge D_k(\graph) \wedge \rightplan_\ell.a_k(b) \big)
  \rightarrow \ltlX(\rightplan_\ell.w(b)) \Big]
  \end{align}
Based on this we set:
\begin{equation}
    \proptrans := \ltlG \left( \statetrans \wedge \weighttrans \right)\enspace.
\end{equation}

We can now show the following:

\begin{lemma}[\leanproof{\leanurl{ProperTransition}}]\label{lem:prop-trans}
    Let~$f$ realize $\proptrans$ and~$\posord$.
    The view complexes of~$f$ satisfy (C5).
\end{lemma}

\begin{proof}
Let~$r\geq 1$, with $q(r,0)=q\in Q$, and $\zero(r,0)=z\in\{\top,\bot\}$ in the view complex of~$f$.
Since~$f$ realizes $\proptrans$, we have $f_1(v_1(\rlink^{r+1}))_{q}=\operatorname{next}(q,z)$; hence
$q(r+1,0)=\operatorname{next}(q(r,0),\zero(r,0))$. 

Let~$h\in\Gamma^r$ and~$\graph\in\Gamma$, and let~$h'\in\Gamma^\omega$ be a continuation of~$h\graph$.
Assume that the left endpoint of~$h$ in the inputless view complex is the~$i^\textit{th}$ vertex.
Following the description of the mapping of the inputless view complex
into~$[0,1]$ in
Section~\ref{sec:graph-history-interval}, we know that if~$p$ is the left endpoint of~$h$, then~$\bar{p}$ is the left endpoint of~$h\graph$
if and only if~$\graph=p\leftrightarrow\bar{p}$. Moreover, condition (C5) requires that
\begin{align*}
    \text{if }\graph=p\rightarrow\bar p,\qquad \rightplan_p(r,i).a_0 &= \rightplan_p(r+1,3i).w, \\
    \text{if }\graph=p\leftrightarrow\bar p,\qquad \rightplan_p(r,i).a_1 &= \rightplan_{\bar p}(r+1,3i+1).w, \\
    \text{otherwise if }\graph=p\leftarrow\bar p,\qquad \rightplan_p(r,i).a_2 &= \rightplan_p(r+1,3i+2).w\enspace.
\end{align*}

Assume that
process 1 owns the left endpoint of~$h$ in the view complex of~$f$. Then, since~$f$ realizes~$\posord$,
by Lemma~\ref{lem:posord}, $\exec(f,h')[r\dots]\models \oneisleft$. Moreover, if~$\graph=\rlink$,
$D_0(\graph)$ holds. If $f_1(v_1(h))_{\rightplan}.a_0=b$, then, since $\exec(f,h')[r\dots]\models \weighttrans$,
we conclude that $f_1(v_1(h\graph))_{\rightplan}.w=b$. By construction of the mapping, this implies that $w(r+1, 3i)=b$.
If~$\graph=1\leftrightarrow 2$, then, $D_1(\graph)$ is true. Moreover, again by Lemma~\ref{lem:posord},
$\exec(f,h')[r+1\dots]\models\neg \oneisleft$. Then, if $a_1(r,i)=b$, then $w(r+1, 3i+1)=b$.
We can repeat this reasoning for all the other graphs, and similarly if process~2 owns the left endpoint of~$h$.
\end{proof}

Define the formula
\begin{equation}
    \phi(M_\single) := \posord \wedge \pathagr \wedge \localsum \wedge \zeroclaim \wedge \initcheck \wedge \proptrans\wedge \ltlG\left(\bigwedge_{q \in F} q_1 \ne q\right)\enspace.
\end{equation}
We next show:

\begin{lemma}[\leanproof{\leanurl{RowConditions}}]
    \label{lem:conditions<=>realizability}
    Algorithm~$f$ realizes $\phi(M_\single)$ if and only if it satisfies conditions (C0)--(C5)
    and no halting state is output.
\end{lemma}

\begin{proof}
($\Rightarrow$):
Assume that~$f$ realizes $\phi(M_\single)$. Then, Lemmas~\ref{lem:posord}, \ref{lem:pathagr},\ref{lem:local-sum}, \ref{lem:zero-claim}, \ref{lem:init-check}, and \ref{lem:prop-trans} ensure that the view complex of~$f$ meets the
requirements (C0)--(C5).
Moreover, the conjunct $\ltlG(\bigwedge_{q\in F} q_1\ne q)$ excludes halting states from process~1's outputs,
and~$\pathagr$ gives the same conclusion for process~2.

($\Leftarrow$):
Suppose that~$f$ satisfies (C0)--(C5) and outputs no halting state.
We next show that~$f$ realizes each conjunct of~$\phi(M_\single)$.

\smallskip
\noindent\textbf{$f$ realizes~$\posord$.} Indeed, such a view complex implies that
\begin{align*}
    f_1(v_1(\rlink))_{\pos} &= 0, \\
    f_2(v_2(\rlink))_{\pos} &= f_2(v_2(\dlink))_{\pos}=1, \\
    f_1(v_1(\dlink))_{\pos} &= f_1(v_1(\llink))_{\pos}=2, \\
    \text{and}\qquad f_2(v_2(\llink))_{\pos} &= 0\enspace.
\end{align*}
So, for all~$h\in\Gamma^\omega$, $\exec(f,h)\models \bigwedge_{\graph\in\Gamma} \bigl(\link=\graph \rightarrow (\pos_1, \pos_2) = \childpos_\graph(0, 1)\bigr)$.

Let~$h\in\Gamma^r$ have its left endpoint at vertex~$i$, and set
$u:=f_1(v_1(h))_{\pos}$ and $v:=f_2(v_2(h))_{\pos}$.
If the left endpoint is process~1's view, then (C0) gives~%
$u=i\bmod 3$ and $v=(i+1)\bmod 3$.
The pairs of integer positions of processes~1 and~2 in~%
$h\rlink$, $h\dlink$, and~$h\llink$ are, respectively,
$(3i,3i+1)$, $(3i+2,3i+1)$, and $(3i+2,3i+3)$.
Reducing each coordinate modulo~3 gives
$\childpos_{\rlink}(u,v)$, $\childpos_{\dlink}(u,v)$, and
$\childpos_{\llink}(u,v)$, respectively.

If the left endpoint is process~2's view, then (C0) gives
$u=(i+1)\bmod 3$ and~$v=i\bmod 3$.
The pairs of integer positions in~$h\rlink$, $h\dlink$, and~$h\llink$
are now $(3i+3,3i+2)$, $(3i+1,3i+2)$, and $(3i+1,3i)$.
Again, reducing each coordinate modulo~3 gives the corresponding
value of $\childpos_\graph(u,v)$.
Thus, in both cases, (C0) yields, for every~$\graph\in\Gamma$,
\[
    \bigl(f_1(v_1(h\graph))_{\pos}, f_2(v_2(h\graph))_{\pos}\bigr)
    = \childpos_\graph(u,v)\enspace.
\]
So, for all~$h'\in\Gamma^\omega$, $\exec(f,h')\models \ltlG
    \left( \bigwedge_{u\in\ZZ_3}
        \bigwedge_{v\in\ZZ_3}
        \bigwedge_{\graph\in\Gamma}
    \phi(u,v, \graph)\right)$. Hence, $f$ realizes~$\posord$.
    
    \smallskip
    \noindent\textbf{$f$ realizes~$\pathagr$.} Since this view complex also satisfies (C1),~$f$ realizes~$\pathagr$. Indeed, otherwise there
    exist~$r\geq 1$ and~$h\in \Gamma^r$ such that $f_1(v_1(h))_q\neq f_2(v_2(h))_q$ or $f_1(v_1(h))_\zero\neq
    f_2(v_2(h))_\zero$. But then, the two endpoints of~$h$ are labeled by two different values on their~$q$ or
    $\zero$-components, contradicting (C1). 
    
    \smallskip
    \noindent\textbf{$f$ realizes~$\localsum$.}
    Let~$r\geq 1$, $h\in\Gamma^r$, and~$h'\in\Gamma^\omega$ a continuation of~$h$.
     Assume that the left endpoint of~$h$ is process 1's view (then its right endpoint is process 2's). Since~$f$
    realizes~$\posord$, $f_2(v_2(h))_\pos \equiv f_1(v_1(h))_\pos +1 \mod 3$, by Lemma~\ref{lem:posord}. Hence,
    $\exec(f,h')[r\dots]\models\oneisleft$. Moreover, if $f_1(v_1(h))_q=q$ and $f_1(v_1(h))_\zero = z$,
    condition (C2) ensures that
    \[
        \sum_{k=0}^2f_1(v_1(h))_\rightplan.a_k + f_1(v_1(h))_\carry
        = m(q,z)\cdot f_1(v_1(h))_{\rightplan}.w + f_2(v_2(h))_\carry\enspace,
    \]
    and $\exec(f,h')[r\dots] \models \psi_1(q,z)$. If
    its left endpoint is process 2's view, then $\exec(f,h')[r\dots]\models\neg \oneisleft$ and again,
    one can see that (C2) yields that $\exec(f,h')[r\dots] \models \psi_2(q,z)$. Hence~$f$ realizes~$\localsum$.
    
    \smallskip
    \noindent\textbf{$f$ realizes $\zeroclaim$.} Let~$r\geq 1$ such that $\zero(r,0)=\top$. Let~$h\in\Gamma^r$, and~$h'\in\Gamma^\omega$ be
    a continuation of~$h$. Assume that the left endpoint of~%
    $h$ is process 1's view. Then again, since the view complex satisfies (C0), 
    $\exec(f,h')[r\dots]\models\oneisleft$. Since the view complex satisfies (C1), $f_1(v_1(h))_\zero=\zero(r,0)=\top$. Since the view complex satisfies (C3), we know that $f_1(v_1(h))_{\rightplan}.w=1$,
    hence
    \[
        \exec(f,h')[r\dots] \models \oneisleft \wedge (\zero_1=\top\rightarrow \rightplan_1.w=1)\enspace.
    \]
    If the left endpoint of~$h$ is process 2's view, we have
    \[
        \exec(f,h')[r\dots]\models\neg\oneisleft\wedge (\zero_2=\top\rightarrow \rightplan_2.w=1)\enspace,
    \]
    and~$f$ realizes~$\zeroind$.

    Let~$h:=(\rlink)^r$, and~%
    $h'\in\Gamma^\omega$ a continuation of~$h$. By construction, the view of process 1 of~%
    $h$ is the leftmost vertex of the view complex at round~$r$. Since $\carry(r,0)=0$,
    $\exec(f,h')[r\dots]\models\carry_1=0$. Moreover, if $\zero(r,0)=f_1(v_1(h))_\zero=\bot$,
    $\rightplan(r,0).w=f_1(v_1(h))_\rightplan.w=0$, hence $\exec(f,h')[r\dots]\models \zero_1=\bot\rightarrow
    (\rightplan_1.w=0)$. Since this holds for all~$r\geq 1$,
    \[
        \exec(f,h')\models\left[ \carry_1 = 0 \wedge \left( \zero_1 = \bot \rightarrow \rightplan_1.w = 0 \right) \right] \ltlW (\link \ne \rlink)\enspace.
    \]

    Similarly, let~$h:=(\llink)^r$, and~%
    $h'\in\Gamma^\omega$ a continuation of~$h$. By construction, the view of process 2 of~%
    $h$ is the rightmost vertex of the view complex at round~$r$. Since $\carry(r,3^r)=0$,
    $\exec(f,h')[r\dots]\models\carry_2=0$. This being true for any~$r\geq 1$, we have that
    \[
        \exec(f,h')\models (\carry_2=0)\ltlW (\link\neq\llink)\enspace.
    \]
     
    \smallskip
    \noindent\textbf{$f$ realizes $\initcheck$.} Condition (C4) implies that $f_p(v_p(\graph))_q=q_\init$, for~$p\in\{1,2\}$ and~%
    $\graph\in\Gamma$. Moreover, (C4) gives $f_1(v_1(\rlink))_\rightplan.w=1$.
    If $f_1(v_1(\graph))_\zero=\bot$, then, since by Path Agreement condition (C1), $f_1(v_1(\rlink))_\zero=\bot$,
    Zero Claim Check condition (C3) states that $f_1(v_1(\rlink))_\rightplan.w=0$, which is a contradiction.
    Hence $f_1(v_1(\graph))_\zero=\top$, for all~$\graph\in\Gamma$, and for all~$h\in\Gamma^\omega$,
    $\exec(f,h)\models\initcheck$.
    
    \smallskip
    \noindent\textbf{$f$ realizes $\proptrans$.} The view complex of~$f$ satisfies Proper Transition condition (C5), hence
    at every round~$r$ of the view complex, $q(r+1,0)=\operatorname{next}(q(r,0), \zero(r,0))$. Moreover, since it also satisfies
    Path Agreement condition (C1), for all~%
    $h\in\Gamma^r$ and~$\graph\in\Gamma$,
    \begin{align*}
        f_1(v_1(h\graph))_q &= f_1(v_1(\rlink^{r+1}))_q \\
        &= \operatorname{next}(f_1(v_1(\rlink^r))_q, f_1(v_1(\rlink^r))_\zero)\enspace.
    \end{align*}
    Again, by the Path Agreement condition,
    \begin{align*}
        f_1(v_1(\rlink^r))_q &= f_1(v_1(h))_q \\
        \text{and}\qquad f_1(v_1(\rlink^r))_\zero &= f_1(v_1(h))_\zero\enspace.
    \end{align*}
    Hence, $f$ realizes $\ltlG\statetrans$.

    Assume now that~$f$ does not realize $\ltlG \weighttrans$. There exist~$r\geq 1$, $h\in\Gamma^r$, $\graph\in \Gamma$, $0\leq k \leq 2$, $0\leq b\leq 1$ such that
    \begin{align*}
        f_2(v_2(h))_\pos &\equiv f_1(v_1(h))_\pos + 1 \mod 3, \\
        f_1(v_1(h))_\rightplan.a_k &= b, \\
        D(\top, \graph) &= k \\
        \text{and}\qquad f_p(v_p(h\graph))_\rightplan.w &\neq b
    \end{align*}
    with~$p$ the process
    whose view is the left endpoint of~$h\graph$ (the case where $f_2(v_2(h))_\pos \not\equiv f_1(v_1(h))_\pos + 1 \mod 3$
    is symmetric). By Lemma~\ref{lem:j}, if the left endpoint of~$h$ is~$i$ (even), the left endpoint of~$h\graph$
    is the $(3i+ D(\top, \graph))^{\textit{th}}$ vertex at round~$r+1$. But Proper Transition condition (C5) ensures that
    \[
        \rightplan_p(r+1, 3i+k).w=\rightplan_1(r,i).a_k\enspace,
    \]
    which is a contradiction.
    Hence~$f$ realizes $\ltlG \weighttrans$ and $\ltlG\statetrans$, and therefore $\proptrans$.

    Finally, since no halting state is output, $f$ realizes $\ltlG(\bigwedge_{q\in F} q_1\ne q)$.
    Together with the preceding conjuncts, this shows that~$f$ realizes $\phi(M_\single)$.
\end{proof}

\paragraph{Correctness of the reduction.}

We first recover a non-halting run of~$M_\single$ from any algorithm realizing~$\phi(M_\single)$.
\begin{lemma}[\leanproof{\leanurl{CounterSoundness}}]
    Suppose that there exists~$f$ that realizes $\phi(M_\single)$.
    Then~$M_\single$ does not halt.
    \label{lem:soundness}
\end{lemma}
\begin{proof}
    Lemma~\ref{lem:conditions<=>realizability} gives that~$f$ satisfies the conditions (C0)--(C5).
    For~$r \ge 1$, Path Agreement gives that~$q(r, i)$ and $\zero(r, i)$ are constant over~$i \le 3^r$.
    We denote these constants by~$q[r]$ and~$\zero[r]$, using square brackets for the resulting sequences indexed by round.
    For~$i \le 3^r - 1$, define
    \[
        w(r, i) := \rightplan(r, i).w, \quad \pathsum[r] := \sum_{i = 0}^{3^r - 1} w(r, i), \quad C[r] := -\log_3(\pathsum[r] \cdot 3^{-r}).
    \] 
    From Local Sum Check, the boundary carry conditions in Zero Claim Check, and Proper Transition, we have
    \[
    \pathsum[r + 1] = m(q[r], \zero[r]) \cdot \pathsum[r].
    \] 
    As~$\pathsum[1] = 3$ and $m(q[r], \zero[r]) \in \left\{ 1, 3, 9 \right\}$, we get that~$\pathsum[r] \ge 3$ and hence~$C[r] \ge 0$ for all~$r \ge 1$.

    For~$r \ge 1$, Zero Claim Check gives that $\zero[r] = \top$ if and only if $\forall\,i \le 3^{r} - 1, \,w(r, i) = 1$, which holds if and only if~$C[r] = 0$.

    We prove the following by induction: $(q[1], C[1]), (q[2], C[2]), \dots$ is the non-halting run of~$M_\single$.
    The Initial Check condition gives that $q[1] = q_{\init}$ and~$C[1] = 0$, so the statement holds for~$r = 1$.
    Suppose that $( q[r], C[r])$ is the configuration of~$M_\single$ at step~$r$ of its run.
    The Proper Transition condition gives that $q[r + 1] = \operatorname{next}(q[r], \zero[r])$.
    As~$C[r] = 0$ if and only if $\zero[r] = \top$, the state of~$M_\single$ at step~$r + 1$ is equal to~$q[r + 1]$.

    We show the counter value is correctly simulated.
    We note that:
    \begin{align*}
        3^{-C[r + 1]} = \pathsum[r + 1] \cdot 3^{-(r + 1)} &= \frac{m(q[r], \zero[r])}{3} \cdot \pathsum[r] \cdot 3^{-r} = \frac{m(q[r], \zero[r])}{3} \cdot 3^{-C[r]}\enspace.
    \end{align*}
    Doing casework on~$q[r]$ and~$\zero[r]$ gives:
    \begin{itemize}
        \item If~$\Delta(q[r])$ is an increment operation, then $m(q[r], \zero[r]) = 1$, so $C[r + 1] = C[r] + 1$,
        \item if $\zero[r] = \top$ and~$\Delta(q[r])$ is a decrement operation, then $m(q[r], \zero[r]) = 3$, so $C[r + 1] = C[r]$,
        \item if $\zero[r] = \bot$ and~$\Delta(q[r])$ is a decrement operation, then $m(q[r], \zero[r]) = 9$, so $C[r + 1] = C[r] - 1$.
    \end{itemize}
    Thus the value of~$C[r + 1]$ corresponds to the counter value of~$M_\single$ in step~$r + 1$.
    This concludes the inductive step.
    Thus $(q[1], C[1]), (q[2], C[2]), \dots$ is the run of~$M_\single$.
    As~$f$ realizes $\ltlG(\bigwedge_{q \in F} q_1 \ne q)$, we have that~$q[r] \notin F$ for all~$r \ge 1$.%
    Thus the run is non-halting.%
\end{proof}

It remains to show that there exists a distributed algorithm whose view complex satisfies (C0)--(C5).
The following lemma gives a recursive procedure for a process to compute the position of its view in its view complex.
\begin{lemma}[\leanproof{\leanurl{FindIndex}}]
    \label{lem:find_index}
    Let~$\ind(v)$ denote the position of view~$v$ in its view complex.
    For~$h \in \Gamma^*$, denoting $v_p := v_p(h)$ and~$v_{\bar{p}} := v_{\bar{p}}(h)$,
    \begin{equation}
        \begin{aligned}
            &\ind(\mathsf{I}_1) = 0, \quad &\ind(\mathsf{N}_p(v_p)) = 3 \cdot \ind(v_p), \quad &\ind(\mathsf{C}_p(v_p, v_{\bar{p}})) = \ind(v_p) + 2 \cdot \ind(v_{\bar{p}})\enspace, \\
            &\ind(\mathsf{I}_2) = 1, \quad  &\ind(\mathsf{N}_{\bar{p}}(v_{\bar{p}})) = 3 \cdot \ind(v_{\bar{p}}), \quad &\ind(\mathsf{C}_{\bar{p}}(v_{\bar{p}}, v_p)) = \ind(v_{\bar{p}}) + 2 \cdot \ind(v_{p})\enspace.
        \end{aligned}
    \end{equation}
\end{lemma}
\begin{proof}
    Views~$\mathsf{I}_1$ and~$\mathsf{I}_2$ have positions 0 and 1 by construction.
    Without loss of generality, assume that~$v_p$ is on the left vertex of the edge corresponding to~$h$, so $\ind(v_{\bar{p}}) = \ind(v_p) + 1$.
    From Figure~\ref{fig:SubdivLabels} it is straightforward to see that $\ind(\mathsf{N}_p(v_p)) = 3 \cdot \ind(v_p)$ and $\ind(\mathsf{N}_{\bar{p}}(v_{\bar{p}})) = 3 \cdot \ind(v_{\bar{p}})$.
    It also follows from the same figure that $\ind(\mathsf{C}_{\bar{p}}(v_{\bar{p}}, v_{p})) = 3 \cdot \ind(v_p) + 1 = \ind(v_{\bar{p}}) + 2 \cdot \ind(v_p)$ and $\ind(\mathsf{C}_p(v_p, v_{\bar{p}})) = 3 \cdot \ind(v_p) + 2 = \ind(v_{p}) + 2 \cdot \ind(v_{\bar{p}})$.
\end{proof}
Thus at each round of a history, the processes can compute the position of their current views using~$\ind$.

\begin{algorithm}
    \label{algo:completeness}
    \caption{Algorithm~$f$ at process~$p$}
    \For{round $r = 1, 2, \dots$}{
        Send and receive messages, update view accordingly\;\label{algo:start_round}
        \leIf{r = 1}{$(q, \counter) \gets (q_{\init}, 0)$}{$(q, \counter) \gets$ successor configuration of~$(q, \counter)$ under~$\Delta(q)$} \label{line:initialstate}
        $\view \gets$ view of~$p$\;
        $\realpos \gets \ind(\view)$\;
        \leIf{$\counter = 0$}{$\zero \gets \top$}{$\zero \gets \bot$} \label{line:zero}
        $\mul \gets m(q, \zero)$\;
        $w \gets \left\lfloor (\realpos + 1)/3^{\counter} \right\rfloor - \left\lfloor \realpos/3^\counter \right\rfloor$\; \label{line:weight}
        $\carry \gets \left\lfloor \mul \cdot \realpos/3^\counter \right\rfloor - \mul \left\lfloor \realpos/3^{\counter} \right\rfloor$\;
        $\counter' \gets $ value of counter obtained from successor configuration of~$(q, \counter)$ under~$\Delta(q)$\;
        \For{k = 0, 1, 2}{
            $a_k \gets \left\lfloor (3 \cdot \realpos + k + 1)/3^{\counter'} \right\rfloor - \left\lfloor (3 \cdot \realpos + k)/3^{\counter'} \right\rfloor$\;\label{line:promise}
        }
        $\rightplan \gets (w, a_0, a_1, a_2)$\;
        \Return{$(q, \zero, \rightplan, \carry, \realpos \bmod 3)$}\;\label{algo:output_tuple}
    }
\end{algorithm}

\begin{lemma}[\leanproof{\leanurl{CounterCompleteness}}]
    \label{lem:completeness}
    If~$M_\single$ does not halt, then~$f$ given by Algorithm~\ref{algo:completeness} realizes $\phi(M_\single)$.
\end{lemma}
\begin{proof}
    In the view complex of~$f$ at round~$r$, at the $i$-th vertex whose view is that of~$p$, for every variable $X \in \left\{ q, \counter, \view, \realpos, \mul, \carry, \counter', w, a_0, a_1, a_2 \right\}$, we let~$X(r, i)$ denote the value of~$X$ at~$p$ in round~$r$.

    Using the fact that $x - 1 < \left\lfloor x \right\rfloor \le x$ gives that in any round, $w = \left\lfloor (\realpos + 1)/3^{\counter} \right\rfloor - \left\lfloor \realpos/3^\counter \right\rfloor < (\realpos + 1)/3^{\counter} - \realpos/3^\counter + 1 \le 2$, implying that~$w \in \left\{ 0, 1 \right\}$, and $\carry = \left\lfloor \mul \cdot \realpos/3^\counter \right\rfloor - \mul \left\lfloor \realpos/3^{\counter} \right\rfloor < \mul \cdot \realpos/3^\counter - \mul(\realpos/3^\counter - 1) = \mul \le 9$, implying that $\carry \in \left\{ 0, \dots, 8 \right\}$.

    Lemma~\ref{lem:find_index} implies that $\realpos(r, i) = \ind(\view) = i$, so~$f$ satisfies the Position Order condition.

    Let the run of~$M_\single$ be $(q_1, c_1), (q_2, c_2), \dots$ where $(q_1, c_1) = (q_{\init}, 0)$.
    We prove by induction that $q(r, i) = q_r$ and $\counter(r, i) = c_r$ for all~$i \le 3^r$.
    By line~\ref{line:initialstate}, $q(1, i) = q_{\init}$ and $\counter(1, i) = 0$ for~$0 \le i \le 3$.
    In every round~$r \ge 2$, the values of~$q$ and~$\counter$ computed in line~\ref{line:initialstate} depend only on their values in the previous round.
    So if a process always has~$q = q_r$ and~$\counter = c_r$ in round~$r$ of any run, then it always has~$q = q_{r + 1}$ and~$\counter = c_{r + 1}$ in round~$r + 1$, which completes the inductive step.
    From line~\ref{line:zero}, we deduce that $\zero(r, i)$ is constant over all~$i \le 3^{r}$.
    Thus~$f$ satisfies Path Agreement.
    We henceforth write~$q[r]$, $\counter[r]$, $\counter'[r]$, and~$\zero[r]$ for the common values at round~$r$.

    We now prove that $\sum_{k = 0}^2 a_k(r, i) + \carry(r, i) = m(q[r], \zero[r]) \cdot w(r, i) + \carry(r, i + 1)$.
    This can be derived from the following formulas, thus showing that~$f$ satisfies the Local Sum Check condition.

    For all~$k\in\{0,1,2\}$ and~$j\in\{i,i+1\}$, we have
    \begin{align*}
        a_k(r, i) &= \left\lfloor \frac{3 \cdot \realpos(r, i) + k + 1}{3^{\counter'[r]}} \right\rfloor - \left\lfloor \frac{3 \cdot \realpos(r, i) + k}{3^{\counter'[r]}} \right\rfloor \\
        w(r, i) &= \left\lfloor \frac{\realpos(r, i) + 1}{3^{\counter[r]}} \right\rfloor - \left\lfloor \frac{\realpos(r, i)}{3^{\counter[r]}} \right\rfloor \\
        \carry(r, j) &= \left\lfloor \frac{mul(r, j) \cdot \realpos(r, j)}{3^{\counter[r]}} \right\rfloor - mul(r, j) \left\lfloor \frac{\realpos(r, j)}{3^{\counter[r]}} \right\rfloor \\
        \counter'[r] &= \counter[r] + 1 - \log_3 m(q[r], \zero[r]) \\
        mul(r, i) &= m(q[r], \zero[r])\enspace.
    \end{align*}

    If $\zero[r] = \top$, then~$\counter[r] = 0$ and so for all~$i \le 3^r - 1$, we have $w(r, i) = \left\lfloor \realpos + 1 \right\rfloor - \left\lfloor \realpos \right\rfloor = 1$.
    If $\zero[r] = \bot$, then~$\counter[r] \ge 1$ and so $w(r, 0) = \left\lfloor 1/3^{\counter[r]} \right\rfloor = 0$.
    As~$\realpos(r, 0) = 0$, using the above formula for $\carry(r, j)$ with~$j = 0$ gives $\carry(r, 0) = 0$.
    Finally, for~$r \ge 1$, since~$\counter[r] < r$, we have that~$3^{\counter[r]}$ divides $\realpos(r, 3^r) = 3^r$, giving that $\carry(r, 3^r) = 0$.
    Thus~$f$ satisfies Zero Claim Check.

    In round~$r = 1$, $q = q_{\init}$ and~$\counter = 0$, so~$w(1, i) = 1$ for~$0 \le i \le 2$.
    Thus~$f$ satisfies Initial Check.

    From lines~\ref{line:initialstate} and~\ref{line:zero}, we have that $q[r + 1] = \operatorname{next}(q[r], \zero[r])$. %
    Lastly, by line~\ref{line:promise} in round~$r$ we have $a_k(r, i) = \left\lfloor (3 \cdot \realpos(r, i) + k + 1)3^{-\counter'[r]} \right\rfloor - \left\lfloor (3 \cdot \realpos(r, i) + k)3^{-\counter'[r]} \right\rfloor$ and by line~\ref{line:weight} in round~$r + 1$ we have $w(r + 1, 3i + k) = \left\lfloor (3 \cdot \realpos(r, i) + k + 1)3^{-\counter[r + 1]} \right\rfloor - \left\lfloor (3 \cdot \realpos(r, i) + k)3^{-\counter[r + 1]} \right\rfloor$.
    Both~$\counter'[r]$ and~$\counter[r + 1]$ are equal to the counter value in the successor configuration of $(q[r], \counter[r])$ under~$\Delta(q[r])$, thus $a_k(r, i) = w(r + 1, 3i + k)$.
    Thus~$f$ satisfies the Proper Transition condition.
    Since~$M_\single$ does not halt and $q(r,i)=q_r$, no halting state is output.
    By Lemma~\ref{lem:conditions<=>realizability}, $f$ realizes $\phi(M_\single)$.
\end{proof}

Finally, Lemmas~\ref{lem:soundness} and~\ref{lem:completeness} yield the following characterization.
\begin{theorem}[\leanproof{\leanurl{CounterEquivalence}}]
    \label{thm:nonhalt<=>realizability}
    Machine~$M_\single$ does not halt if and only if $\phi(M_\single)$ is realizable.
\end{theorem}

\subsection{Extension to General Not-Nowhere-Dense Adversaries}

We first need a general observation on safety properties. 

\begin{lemma}[\leanproof{\leanurl{SafetyClosure}}]\label{lem:safety:is:safe}
    Let~$\phi$ be a safety formula and let~$f$ be a distributed algorithm.
    Then~$f$ realizes~$\phi$ under~$A$ if and only if it realizes~$\phi$ under~$\overline{A}$.
\end{lemma}
\begin{proof}
    ($\Rightarrow$):
    Assume by contradiction that~$f$ realizes~$\phi$ under~$A$ and that it does not realize~$\phi$ under~$\overline{A}$.
    In particular, there exists a graph history~$h\in \overline{A}\setminus A$ such that $\exec(f,h) \not\in L(\phi)$.
    Since~$L(\phi)$ is a safety property, there exists a finite round number~$R$ such that $\exec(f,h') \not\in L(\phi)$ whenever $\exec(f,h')$ agrees with $\exec(f,h)$ on the first~$R$ rounds.
    But since the round-$R$ prefix of $\exec(f,h')$ only depends on the round-$R$ prefix of~$h'$, we deduce that $\exec(f,h') \not\in L(\phi)$ whenever the round-$R$ prefixes of~$h$ and~$h'$ are equal, \emph{i.e.}, when $d(h,h') < 2^{-R}$.
    But since~$A$ is dense in~$\overline{A}$, there is an~$h'\in A$ with $d(h,h') < 2^{-R}$.
    Hence $\exec(f,h') \not\in L(\phi)$ for some~$h'\in A$, a contradiction.

    ($\Leftarrow$):
    This is trivial since~$A\subseteq \overline{A}$.
\end{proof}

We now show how we can exploit the fact that~$A$ is not nowhere-dense to obtain undecidability.

\begin{definition}[\leanformalized{\leanline{Library/Distributed/GraphAdversaries/Cells}{101}}]\label{def:live-history}
    Let~$A\subseteq\Gamma^\omega$ be a message adversary.
    A finite graph history~$h\in\Gamma^*$ is \emph{live} under~$A$ if $h\in\Pref(A)$, \emph{i.e.}, if there exists~$k\in\Gamma^\omega$ such that~$h\cdot k\in A$.
\end{definition}
An edge of the full inputless view complex is called \emph{live} under~$A$ if its corresponding graph history is live under~$A$.

\begin{lemma}[\leanproof{\leanurl{MissingExtension}}]\label{lem:nowhere-dense-live}
    Let~$A\subseteq\Gamma^\omega$ be a message adversary.
    Then~$\pi(A)$ is nowhere dense in~$[0,1]$ if and only if, for every~$h\in\Gamma^*$, there exists~$k\in\Gamma^*$ such that~$h\cdot k$ is not live under~$A$.
\end{lemma}
\begin{proof}
    By \Cref{lem:j}, the intervals~$J(u)$ with~$u\in\Gamma^r$ form a uniform subdivision of~$[0,1]$ into intervals of length~$3^{-r}$.
    Hence every nonempty open interval in~$[0,1]$ contains some interval~$J(u)$.

    ($\Rightarrow$):
    Let~$h\in\Gamma^*$.
    Since~$\pi(A)$ is nowhere dense, there exists a nonempty open interval~$U\subseteq J(h)^\circ$ disjoint from~$\pi(A)$.
    Choose~$u\in\Gamma^*$ such that~$J(u)\subseteq U$.
    By \Cref{J:inclusion:iff:prefix}, we have~$u=h\cdot k$ for some~$k\in\Gamma^*$.
    If~$h\cdot k$ were live, an infinite extension belonging to~$A$ would have its image under~$\pi$ in~$J(h\cdot k)\subseteq U$, a contradiction.

    ($\Leftarrow$):
    Let~$0\leq a<b\leq 1$.
    Choose~$h\in\Gamma^*$ such that $J(h)\subseteq(a,b)$, and then~$k\in\Gamma^*$ such that~$h\cdot k$ is not live.
    By \Cref{pi:inverse:J:interior},
    \begin{equation}
        J(h\cdot k)^\circ\cap\pi(A)
        \subseteq \pi\bigl(A\cap h\cdot k\cdot\Gamma^\omega\bigr)
        =\emptyset
        \enspace.
    \end{equation}
    By \Cref{J:inclusion:iff:prefix}, the nonempty open interval~$J(h\cdot k)^\circ$ lies in~$(a,b)$.
    Thus every nonempty open interval contains a nonempty open subinterval disjoint from~$\pi(A)$, proving nowhere denseness.
\end{proof}

\begin{lemma}[\leanproof{\leanurl{DenseResidual}}]\label{lem:not:nowhere:dense:is:dense:somewhere}
    If~$\pi(A)$ is not nowhere dense, then there exists a finite graph history~$h\in\Gamma^*$ such that~$h^{-1}A$ is dense in~$\Gamma^\omega$.
\end{lemma}
\begin{proof}
    By \Cref{lem:nowhere-dense-live}, there exists an~$h\in\Gamma^*$ such that~$h\cdot k$ is live for every~$k\in\Gamma^*$.
    Let~$k\in\Gamma^\omega$.
    Denote by~$k[\dots r]$ the round-$r$ prefix of~$k$.
    Since $h\cdot k[\dots r]$ is live, there exists an~$m_r\in\Gamma^\omega$ such that $h\cdot k[\dots r]\cdot m_r \in A$.
    We have $k[\dots r] \cdot m_r \in h^{-1}A$ and $d(k , k[\dots r]\cdot m_r) < 2^{-r}$.
    Hence $k[\dots r]\cdot m_r \to k$ as~$r\to\infty$.
    This shows that~$k\in \overline{h^{-1}A}$ and concludes the proof.
\end{proof}

For the next lemma, use the history~$h$ and formula~$\widehat{\phi}$ constructed at the end of \Cref{sec:undecidability}.

\begin{lemma}[\leanproof{\leanurl{GuardedReduction}}]\label{lem:guarded-safety-reduction}
    A safety formula~$\phi$ is realizable under~$\Gamma^\omega$ if and only if~$\widehat{\phi}$ is realizable under~$A$.
\end{lemma}
\begin{proof}
    ($\Rightarrow$):
    Let~$f$ be a distributed algorithm that realizes~$\phi$ under~$\Gamma^\omega$.
    Define algorithm~$\hat{f}$ by outputting arbitrarily during the first~$\lvert h\rvert$ rounds and then following~$f$ after compressing the initial round-$|h|$ views into a single initial view for each process.
    Then~$\hat{f}$ realizes~$\ltlX^{\lvert h\rvert} \phi$ under~$h\cdot\Gamma^\omega$.
    Now let~$k\in A$.
    If~$k \not\in h\cdot\Gamma^\omega$, then $\exec(\hat{f}, k) \not\models \operatorname{Prefix}_h$, so $\exec(\hat{f}, k) \models \widehat{\phi}$.
    If~$k \in h\cdot\Gamma^\omega$, then $\exec(\hat{f}, k) \models \operatorname{Prefix}_h$ and $\exec(\hat{f}, k) \models \ltlX^{|h|} \phi$, so $\exec(\hat{f}, k) \models \widehat{\phi}$ as well.

    ($\Leftarrow$):
    Let~$\hat{f}$ be a distributed algorithm that realizes~$\widehat{\phi}$ under~$A$.
    The formula~$\widehat{\phi}$ is a safety formula.
    Indeed, if an execution~$e$ violates~$\widehat{\phi}$, its graph prefix is~$h$ and its suffix $e[\lvert h\rvert+1\dots]$ violates~$\phi$.
    A finite bad prefix of this suffix, together with the first~$\lvert h\rvert$ rounds, forms a finite bad prefix for~$\widehat{\phi}$.
    By \Cref{lem:safety:is:safe}, algorithm~$\hat{f}$ realizes~$\widehat{\phi}$ under~$\overline{A}$.
    Since~$h^{-1}A$ is dense in~$\Gamma^\omega$, every~$h\cdot k \in h\cdot\Gamma^\omega$ is a limit of graph histories in~$A$.
    Thus~$h\cdot \Gamma^\omega \subseteq \overline{A}$, and~$\hat{f}$ realizes~$\ltlX^{\lvert h\rvert}\phi$ under~$h\cdot \Gamma^\omega$.
    Now define~$f$ by replacing the initial view of process~$p$ by~$v_p(h)$ and applying~$\hat{f}$.
    Then~$f$ realizes~$\phi$ under~$\Gamma^\omega$.
\end{proof}

Fix~$A$ such that~$\pi(A)$ is not nowhere dense.
By \Cref{lem:not:nowhere:dense:is:dense:somewhere}, choose a finite graph
  history~$h$ such that~$h^{-1}A$ is dense in~$\Gamma^\omega$.
For a 2-counter machine~$M$, apply the counter encoding to both counters,
  sharing the control state.
The resulting safety formula~$\phi(M)$ is realizable under~$\Gamma^\omega$
  if and only if~$M$ does not halt, by \Cref{thm:nonhalt<=>realizability}.
By \Cref{lem:guarded-safety-reduction}, this holds if and only if
  $\widehat{\phi(M)}$ is realizable under~$A$.
Since~$h$ is fixed independently of~$M$, the map
  $M\mapsto\widehat{\phi(M)}$ is computable.
Thus the undecidable non-halting problem for 2-counter machines reduces to
  $\textsc{LL-Synthesis}(A)$, proving the undecidability direction of
  \Cref{thm:undecidability:adversary}.
Taking~$A=\Gamma^\omega$ also proves \Cref{thm:undecidability}.

\section{Decidability via Bounded Component Trees}
\label{sec:tree-decidability}

We prove \Cref{thm:nowhere-dense-decision-procedure} using the component trees
  of \Cref{sec:decidability}.
We first compute a uniform consensus bound and establish the correspondence
  between component trees and distributed algorithms.
We then bound the number of edges at each node by compression and represent
  the compressed trees as symbolic trees over a finite alphabet.
Finally, we describe the accepting symbolic trees by an \ac{MSO} formula.

\subsection{The Block Automaton and Component Trees}

We first compute the consensus bound and construct the block automaton, then
  prove the correspondence between uncompressed component trees and distributed algorithms.

\begin{proof}[Proof of \Cref{lem:computable-consensus-bound}]
Construct the automaton $\mathcal A_\text{live}$ recognizing~$\Pref(A)$ as in \Cref{sec:decidability}.
By \Cref{lem:repeated_bounded_time_consensus}, for every graph history $h \in \Pref(A)$,
  consensus started anew after~$h$ with fresh proposals is solvable within a bounded
  number of further rounds under the allowed graph continuations.
Because $\mathcal A_\text{adv}$ is deterministic, these continuations depend only on the state
  reached after~$h$.
There are finitely many live states reachable from $s_\text{init}$, so choosing one
  such algorithm for each state and taking the maximum of their round bounds gives
  a uniform bound.

Iterating over~$R \geq 1$, compute the following:
For each live state~$s$ reachable from $s_\text{init}$, compute all live graph histories
  of length~$R$ starting from~$s$.
Since there are finitely many local views after~$R$ rounds, we can enumerate all
  distributed algorithms that decide at round~$R$.
For each such algorithm, check whether it solves consensus for every live graph history
  of length~$R$ starting from~$s$ and every pair of binary proposals.
On arbitrary later inputs, the processes simulate these views using their initial
  proposals and keep their decisions after round~$R$.
Conversely, any consensus algorithm deciding within~$R$ rounds yields one of these
  algorithms by fixing each process's inputs to its proposal and delaying decisions
  until round~$R$.
Thus the test succeeds exactly when consensus is solvable within~$R$ rounds
  under the graph continuations allowed from that state.
Stop when the test succeeds for every reachable live state and return~$D=R$.
The existence of a uniform bound guarantees that the above search terminates.
\end{proof}

\begin{proof}[Proof of \Cref{lem:tree-combined-automaton}]
Use the automata $\mathcal A_\text{live}$ and~$\mathcal A_\phi$ constructed in \Cref{sec:decidability}.
Put $M:=\chi_\phi(Q_\phi)\cup\{0\}$ and take
\begin{align*}
  Q&:=S\times Q_\phi\times M,&
  q_\text{init}&:=(s_\text{init},q_\phi,0)\enspace,\\
  Q_\text{live}&:=F_\text{live}\times Q_\phi\times M,&
  \chi(s,z,m)&:=m \enspace.
\end{align*}
For a block $(u,o)\in\Sigma^D\times\Omega^D$, let $z_0=z,z_1,\ldots,z_D$
  be the states of~$\mathcal A_\phi$ along that block, and define
\[
  \delta((s,z,m),(u,o))
    :=\left(\tau^*(s,\Graph(u)),z_D,
      \min_{0\leq r<D}\chi_\phi(z_r)\right) \enspace.
\]
The successor state stores the minimum priority of the completed block.
The minimum includes the starting state and excludes the endpoint, which
  contributes to the next block.
Since blocks have fixed finite length, their minima have the same least
  infinitely recurring priority as the original run.
The extra initial priority~$0$ does not affect acceptance.
The first coordinate gives the required test for allowed prefixes.
\end{proof}

\begin{proof}[Proof of \Cref{lem:tree-uncompressed-equivalence}]
($\Rightarrow$):
Suppose that~$f$ realizes~$\phi$ under~$A$.
Build an uncompressed component tree by labeling every process-$p$ vertex~$v$
  with~$f_p(v)$.
Inductively, the edge label~$q(e(h'))$ at each node is the state reached by~%
  $\mathcal B$ on~$h'$ with the outputs of~$f$.
Every run follows an infinite input signal history~$h\in\Sigma^\omega$.
If $\Graph(h)\in A$, its execution satisfies~$\phi$ by assumption.
Otherwise, $\mathcal B$ accepts the execution by \Cref{lem:tree-combined-automaton}.
This also covers graph histories outside~$A$ whose every finite prefix is allowed under~$A$.
Thus the tree is accepting.

($\Leftarrow$):
Fix an accepting uncompressed component tree.
Every edge of the round-$nD$ view complex under~$A$ belongs to exactly one
  depth-$n$ node: the assertion holds at the root, and the children partition
  the allowed extensions of their parent's edges.
Moreover, edges sharing a vertex belong to the same node.
Indeed, their preceding block prefixes share a vertex by
  \Cref{lem:IndistDownClosed}.
Inductively, those prefixes lie at the same parent, and the two extensions
  then lie in the same child component.

At every node carrying~$(K,q)$, assign its chosen output to each vertex
  of $K^{(1)},\ldots,K^{(D)}$ that occurs under~$A$.
Every possible view at a positive round receives an output.
If two occurrences of a view receive outputs from depth-$n$ nodes, their
  prefixes at round~$nD$ share a vertex by \Cref{lem:IndistDownClosed}.
They therefore belong to the same node, whose labeling assigns one output
  to their common view.
Assign a fixed default output to initial views and to views that do not occur under~$A$.
These assignments define a distributed algorithm~$f$.

For~$h\in\Sigma^\omega$ with $\Graph(h)\in A$, the edges
  $e_n:=e(h[\dots nD])$ give a run of the tree.
Induction on blocks shows that its state labels are the states reached by~%
  $\mathcal B$ under~$f$.
Acceptance therefore implies that $\exec(f,h)\models\phi$.

If the tree is effectively given, compute an output from a view by enumerating
  the finitely many allowed edges at its round.
For an incident edge, follow its prefixes through the component tree and read
  the prescribed output; if there is no such edge, use the default output.
The consistency just proved makes the result independent of the chosen edge.
\end{proof}

\subsection{Compressing Components}

Let~$K$ and~$L$ be subgraphs of view complexes, possibly at different rounds.
A \emph{map of view complexes}~$T\colon K\to L$ is a map from the vertices of~$K$
  to those of~$L$ that preserves the owning process and sends edges to edges.
For an edge $e=\{v_1,v_2\}$, we write $T(e):=\{T(v_1),T(v_2)\}$
  for its induced image in~$E(L)$.
The following lemma extends such maps to later rounds.

\begin{lemma}[\leanproof{\leanurl{ViewExtension}}]\label{lem:view-extension}
Let~$T\colon K\to L$ be a map of view complexes as above.
For every~$t\geq0$, there is a map $T^{(t)}\colon K^{(t)}\to L^{(t)}$ such that
\[
  T^{(t)}(ew)=T(e)w
\]
for all $e\in E(K)$ and $w\in\Sigma^t$.
These extensions preserve identities and composition.
\end{lemma}
\begin{proof}
Set~$T^{(0)}:=T$.
Given~$T^{(t)}$, define the image of a vertex at the next round by
\begin{align*}
  \mathsf N_p(x,u)&\longmapsto\mathsf N_p(x,T^{(t)}(u))\enspace,\\
  \mathsf C_p(x,u,y,v)&\longmapsto
    \mathsf C_p(x,T^{(t)}(u),y,T^{(t)}(v))\enspace.
\end{align*}
The constructors are disjoint and injective, so these prescriptions give
  a well-defined map on vertices.
By \Cref{tab:ViewDefn}, the endpoints of an edge~$ewa$ are sent to the
  endpoints of~$T(e)wa$. Thus the map preserves edges and vertex owners.
This proves the assertion by induction on~$t$.
The same recursion shows that extending an identity gives an identity and
  that $(U\circ T)^{(t)}=U^{(t)}\circ T^{(t)}$.
\end{proof}

A \emph{reduction} at a position~$(K,q)$ is a map of view complexes~$r\colon K\to K$
  such that
\[
  r\circ r=r,\qquad q(r(e))=q(e)
\]
for all $e\in E(K)$.
Thus~$r$ fixes its image and preserves the automaton state on each edge,
  and hence its priority.
Its image~$r(K)$ is a nonempty connected subgraph of~$K$.

The uniform consensus bound gives the following equality of earlier views
  within every connected subgraph under~$A$.

\begin{lemma}[\leanproof{\leanurl{CommonView}}]\label{lem:bounded-common-view}
Let~$A$ be a message adversary under which consensus is repeatedly solvable
  in uniformly bounded time~$D$.
Let~$r\geq D$, and let~$K$ be a nonempty connected subgraph of the
  round-$r$ view complex under~$A$.
There exists a process~$p\in{\cal P}$ such that
\[
  v_p(h[\dots r-D])=v_p(h'[\dots r-D])
\]
for all~$h,h'\in\Sigma^r$ with $e(h),e(h')\in E(K)$.
\end{lemma}
\begin{proof}
Let~$h\in\Sigma^{r-D}$ satisfy $\Graph(h)\in\Pref(A)$.
Under $\Graph(h)^{-1}A$, consensus is solvable within~$D$ rounds.
By \Cref{lem:cons:means:broadcaster}, choose an algorithm~$f_h$ that makes both
  processes agree at round~$D$ on a broadcaster in these rounds.
Simulate~$f_h$ with fixed local inputs on the communication graphs of the next~%
  $D$ rounds.

For an edge~$e(hu)$ of~$K$, where~$|h|=r-D$ and~$|u|=D$, let~$p$ be the
  common decision of this simulation and label the edge by $(p,v_p(h))$.
We show that the label depends only on either endpoint.

Consider the endpoint~$v_i(hu)$ of process~$i$.
If~$i$ received a message in these~$D$ rounds, its view contains both
  round-$(r-D)$ views: its own, and the other process's view recovered from
  the received message using \Cref{lem:IndistDownClosed}.
These two views determine~$h$ by \Cref{sec:view-complexes}.
The endpoint therefore determines which algorithm~$f_h$ is simulated,
  its local decision~$p$, and the earlier view~$v_p(h)$.
If~$i$ received no message in these rounds, the other process was not a
  broadcaster, so the common decision must be~$i$.
The edge label is then $(i,v_i(h))$, which is again determined by~$v_i(hu)$.

Thus edges sharing a vertex have the same label.
Connectivity of~$K$ makes this label constant, proving the claim.
\end{proof}

The number of edges in a component can grow without bound.
For example, under~$A=\{\rlink^\omega\}$ and a fixed input sequence of
  process~1, the round-$r$ component is a star with one edge for every input
  sequence of process~2.
We use the block that produces a component to construct a reduction with
  bounded image.

Fix a position~$(K,q)$, an output labeling~$\lambda$ from~$K$, and a connected
  component~$K'$ of~$\liveext{K}$.
For~$eu\in E(K')$, let $q'(eu):=q_\lambda(e,u)$.
Partition the edges of~$K$ that have an extension in~$K'$ into classes:
  $e$ and~$e'$ belong to the same class if the following two conditions hold.
For every~$u\in\Sigma^D$,
\[
  eu\in E(K')\ \Longleftrightarrow\ e'u\in E(K') \enspace.
\]
For every~$u\in\Sigma^D$ such that $eu\in E(K')$, we also require
\[
  q_\lambda(e,u)=q_\lambda(e',u) \enspace.
\]

Put
\begin{equation}\label{eq:tree-bounds}
  a:=|\Sigma|^D,\qquad N:=(1+|Q|)^a,\qquad B:=Na,\qquad b:=Ba \enspace.
\end{equation}
For each block, either the extension is outside~$K'$ or it has one of~$|Q|$
  state labels. Thus there are at most~$N$ classes, and each edge has at most~%
  $a$ extensions.

\begin{lemma}[\leanproof{\leanurl{Compression}}]\label{lem:tree-compression}
Choose one representative of each class.
There is a reduction~$r\colon K'\to K'$ at~$(K',q')$ such that
\begin{equation}\label{eq:tree-compression-map}
  r(eu)=e'u
\end{equation}
for all $eu\in E(K')$, where~$e'$ is the chosen representative of~$e$.
Its image has at most~$B$ edges.
If~$|E(K)|\leq B$, then $|E(\liveext{K})|\leq b$.
\end{lemma}
\begin{proof}
Let~$K$ be a subgraph of the round-$nD$ view complex under~$A$.
By \Cref{lem:bounded-common-view}, all edges of~$K$ with an extension in~$K'$
  share one process-$p$ vertex.
Since distinct edges have distinct pairs of endpoints, their process-$\bar p$
  endpoints are distinct.
Fix the common vertex and send each other endpoint to that of its representative
  edge. This defines a map~$T$ from the subgraph formed by these edges to itself.
Its extension~$T^{(D)}$ exists by \Cref{lem:view-extension}.

By the definition of the classes, $eu\in E(K')$ implies
  $T(e)u\in E(K')$.
The extension therefore restricts to a map~$r\colon K'\to K'$ satisfying
  \eqref{eq:tree-compression-map}.
Representatives are fixed by~$T$, so~$r$ is idempotent on edges and
  hence on vertices, since the graphs have no isolated vertices.
The same definition gives equality of successor states,
  and hence of their block priorities.
Thus~$r$ is a reduction at~$(K',q')$.

Its image has at most~$N$ representative prefixes, each with at most~$a$
  extensions, hence at most~$Na=B$ edges.
If~$|E(K)|\leq B$, then $|E(\liveext{K})|\leq |E(K)|a\leq b$.
\end{proof}

\begin{definition}[\leanformalized{\leanline{Formalizations/ComponentTrees/Trees}{78}}]\label{def:tree-component-tree}
A \emph{compressed component tree} carries a position~$(K,q)$, a reduction~%
  $r\colon K\to K$ with $|E(r(K))|\leq B$, and an output labeling~$\lambda$ from~$r(K)$
  at each node.
The root carries~$(K_0,q_0)$.
Use the restriction of~$q$ to~$r(K)$ to compute successor states.
Every connected component~$K'$ of~$\liveext{(r(K))}$ gives exactly one child carrying~%
  $(K',q')$, where
\[
  q'(eu):=q_\lambda(e,u)
\]
for all $e\in E(r(K))$ and $u\in\Sigma^D$ such that $eu\in E(K')$.
These are all its children.

A \emph{compatible run} consists of an infinite branch from the root with
  positions~$(K_n,q_n)$ and reductions~$r_n$, together with edges~$e_n\in E(K_n)$
  and blocks~$u_n\in\Sigma^D$ such that
\begin{equation}\label{eq:tree-compatible-run}
  e_0=e_\text{init},\qquad e_{n+1}=r_n(e_n)u_n\in E(K_{n+1}) \enspace.
\end{equation}
The tree is \emph{accepting} if $(q_n(e_n))_{n\geq0}$ satisfies parity for every
  compatible run.
\end{definition}

\begin{figure}
  \centering
  \includegraphics[width=\textwidth]{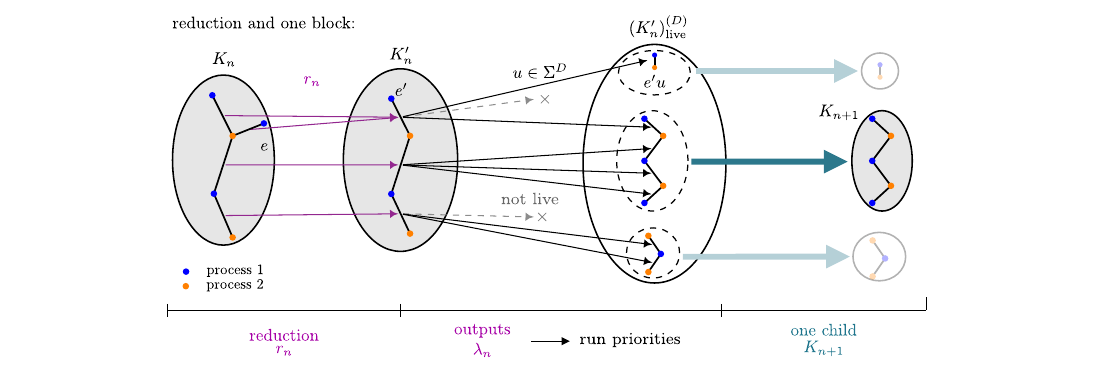}
  \caption{A reduction and one block along a branch.
    The reduction~$r_n$ maps~$K_n$ onto~$r_n(K_n)$, labelled~$K'_n$ in the figure.
    Purple arrows show its action on edges, with~$e'=r_n(e)$.
    An output labeling from~$r_n(K_n)$ gives the allowed extensions~$\liveext{(r_n(K_n))}$.
    Each connected component gives a child~$K_{n+1}$; the middle child is highlighted.
    Colored vertices represent process views.}
  \label{fig:compressed-component-tree}
\end{figure}

Every reduced graph has at most~$B$ edges, so its allowed extensions have at
  most~$b$ edges. Thus every node has at most~$b$ edges and~$b$ children.
The priority at~$e_{n+1}$ is the priority of block~$n$.
\Cref{fig:compressed-component-tree} shows one step of the construction.

Preserving state labels alone does not guarantee acceptance: repeated
  compressions can give compatible runs that do not follow any execution
  of the original tree.
We construct the reductions using an ordinal progress measure to preserve parity.

\begin{theorem}[\leanproof{\leanurl{BoundedRepresentation}}]\label{thm:tree-bounded-representation}
An accepting uncompressed component tree exists if and only if an accepting
  compressed component tree exists.
From an effectively given accepting compressed tree, one can effectively
  construct an accepting uncompressed tree.
\end{theorem}

\begin{proof}
($\Rightarrow$):
Fix an accepting uncompressed tree~$\mathcal U$.
Every edge under~$A$ at a round divisible by~$D$ occurs in exactly one node.
Write~$q_{\mathcal U}(e)$ for its state label.
Consider the rooted tree whose vertices are these edges and whose transitions
  are~$e\to eu$ for allowed extensions~$eu$ with~$u\in\Sigma^D$.
Give this transition the block priority $\chi(q_{\mathcal U}(eu))$.
Every infinite path gives a run of~$\mathcal U$ and satisfies parity.

Following the ordinal-rank construction of Berwanger, Mathew, and
  Ramanujam~\cite[proof of Theorem~5]{berwanger2016retraction}, we assign to each~$e$
  a tuple~$\mu(e)$ of ordinals indexed by the odd priorities in increasing order.
For an odd priority~$d$, write~$f\prec_d e$ when~$f$ properly extends~$e$,
  all transition priorities between them are at least~$d$, and~$d$ occurs among them.
An infinite descending chain would concatenate to a rejecting path with
  least infinitely recurring priority~$d$.
Thus~$\prec_d$ is well founded and has ordinal rank
\[
  \mu_d(e):=\sup\{\mu_d(f)+1\mid f\prec_d e\} \enspace,
\]
where the supremum of the empty set is zero.
Suppose that a nonempty segment from~$e$ to~$f$ has minimum priority~$c$.
For every odd~$d\leq c$, each $\prec_d$-successor of~$f$ is also a
  $\prec_d$-successor of~$e$, so $\mu_d(f)\leq\mu_d(e)$.
If~$c$ is odd, then~$f\prec_c e$, giving a strict decrease in coordinate~$c$.
Writing~$\mu_{\leq c}$ for the coordinates at most~$c$, we obtain
\begin{equation}\label{eq:tree-progress}
  \mu_{\leq c}(e)\geq_{\mathrm{lex}}\mu_{\leq c}(f)\enspace,
\end{equation}
with strict inequality if~$c$ is odd.
If there are no odd priorities, the tuple is empty.

Construct the compressed tree recursively.
Associate each of its nodes~$v$ with a node~$w$ of~$\mathcal U$ at the same depth,
  maintaining
\[
  K_v\subseteq U_w,\qquad q_v=\widehat q_w|_{E(K_v)} \enspace,
\]
where~$(U_w,\widehat q_w)$ is the position at~$w$.
The root positions agree, and its reduction is the identity.
Suppose that~$v$ and its reduction~$r_v$ have been constructed.
Restrict the output labeling at~$w$ to the extensions of~$r_v(K_v)$.
For each component~$K'$ of $\liveext{(r_v(K_v))}$, use the classes of
  \Cref{lem:tree-compression} and choose a representative with minimum
  lexicographic $\mu$-value in each class.
Each class is finite and nonempty.
The connected graph~$K'$ lies in a unique child component~$U_{w'}$ of~$w$.
Associate the child~$v'$ carrying~$K'$ with~$w'$, and give it the reduction
  $r_{v'}\colon K'\to K'$ supplied by \Cref{lem:tree-compression}.
Every edge of~$K'$ has its original state label, preserving the invariant.
Several compressed children may be associated with the same child of~$\mathcal U$.

Consider a compatible step from~$v$ to~$v'$ using an edge~$e$ and a block~$u$.
Put~$f:=r_v(e)$, so the next edge is~$fu$, and put $g:=r_{v'}(fu)$.
If~$f'$ is the representative of~$f$ chosen for~$K'$, then~$g=f'u$.
Since~$f$ and~$f'$ belong to the same class,
\[
  q_{v'}(fu)=q_\lambda(f,u)=q_\lambda(f',u)=q_{\mathcal U}(g) \enspace.
\]
Put $c:=\chi(q_{v'}(fu))$.
The minimum choice gives $\mu(f')\leq_{\mathrm{lex}}\mu(f)$, also after
  restriction to any initial segment of coordinates.
Applying \eqref{eq:tree-progress} to the transition~$f'\to f'u$ yields
\[
  \mu_{\leq c}(f)\geq_{\mathrm{lex}}\mu_{\leq c}(f')
    \geq_{\mathrm{lex}}\mu_{\leq c}(g)\enspace,
\]
with the second inequality strict when~$c$ is odd.
If a compatible run had odd least infinitely recurring priority~$d$, then
  eventually all its priorities would be at least~$d$, with~$d$ occurring infinitely often.
The $\mu_{\leq d}$-values of its reduced edges~$r_n(e_n)$ would then be
  nonincreasing with infinitely many
  strict decreases, contradicting well-foundedness of lexicographic order
  on finite tuples of ordinals.
The compressed tree is accepting.

($\Leftarrow$):
Fix an accepting compressed tree, with positions~$(K_v,q_v)$, reductions~$r_v$,
  and output labelings~$\lambda_v$.
Construct an uncompressed tree whose nodes~$w$ carry positions~%
  $(U_w,\widehat q_w)$.
Associate each~$w$ with a compressed node~$v$ at the same depth and a map
  of view complexes~$T_w\colon U_w\to K_v$ satisfying
\begin{equation}\label{eq:tree-simulation}
  \widehat q_w(e)=q_v(T_w(e))
\end{equation}
for all $e\in E(U_w)$.
At the root, take the identity map.
Put $T'_w:=r_v\circ T_w$, with image in~$r_v(K_v)$.
For each process-$p$ vertex~$z$ of~$U_w^{(r)}$, $1\leq r\leq D$, set
\[
  (\widehat\lambda_w)_p(z):=(\lambda_v)_p((T'_w)^{(r)}(z)) \enspace.
\]
The extensions exist by \Cref{lem:view-extension}, so this defines one
  output per vertex.
The reduction preserves starting states, so copied outputs give
\[
  q_{\widehat\lambda_w}(e,u)=q_{\lambda_v}(T'_w(e),u)
\]
for all $e\in E(U_w)$ and $u\in\Sigma^D$.
This preserves both the test for allowed extensions and the block priority.

For each component~$U'$ of~$\liveext{U_w}$, its image under~$(T'_w)^{(D)}$
  is a connected subgraph of $\liveext{(r_v(K_v))}$.
It therefore lies in a single component~$K'$ at~$v$.
Associate the child~$w'$ carrying~$U'$ with the compressed child~$v'$ carrying~%
  $K'$, and define
\[
  T_{w'}:=(T'_w)^{(D)}|_{U'} \enspace.
\]
This map preserves vertex owners, edges, and successor states, so
  \eqref{eq:tree-simulation} holds at~$w'$.
On edges this map satisfies
\[
  T_{w'}(eu)=r_v(T_w(e))u \enspace.
\]
Thus every run of the constructed uncompressed tree gives a compatible run
  of the compressed tree with the same state labels.
It is accepting.
All steps use finite graphs, output labelings, and the data of the associated
  compressed node, so the construction is effective when that tree is effectively given.
\end{proof}

\subsection{Symbolic Component Trees}

The graphs in a compressed component tree have bounded size.
Up to isomorphism, their reductions, output labelings, and successor components
  therefore range over finite sets.
We use these finite descriptions as the labels of a symbolic component tree.

We proceed as follows.
Forget the views naming the vertices, retaining each vertex's owning process
  and each edge's automaton state.
Choose one representative of each isomorphism class of nonempty connected
  bipartite graphs with at most~$b$ edges, one part per process, and edge labels
  in~$Q_\text{live}$.
Choose their vertex sets disjoint, with~$(K_0,q_0)$ representing its class.
There are only finitely many representatives.

Let~$\mathcal E$ be the union of the edge sets of the chosen representatives.
Each node will record its component, how to continue from it, and how its
  edges were generated from its parent.
The alphabet~$\Theta$ consists of all tuples
\[
  \ell=(K,q,r,\lambda,\operatorname{gen})
\]
such that:
\begin{itemize}
\item $(K,q)$ is one of the chosen representatives;
\item $r\colon K\to K$ is a reduction at~$(K,q)$ with $|E(r(K))|\leq B$;
\item $\lambda$ is an output labeling from~$r(K)$;
\item $\operatorname{gen}$ is either the dummy symbol~$\ast$ for the root or a map
  $E(K)\to\mathcal E\times\Sigma^D$.
\end{itemize}
For a non-root node~$w$, an entry
\[
  \operatorname{gen}_w(e')=(\widehat e,u)
\]
means that the child edge~$e'$ represents the extension~$\widehat e u$, where~%
  $\widehat e$ is an edge of the reduced parent graph.
Thus~$\operatorname{gen}_w$ describes how the node was reached, while~%
  $r_w,\lambda_w$ describe how to continue from it.

The alphabet~$\Theta$ is finite and effectively computable: all its fields
  range over finite, effectively enumerable sets.
To check how parent and child labels fit together, apply the view constructors
  for~$D$ rounds, starting from distinct formal views at the parent's vertices.
The constructors are disjoint and injective, so this gives exactly the same
  view equalities as at any concrete realization of the graph.
Automaton transitions determine the allowed extensions and their state labels.
There are at most~$b$ such extensions from the reduced graph, so~$b$ child
  directions suffice.
Add a padding letter~$\#$ for unused child directions.

\begin{definition}[\leanformalized{\leanline{Formalizations/ComponentTrees/GeneratingAlphabet}{27}}]\label{def:tree-symbolic}
A \emph{symbolic component tree} is a labeling of the full $b$-ary tree over~%
  $\Theta\cup\{\#\}$ satisfying the following conditions.
The root carries~$(K_0,q_0)$ and the dummy predecessor field~$\ast$.
Padding nodes have only padding children.
At each non-padding node~$v$, write the label as
  $(K_v,q_v,r_v,\lambda_v,\operatorname{gen}_v)$.
\begin{enumerate}
\item For each non-padding child~$w$ of~$v$, the field~$\operatorname{gen}_w$ is a map
  into $E(r_v(K_v))\times\Sigma^D$.
  Define $\beta_w(e'):=\widehat e u$ whenever
  $\operatorname{gen}_w(e')=(\widehat e,u)$.
  There must be a connected component~$K'$ of $\liveext{(r_v(K_v))}$ such that
  $\beta_w\colon E(K_w)\to E(K')$ is a bijection and, for all
  $e',e''\in E(K_w)$ and~$p\in\mathcal P$,
  \[
    e'_p=e''_p
    \quad\Longleftrightarrow\quad
    (\beta_w(e'))_p=(\beta_w(e''))_p \enspace.
  \]
  Thus~$\beta_w$ induces an owner-preserving graph isomorphism~$K_w\to K'$.
\item Each such entry satisfies
  $q_w(e')=q_{\lambda_v}(\widehat e,u)$.
\item The sets $\beta_w(E(K_w))$, as~$w$ ranges over the non-padding children
  of~$v$, are exactly the edge sets of the connected components of
  $\liveext{(r_v(K_v))}$, each occurring once.
\end{enumerate}

A \emph{compatible run} consists of an infinite branch $v_0v_1\cdots$ from
  the root through non-padding nodes, edges $e_n\in E(K_{v_n})$ with
  $e_0=e_\text{init}$, and blocks~$u_n\in\Sigma^D$ such that, for every~$n\geq0$,
\begin{equation}\label{eq:tree-symbolic-step}
  \operatorname{gen}_{v_{n+1}}(e_{n+1})=(r_{v_n}(e_n),u_n) \enspace.
\end{equation}
Its state sequence is
\[
  \bigl(q_{v_n}(e_n)\bigr)_{n\geq0} \enspace.
\]
The first state is $q_{v_0}(e_0)=q_\text{init}$, and the priority of block~$u_n$
  is $\chi(q_{v_{n+1}}(e_{n+1}))$.
The tree is \emph{accepting} if, for every compatible run, the least priority
  occurring infinitely often in $\bigl(\chi(q_{v_n}(e_n))\bigr)_{n\geq0}$ is even.
\end{definition}

The symbolic encoding preserves acceptance and allows us to reconstruct a compressed component tree.
\begin{lemma}[\leanproof{\leanurl{SymbolicTrees}}]\label{lem:tree-symbolic-equivalence}
An accepting compressed component tree exists if and only if an accepting
  symbolic component tree exists.
From an effectively given accepting symbolic component tree, one can
  effectively construct an accepting compressed component tree.
\end{lemma}
\begin{proof}
By \Cref{lem:view-extension}, an isomorphism~$\beta$ extends to an isomorphism
  at every subsequent round, since the extensions of~$\beta$ and~$\beta^{-1}$
  remain inverse maps.
It sends a reduction~$r$ to~$\beta\circ r\circ\beta^{-1}$ and transports
  output labelings and successor components.
Corresponding edges read the same signals and outputs from equal states,
  so their successor states agree.

($\Rightarrow$):
Given a compressed component tree, identify each node's labeled graph with
  its chosen representative, using the identity at the root.
Transport the reductions and output labelings along these isomorphisms,
  order the children arbitrarily, and add padding.
For a child edge~$e'$, apply the inverse child isomorphism, followed by the
  extended parent isomorphism, to obtain a formal extension~$\widehat e u$.
Set $\operatorname{gen}_w(e'):=(\widehat e,u)$ and give the root the dummy field.
These tables satisfy the local conditions, since the concrete children are
  exactly the full components of the allowed extensions.

($\Leftarrow$):
Start with the concrete initial position.
At each node, transport the recorded reduction and output labeling back
  along the current isomorphism.
For each non-padding child, its generating table induces an isomorphism
  to one formal successor component by \Cref{def:tree-symbolic}.
Compose it with the inverse of the extended parent isomorphism to identify
  the child's representative with a concrete successor component.
The local conditions preserve successor states and include every component
  exactly once. Continue recursively at these children.

Under these identifications, the concrete compatibility condition
  $e_{n+1}=r_n(e_n)u_n$ is exactly \eqref{eq:tree-symbolic-step}.
Thus compatible runs correspond in both directions with the same state labels,
  and acceptance is preserved.
Every finite part of the reconstructed compressed tree is computable from
  the symbolic labels, which include the generating tables.
\end{proof}

\subsection{Acceptance and the Decision Procedure}

It remains to express acceptance of symbolic component trees in \ac{MSO}.

\begin{lemma}[\leanproof{\leanurl{MSO}}]\label{lem:tree-mso}
One can construct an \ac{MSO} formula~$\operatorname{Acc}(\mathbf X)$ over the
  full $b$-ary tree whose models are exactly the accepting symbolic
  component trees.
\end{lemma}
\begin{proof}
Let $\mathbf X=(X_\ell)_{\ell\in\Theta\cup\{\#\}}$ specify the node labels.
Let~$\operatorname{Local}(\mathbf X)$ require these sets to partition the nodes
  and satisfy the root, child, and padding conditions of \Cref{def:tree-symbolic}.
The conditions on a node and its children are finite tests on their labels:
  the generating tables must induce the required isomorphisms, preserve
  successor states, and cover all successor components exactly once.
Since~$b$ and the alphabet are finite, these tests are first-order expressible.

Describe a compatible run by a set~$P$ and a tuple~$\mathbf H$ containing one
  set~$H_e$ for each edge~$e$ of a representative.
Require~$P$ to contain the root, be closed under ancestors, and contain exactly
  one child of each of its nodes.
The ancestor relation is \ac{MSO}-definable, so these conditions say that~$P$ is
  an infinite branch.
Require the sets~$H_e$ to partition~$P$, with~$H_e(v)$ indicating that the run
  uses edge~$e$ at node~$v$.
The root must use~$e_\text{init}$, and every chosen edge must belong to the
  representative at its node.
For each step along~$P$, require \eqref{eq:tree-symbolic-step}.
This is a finite test on the parent and child labels and the two chosen edges.
Together these conditions define an \ac{MSO} formula
  $\operatorname{Run}(\mathbf X,P,\mathbf H)$ for compatible runs.

For each~$c\in\chi(Q)$, let~$\operatorname{Pri}_c(v)$ be the finite disjunction
  of $X_\ell(v)\wedge H_e(v)$ over letters~$\ell$ and edges~$e$ whose
  recorded state has priority~$c$.
With~$x\prec y$ denoting strict ancestry, put
\[
  \operatorname{Inf}_c(\mathbf X,P,\mathbf H)
    :=\forall x\bigl(x\in P\ \rightarrow\
       \exists y\,(y\in P\wedge x\prec y\wedge\operatorname{Pri}_c(y))\bigr) \enspace.
\]
On a compatible run, this says that priority~$c$ occurs infinitely often.
Thus the parity condition is expressed by
\[
  \operatorname{Parity}(\mathbf X,P,\mathbf H)
    :=\bigvee_{\substack{d\in\chi(Q)\\d\text{ even}}}
       \left(\operatorname{Inf}_d
         \wedge\bigwedge_{\substack{c\in\chi(Q)\\c<d}}
                    \neg\operatorname{Inf}_c\right)\enspace,
\]
where the arguments of~$\operatorname{Inf}_c$ are suppressed on the right.
The required formula is
\begin{equation}\label{eq:tree-mso}
  \operatorname{Acc}(\mathbf X)
    :=\operatorname{Local}(\mathbf X)\ \wedge\
      \forall P\,\forall\mathbf H\,
        \bigl(\operatorname{Run}(\mathbf X,P,\mathbf H)
          \rightarrow\operatorname{Parity}(\mathbf X,P,\mathbf H)\bigr) \enspace.
\end{equation}
The generating tables are part of~$\mathbf X$, so all quantified runs use the
  same tables, reductions, and output labelings at a node.
Padding carries no edges, so compatible runs cannot enter it.

Thus~$\operatorname{Acc}$ holds exactly when the labeling is a symbolic
  component tree and all its compatible runs satisfy the parity condition.
\end{proof}

By Rabin's theorem~\cite{rabin1969decidability}, the models of~$\operatorname{Acc}$
  form an effectively regular tree language with decidable nonemptiness.
When nonempty, this language contains a tree that can be effectively generated
  by a finite labeled graph with one successor per direction.
Its labels include the generating tables, so the generator also supplies the
  correspondence data needed for reconstruction.

\begin{proof}[Proof of \Cref{thm:nowhere-dense-decision-procedure}]
Emptiness of~$A$ is decidable; if~$A=\emptyset$, any distributed algorithm suffices.
Otherwise, compute~$D$ by \Cref{lem:computable-consensus-bound}, construct
  the block automaton~$\mathcal B$, and compute the bounds in \eqref{eq:tree-bounds}.
Construct the formula~$\operatorname{Acc}$ of \Cref{lem:tree-mso}.
The preceding results give
\[
  \begin{split}
  \phi\text{ is realizable under }A
    &\ \Longleftrightarrow\ \text{an accepting uncompressed component tree exists}\\
    &\ \Longleftrightarrow\ \text{an accepting compressed component tree exists}\\
    &\ \Longleftrightarrow\ \text{an accepting symbolic component tree exists}\\
    &\ \Longleftrightarrow\ \operatorname{Acc}\text{ has a model} \enspace.
  \end{split}
\]
The last condition is decidable.
In the positive case, construct a finite generator of a model.
The effective constructions of
  \Cref{lem:tree-symbolic-equivalence,thm:tree-bounded-representation,lem:tree-uncompressed-equivalence}
  yield a realizing distributed algorithm, reconstructing finite parts of the
  component trees as needed.
\end{proof}

This yields computable local output functions. We do not claim finite-state implementations.

\section{Conclusion}

In this work, we presented a characterization of the communication assumptions
  under which distributed synthesis is decidable for two processes.
Our results shed new light on the limits of distributed synthesis
  by relating decidability to the uncertainty processes have about the system.
In the inputless setting, paths describe chains of indistinguishable histories.
If a path keeps refining without gaps, its growing length allows us to encode
  unbounded counters and also
  prevents consensus from being repeatedly solved in bounded time.
Otherwise, if gaps always appear within a fixed number of rounds, repeated
  consensus becomes possible, which we use to decide synthesis.
The same uncertainty thus links the two previously unrelated problems from formal verification
  and distributed computing.
We conjecture that a link between the two problems also holds in larger systems.

A further natural open question is to characterize the message adversaries for which
  synthesis is decidable when the empty communication graph is also allowed,
  with a promising observation that our technique also proves undecidability for two processes
  with independent inputs and no communication as studied by Pnueli and Rosner~\cite{PnueliRosner90}.
The idea is to use an \ac{LTL} safety assumption to prune the input histories
  so that the view complexes form the same growing paths as in our proof.
We can then directly reuse our counter encoding.

Concerning time complexity, for each fixed $\omega$-regular adversary~$A$
  with nowhere-dense image, our construction places $\textsc{LL-Synthesis}(A)$
  in \textsc{4Exptime}.
Here~$n$ is the total size of the specification and the input and output alphabets.
An automaton for~$A$ and its uniform consensus bound~$D$ are treated as constants.
The size of~$\mathcal B$ and the bounds~$B$ and~$b$ (see~\eqref{eq:tree-bounds})
  are at most doubly exponential in~$n$.
It follows that the size of the symbolic alphabet and its transition tables
  is at most triply exponential in~$n$.
Local consistency and acceptance of every compatible run can be checked by a
  universal parity tree automaton of size polynomial in these tables.
Its nonemptiness can be decided in exponential time~\cite{wilke2001alternating},
  giving the stated upper bound.
A lower bound already holds for the fixed adversary~$A=\{\dlink^\omega\}$:
  this case is equivalent to centralized \ac{LTL} synthesis, which is
  \textsc{2Exptime}-complete~\cite{PnueliRosner88}.
Thus synthesis in the decidable cases is \textsc{2Exptime}-hard.
Determining the exact complexity and improving this upper bound remain open.

Finally, a natural open question is whether binary outputs suffice for the
  undecidability direction of our characterization.

\section*{AI Statement}

The question of synthesis in the \ac{LL} model without acknowledgments arose
  following the 2022 \emph{Information and Computation} paper~\cite{BBBFS22}.
Conjecturing decidability, we worked on algorithms, realizing a close relation to
  consensus to determine a changing leader.
After years of discussions and proofs of decidability for special cases by hand,
  without AI assistance, we started to also work towards undecidability
  for other cases.
We later supplied drafts of these results to AI models, whose initial attempts
  did not resolve the general problem.

We then redirected the prompts toward proving undecidability, suggesting the
  use of indistinguishability chains to encode a computation, initially through
  tilings.
After weeks of discussions with different models in this direction,
  OpenAI's GPT-6 Astra finally produced an undecidability proof based on a
  2-counter machine.
The subsequent development involved extensive collaboration with the model
  through further prompting, revisions, and simplifications of the arguments.
These included replacing a parity-game formulation by a tree-language
  formulation and introducing and using view complexes, among many other changes.

The formulation of the decidability boundary in terms of nowhere dense sets,
  and the results connecting this boundary to consensus, were developed
  entirely by the authors, without AI assistance.

The AI-assisted work initially yielded a Lean formalization and a
  \LaTeX{} draft.
Starting from these results, we wrote a new manuscript from scratch, without
  reusing any text from the AI-generated \LaTeX{} draft.
AI tools were subsequently used to help revise and edit the manuscript.
Finally, we used AI again to formalize in Lean the statements and proofs as
  written in this paper.
The authors take full responsibility for the results and the content of the paper.

\section*{Acknowledgments}

We would like to thank B\'eatrice B\'erard for many fruitful discussions on the synthesis problem and for her helpful comments on a preliminary version of the manuscript.
We also thank Betsalel Guinard for discussion on the ``smile'' algorithm.
This work received funding from ANR through DREAMY (ANR-21-CE48-0003)
  and COSTXPRESS (ANR-23-CE45-0013), and from the France 2030 investment plan
  through SAIF (ANR-23-PEIA-0006), administered by ANR.

\appendix

\section{\acs*{LTL} Syntax and Semantics}
\label{app:ltl}

An execution is a word from~$(\Sigma \times \Omega)^\omega$, which records, apart from the input signals, the outputs at every round.
A convenient specification language to define the \textit{valid} system executions is \ac{LTL} interpreted over words from~$(\Sigma \times  \Omega)^\omega$.
The logic can therefore talk about inputs, outputs, and communication links at a given position.
Moreover, it has the usual temporal modalities.
Formally, the set of \ac{LTL} formulas is given by the following grammar, in which $p \in \mathcal{P}$, $x \in X_p$, $y \in Y_p$, and~$\graph \in \Gamma$:
\begin{align*}
    \phi ::= & (\inp_p = x) | (\outp_p = y) | (\link = \graph) | & \text{atomic formulas} \\
             & \ltlX \phi \mid \ltlF \phi \mid \ltlG \phi \mid \phi \ltlU \phi \mid \phi\ltlW \phi & \text{temporal modalities} \\
             & \neg \phi \mid \phi \vee \phi \mid \phi \wedge \phi \mid \phi \rightarrow \phi \mid \phi \leftrightarrow \phi & \text{Boolean connectives}
\end{align*}
Let~$e\in(\Sigma\times\Omega)^\omega$ be an execution, and let~$h:=\Graph(e)$ be its graph history.
Write the round-$r$ letter of~$e$ as
\[
    e[r] = \left( (x_1[r], x_2[r], h[r]), (y_1[r], y_2[r]) \right)
    \qquad(r\geq1)\enspace.
\]
Boolean connectives are interpreted as usual.
Moreover:
\begin{align*}
    & e \models (\inp_p = x) & \text{if } x_p[1] = x \\
    & e \models (\outp_p = y) & \text{if } y_p[1] = y \\
    & e \models (\link = \graph) & \text{if } h[1] = \graph \\
    & e \models \ltlX\phi & \text{if } e[2\dots] \models \phi \\
    & e \models \ltlF\phi & \text{if } \exists r \ge  1: e[r\dots] \models \phi \\
    & e \models \ltlG\phi & \text{if } \forall r \ge  1: e[r\dots] \models \phi \\
    & e \models \phi\ltlU\psi & \text{if } \exists r \ge  1: \left( e[r\dots] \models \psi \wedge \forall 1 \le  r' < r : e[r'\dots] \models \phi \right) \\
    & e \models \psi\ltlW\phi & \text{if } e \models \ltlG\psi  \vee \psi\ltlU \phi 
    \end{align*}

\section{Lean Formalization}
\label{sec:lean-formalization}

For the Lean formalization, we used a
  \href{https://github.com/BioDisCo/vibe-formalizing}{two-agent workflow}
  \cite{FN26badges}, where one agent proposes the Lean formalization
  and the other checks its correspondence with the natural-language statements
  and proofs.

We reuse definitions and results from Mathlib (v4.30.0) wherever possible, including its
  real numbers and topology, finite-type and encoding infrastructure, and
  computability theory.
For example, primitive recursiveness and algorithmic decidability use
  Mathlib's
  \href{https://github.com/leanprover-community/mathlib4/blob/c5ea00351c28e24afc9f0f84379aa41082b1188f/Mathlib/Computability/Primrec/Basic.lean}{\texttt{Primrec}}
  and
  \href{https://github.com/leanprover-community/mathlib4/blob/c5ea00351c28e24afc9f0f84379aa41082b1188f/Mathlib/Computability/RE.lean}{\texttt{ComputablePred}},
  respectively, and nowhere denseness uses its topological definition
  \href{https://github.com/leanprover-community/mathlib4/blob/c5ea00351c28e24afc9f0f84379aa41082b1188f/Mathlib/Topology/GDelta/Basic.lean\#L198}{\texttt{IsNowhereDense}}
  in the relative topology on~$[0,1]$.
For every $\omega$-regular message adversary~$A$, our main theorem
  (\Cref{thm:undecidability:adversary}) is expressed in the formalization as
\[
  \texttt{ComputablePred (Realizable A)}\;\leftrightarrow\;
  \texttt{IsNowhereDense (geometricMap \textquotesingle\textquotesingle\ A)}
\]
Here, \texttt{geometricMap \textquotesingle\textquotesingle\ A} denotes~$\pi(A)$,
  and realizability quantifies existentially over
  \href{\leanline{Library/Distributed/Execution}{81}}{pairs of local strategies}
  and universally over all \href{\leanline{Library/Distributed/Execution}{544}}{executions}
  whose graph history belongs to
  \href{\leanline{Formalizations/Paper/ProblemStatement}{47}}{adversary~$A$},
  with unrestricted private input streams.
Instances include the finite nonempty input and output alphabets, the
  interpretation of atomic propositions, and the \ac{LTL} formula.
The positive result additionally provides a primitive-recursive Boolean
  decision procedure for each fixed $\omega$-regular adversary with nowhere
  dense geometric image.

\href{\leansourceurl{Formalizations/Definition/DefLTL}}{\ac{LTL} syntax and its
  semantics on infinite words} are defined in our development.
Our \href{\leansourceurl{Library/Logic/MSO}}{\ac{MSO} formalization} extends
  Mathlib's first-order syntax and semantics with quantification over
  arbitrary sets, including infinite sets.
To support the \ac{LTL} and \ac{MSO} translations, our
  \href{\leanrepository/tree/\leanrevision/\leansourceroot/Library/Automata}{automata library}
  provides \href{\leanline{Formalizations/Definition/DefNondeterministicOmegaAutomata}{25}}{B\"uchi}
  and \href{\leanline{Formalizations/Definition/DefPathSummary}{30}}{parity automata on infinite words}
  and \href{\leanline{Formalizations/Theorem/ThmParityTreeEmptinessGame}{272}}{parity automata on infinite trees},
  which are not available in Mathlib at the time of writing.
It includes
  \href{\leansourceurl{Formalizations/Theorem/ThmPitermanDeterminization}}{Piterman's determinization}~\cite{piterman2007},
  whose correctness is proved in the development.
The undecidability proof ultimately rests on Mathlib's halting undecidability
  theorem, through a verified chain of reductions incorporating Lean
  adaptations of counter-machine constructions from the
  \href{https://github.com/uds-psl/coq-library-undecidability}{Rocq library of
  undecidability proofs}~\cite{ForsterEtAl2020CoqLibrary}.

\bibliographystyle{alphaurl}
\bibliography{bibl}

\newcommand{\etalchar}[1]{$^{#1}$}
\begin{thebibliography}{FLWD{\etalchar{+}}20}

\bibitem[ACGN26]{ACGN26}
Hagit Attiya, Armando Casta{\~n}eda, Dhrubajyoti Ghosh, and Thomas Nowak.
\newblock Equivalence and separation between {Heard-Of} and asynchronous
  message-passing models.
\newblock In Chryssis Georgiou, editor, {\em Proceedings of the 33rd
  International Colloquium on Structural Information and Communication
  Complexity (SIROCCO 2026)}, volume 16488 of {\em Lecture Notes in Computer
  Science}, pages 53--72, Cham, 2026. Springer.
\newblock \href {https://doi.org/10.1007/978-3-032-26465-7_4}
  {\path{doi:10.1007/978-3-032-26465-7_4}}.

\bibitem[AG15]{AG15}
Yehuda Afek and Eli Gafni.
\newblock A simple characterization of asynchronous computations.
\newblock {\em Theoretical Computer Science}, 561:88--95, 2015.
\newblock \href {https://doi.org/10.1016/j.tcs.2014.07.022}
  {\path{doi:10.1016/j.tcs.2014.07.022}}.

\bibitem[AS85]{alpern_defining_1985}
Bowen Alpern and Fred~B. Schneider.
\newblock Defining liveness.
\newblock {\em Information Processing Letters}, 21(4):181--185, 1985.
\newblock \href {https://doi.org/10.1016/0020-0190(85)90056-0}
  {\path{doi:10.1016/0020-0190(85)90056-0}}.

\bibitem[AVW03]{arnold&vincent&walukiewicz:2003:games}
Andr\'e Arnold, Aymeric Vincent, and Igor Walukiewicz.
\newblock Games for synthesis of controllers with partial observation.
\newblock {\em Theoretical Computer Science}, 303(1):7--34, 2003.
\newblock \href {https://doi.org/10.1016/S0304-3975(02)00442-5}
  {\path{doi:10.1016/S0304-3975(02)00442-5}}.

\bibitem[BBB{\etalchar{+}}22]{BBBFS22}
B{\'e}atrice B{\'e}rard, Benedikt Bollig, Patricia Bouyer, Matthias F{\"u}gger,
  and Nathalie Sznajder.
\newblock Synthesis in presence of dynamic links.
\newblock {\em Information and Computation}, 289:104856, 2022.
\newblock \href {https://doi.org/10.1016/j.ic.2021.104856}
  {\path{doi:10.1016/j.ic.2021.104856}}.

\bibitem[BCJ18]{BCJ18}
Roderick Bloem, Krishnendu Chatterjee, and Barbara Jobstmann.
\newblock Graph games and reactive synthesis.
\newblock In Edmund~M. Clarke, Thomas~A. Henzinger, Helmut Veith, and Roderick
  Bloem, editors, {\em Handbook of Model Checking}, pages 921--962. Springer,
  Cham, 2018.
\newblock \href {https://doi.org/10.1007/978-3-319-10575-8_27}
  {\path{doi:10.1007/978-3-319-10575-8_27}}.

\bibitem[BL69]{BuchiLandweber69}
J.~Richard B{\"u}chi and Lawrence~H. Landweber.
\newblock Solving sequential conditions by finite-state strategies.
\newblock {\em Transactions of the American Mathematical Society},
  138:295--311, 1969.
\newblock \href {https://doi.org/10.1090/S0002-9947-1969-0280205-0}
  {\path{doi:10.1090/S0002-9947-1969-0280205-0}}.

\bibitem[BMR16]{berwanger2016retraction}
Dietmar Berwanger, Anup~Basil Mathew, and R.~Ramanujam.
\newblock A retraction theorem for distributed synthesis, 2016.
\newblock arXiv:1604.07564 [cs.DC].
\newblock \href {https://doi.org/10.48550/arXiv.1604.07564}
  {\path{doi:10.48550/arXiv.1604.07564}}.

\bibitem[BMvdB15]{BMV15hierarchical}
Dietmar Berwanger, Anup~Basil Mathew, and Marie van~den Bogaard.
\newblock Hierarchical information patterns and distributed strategy synthesis.
\newblock In Bernd Finkbeiner, Geguang Pu, and Lijun Zhang, editors, {\em
  Proceedings of the 13th International Symposium on Automated Technology for
  Verification and Analysis (ATVA 2015)}, volume 9364 of {\em Lecture Notes in
  Computer Science}, pages 378--393, Cham, 2015. Springer.
\newblock \href {https://doi.org/10.1007/978-3-319-24953-7_28}
  {\path{doi:10.1007/978-3-319-24953-7_28}}.

\bibitem[Bou98]{bourbaki_general_topology}
Nicolas Bourbaki.
\newblock {\em General Topology: Chapters 1--4}.
\newblock Elements of Mathematics. Springer, Heidelberg, 1998.
\newblock \href {https://doi.org/10.1007/978-3-642-61701-0}
  {\path{doi:10.1007/978-3-642-61701-0}}.

\bibitem[CBFN15]{CBFN15:icalp}
Bernadette Charron-Bost, Matthias F\"ugger, and Thomas Nowak.
\newblock Approximate consensus in highly dynamic networks: {The} role of
  averaging algorithms.
\newblock In Magn{\'u}s~M. Halld{\'o}rsson, Kazuo Iwama, Naoki Kobayashi, and
  Bettina Speckmann, editors, {\em Proceedings of the 42nd International
  Colloquium on Automata, Languages, and Programming (ICALP 2015)}, volume 9135
  of {\em Lecture Notes in Computer Science}, pages 528--539, Heidelberg, 2015.
  Springer.
\newblock \href {https://doi.org/10.1007/978-3-662-47666-6_42}
  {\path{doi:10.1007/978-3-662-47666-6_42}}.

\bibitem[CBS09]{CBS09}
Bernadette Charron-Bost and Andr{\'e} Schiper.
\newblock The {Heard-Of} model: computing in distributed systems with benign
  faults.
\newblock {\em Distributed Computing}, 22(1):49--71, 2009.
\newblock \href {https://doi.org/10.1007/s00446-009-0084-6}
  {\path{doi:10.1007/s00446-009-0084-6}}.

\bibitem[CFQS12]{Cas12:time}
Arnaud Casteigts, Paola Flocchini, Walter Quattrociocchi, and Nicola Santoro.
\newblock Time-varying graphs and dynamic networks.
\newblock {\em International Journal of Parallel, Emergent and Distributed
  Systems}, 27(5):387--408, 2012.
\newblock \href {https://doi.org/10.1080/17445760.2012.668546}
  {\path{doi:10.1080/17445760.2012.668546}}.

\bibitem[CG23]{coutouly_topology_2023}
Yannis Coutouly and Emmanuel Godard.
\newblock A topology by geometrization for sub-iterated immediate snapshot
  message adversaries and applications to set-agreement.
\newblock In Rotem Oshman, editor, {\em Proceedings of the 37th International
  Symposium on Distributed Computing (DISC 2023)}, volume 281 of {\em Leibniz
  International Proceedings in Informatics}, pages 15:1--15:20, Dagstuhl, 2023.
  Schloss Dagstuhl -- Leibniz-Zentrum f{\"u}r Informatik.
\newblock \href {https://doi.org/10.4230/LIPIcs.DISC.2023.15}
  {\path{doi:10.4230/LIPIcs.DISC.2023.15}}.

\bibitem[CGP15]{CG15}
\'Etienne Coulouma, Emmanuel Godard, and Joseph Peters.
\newblock A characterization of oblivious message adversaries for which
  consensus is solvable.
\newblock {\em Theoretical Computer Science}, 584:80--90, 2015.
\newblock \href {https://doi.org/10.1016/j.tcs.2015.01.024}
  {\path{doi:10.1016/j.tcs.2015.01.024}}.

\bibitem[Chu57]{Church57}
Alonzo Church.
\newblock Application of recursive arithmetic to the problem of circuit
  synthesis.
\newblock In {\em Summaries of Talks Presented at the Summer Institute for
  Symbolic Logic}, volume~1, pages 3--50, Ithaca, 1957. Cornell University.

\bibitem[Chu63]{Church62}
Alonzo Church.
\newblock Logic, arithmetic, and automata.
\newblock In V.~Stenstr{\"o}m, editor, {\em Proceedings of the 14th
  International Congress of Mathematicians (ICM 1962)}, pages 23--35,
  Djursholm, 1963. Institut Mittag-Leffler.

\bibitem[Fij26]{Fijalkow25}
Nathana{\"e}l Fijalkow, editor.
\newblock {\em Games on Graphs: From Logic and Automata to Algorithms}.
\newblock Cambridge University Press, Cambridge, 2026.
\newblock \href {https://doi.org/10.1017/9781009500678}
  {\path{doi:10.1017/9781009500678}}.

\bibitem[FLWD{\etalchar{+}}20]{ForsterEtAl2020CoqLibrary}
Yannick Forster, Dominique Larchey-Wendling, Andrej Dudenhefner, Edith Heiter,
  Dominik Kirst, Fabian Kunze, Gert Smolka, Simon Spies, Dominik Wehr, and
  Maximilian Wuttke.
\newblock A {Coq} library of undecidable problems.
\newblock The Sixth International Workshop on Coq for Programming Languages
  (CoqPL 2020), 2020.
\newblock URL:
  \url{https://popl20.sigplan.org/details/CoqPL-2020-papers/5/A-Coq-Library-of-Undecidable-Problems}.

\bibitem[FN26]{FN26badges}
Matthias F{\"u}gger and Thomas Nowak.
\newblock Adding ``verified'' badges to our paper, one lemma at a time.
\newblock {\em Bulletin of the EATCS}, 149:24--32, 2026.
\newblock URL: \url{https://eatcs.org/images/bulletin/beatcs149.pdf}.

\bibitem[FS05]{fs05}
Bernd Finkbeiner and Sven Schewe.
\newblock Uniform distributed synthesis.
\newblock In Prakash Panangaden, editor, {\em Proceedings of the 20th Annual
  IEEE Symposium on Logic in Computer Science (LICS 2005)}, pages 321--330,
  Piscataway, 2005. IEEE.
\newblock \href {https://doi.org/10.1109/LICS.2005.53}
  {\path{doi:10.1109/LICS.2005.53}}.

\bibitem[GGMW13]{GGMW13}
Blaise Genest, Hugo Gimbert, Anca Muscholl, and Igor Walukiewicz.
\newblock Asynchronous games over tree architectures.
\newblock In Fedor~V. Fomin, R{\=u}si{\c n}{\v s} Freivalds, Marta Kwiatkowska,
  and David Peleg, editors, {\em Proceedings of the 40th International
  Colloquium on Automata, Languages, and Programming (ICALP 2013)}, volume 7966
  of {\em Lecture Notes in Computer Science}, pages 275--286, Heidelberg, 2013.
  Springer.
\newblock \href {https://doi.org/10.1007/978-3-642-39212-2_26}
  {\path{doi:10.1007/978-3-642-39212-2_26}}.

\bibitem[Gim22]{Gimbert22}
Hugo Gimbert.
\newblock Distributed asynchronous games with causal memory are undecidable.
\newblock {\em Logical Methods in Computer Science}, 18(3):30:1--30:22, 2022.
\newblock \href {https://doi.org/10.46298/lmcs-18(3:30)2022}
  {\path{doi:10.46298/lmcs-18(3:30)2022}}.

\bibitem[GLZ04]{gastinlermanzeitoun:04}
Paul Gastin, Benjamin Lerman, and Marc Zeitoun.
\newblock Distributed games with causal memory are decidable for
  series-parallel systems.
\newblock In Kamal Lodaya and Meena Mahajan, editors, {\em Proceedings of the
  24th International Conference on Foundations of Software Technology and
  Theoretical Computer Science (FSTTCS 2004)}, volume 3328 of {\em Lecture
  Notes in Computer Science}, pages 275--286, Heidelberg, 2004. Springer.
\newblock \href {https://doi.org/10.1007/978-3-540-30538-5_23}
  {\path{doi:10.1007/978-3-540-30538-5_23}}.

\bibitem[GP20]{godard_back_2020}
Emmanuel Godard and Eloi Perdereau.
\newblock Back to the coordinated attack problem.
\newblock {\em Mathematical Structures in Computer Science}, 30(10):1089--1113,
  2020.
\newblock \href {https://doi.org/10.1017/S0960129521000037}
  {\path{doi:10.1017/S0960129521000037}}.

\bibitem[GSZ09]{GSZ09FMSD}
Paul Gastin, Nathalie Sznajder, and Marc Zeitoun.
\newblock Distributed synthesis for well-connected architectures.
\newblock {\em Formal Methods in System Design}, 34(3):215--237, 2009.
\newblock \href {https://doi.org/10.1007/s10703-008-0064-7}
  {\path{doi:10.1007/s10703-008-0064-7}}.

\bibitem[HKR14]{HKR-book}
Maurice Herlihy, Dmitry~N. Kozlov, and Sergio Rajsbaum.
\newblock {\em Distributed Computing Through Combinatorial Topology}.
\newblock Morgan Kaufmann, Waltham, 2014.
\newblock \href {https://doi.org/10.1016/C2011-0-07032-1}
  {\path{doi:10.1016/C2011-0-07032-1}}.

\bibitem[KMTZ13]{KMTZ13}
Jonathan Katz, Ueli Maurer, Bj{\"o}rn Tackmann, and Vassilis Zikas.
\newblock Universally composable synchronous computation.
\newblock In Amit Sahai, editor, {\em Proceedings of the 10th Theory of
  Cryptography Conference (TCC 2013)}, volume 7785 of {\em Lecture Notes in
  Computer Science}, pages 477--498, Heidelberg, 2013. Springer.
\newblock \href {https://doi.org/10.1007/978-3-642-36594-2_27}
  {\path{doi:10.1007/978-3-642-36594-2_27}}.

\bibitem[KV97]{kv97}
Orna Kupferman and Moshe~Y. Vardi.
\newblock Synthesis with incomplete information.
\newblock In Howard Barringer, Michael Fisher, Dov Gabbay, and Graham Gough,
  editors, {\em Proceedings of the 2nd International Conference on Temporal
  Logic (ICTL 1997)}, pages 91--106, Manchester, 1997. University of
  Manchester.

\bibitem[KV01]{kv01}
Orna Kupferman and Moshe~Y. Vardi.
\newblock Synthesizing distributed systems.
\newblock In Joseph~Y. Halpern, editor, {\em Proceedings of the 16th Annual
  IEEE Symposium on Logic in Computer Science (LICS 2001)}, pages 389--398,
  Piscataway, 2001. IEEE.
\newblock \href {https://doi.org/10.1109/LICS.2001.932514}
  {\path{doi:10.1109/LICS.2001.932514}}.

\bibitem[Lam98]{Lamport98}
Leslie Lamport.
\newblock The part-time parliament.
\newblock {\em ACM Transactions on Computer Systems}, 16(2):133--169, 1998.
\newblock \href {https://doi.org/10.1145/279227.279229}
  {\path{doi:10.1145/279227.279229}}.

\bibitem[LM95]{lubitch_closed_1995}
Ronit Lubitch and Shlomo Moran.
\newblock Closed schedulers: a novel technique for analyzing asynchronous
  protocols.
\newblock {\em Distributed Computing}, 8(4):203--210, 1995.
\newblock \href {https://doi.org/10.1007/BF02242738}
  {\path{doi:10.1007/BF02242738}}.

\bibitem[Min67]{minsky1967}
Marvin~L. Minsky.
\newblock {\em Computation: Finite and Infinite Machines}.
\newblock Prentice-Hall, Englewood Cliffs, 1967.

\bibitem[MW14]{MuschollWalukiewicz14}
Anca Muscholl and Igor Walukiewicz.
\newblock Distributed synthesis for acyclic architectures.
\newblock In Venkatesh Raman and S.~P. Suresh, editors, {\em Proceedings of the
  34th International Conference on Foundation of Software Technology and
  Theoretical Computer Science (FSTTCS 2014)}, volume~29 of {\em Leibniz
  International Proceedings in Informatics}, pages 639--651, Dagstuhl, 2014.
  Schloss Dagstuhl -- Leibniz-Zentrum f{\"u}r Informatik.
\newblock \href {https://doi.org/10.4230/LIPIcs.FSTTCS.2014.639}
  {\path{doi:10.4230/LIPIcs.FSTTCS.2014.639}}.

\bibitem[NSW24]{NSW24}
Thomas Nowak, Ulrich Schmid, and Kyrill Winkler.
\newblock Topological characterization of consensus in distributed systems.
\newblock {\em Journal of the ACM}, 71(6):1--48, 2024.
\newblock \href {https://doi.org/10.1145/3687302} {\path{doi:10.1145/3687302}}.

\bibitem[Pit07]{piterman2007}
Nir Piterman.
\newblock From nondeterministic {B\"uchi} and {Streett} automata to
  deterministic parity automata.
\newblock {\em Logical Methods in Computer Science}, 3(3:5):1--21, 2007.
\newblock \href {https://doi.org/10.2168/LMCS-3(3:5)2007}
  {\path{doi:10.2168/LMCS-3(3:5)2007}}.

\bibitem[Pnu77]{Pnueli77}
Amir Pnueli.
\newblock The temporal logic of programs.
\newblock In Paul Young, editor, {\em Proceedings of the 18th Annual IEEE
  Symposium on Foundations of Computer Science (FOCS 1977)}, pages 46--57,
  Piscataway, 1977. IEEE.
\newblock \href {https://doi.org/10.1109/SFCS.1977.32}
  {\path{doi:10.1109/SFCS.1977.32}}.

\bibitem[PP04]{perrin_infinite_2004}
Dominique Perrin and Jean-{\'E}ric Pin.
\newblock {\em Infinite Words: Automata, Semigroups, Logic and Games}, volume
  141 of {\em Pure and Applied Mathematics}.
\newblock Elsevier, Amsterdam, 2004.
\newblock \href {https://doi.org/10.1016/S0079-8169(04)X8001-6}
  {\path{doi:10.1016/S0079-8169(04)X8001-6}}.

\bibitem[PR88]{PnueliRosner88}
Amir Pnueli and Roni Rosner.
\newblock A framework for the synthesis of reactive modules.
\newblock In Frederich~H. Vogt, editor, {\em Proceedings of the International
  Conference on Concurrency (CONCURRENCY 1988)}, volume 335 of {\em Lecture
  Notes in Computer Science}, pages 4--17, Heidelberg, 1988. Springer.
\newblock \href {https://doi.org/10.1007/3-540-50403-6_28}
  {\path{doi:10.1007/3-540-50403-6_28}}.

\bibitem[PR90]{PnueliRosner90}
Amir Pnueli and Roni Rosner.
\newblock Distributed reactive systems are hard to synthesize.
\newblock In Mihalis Yannakakis, editor, {\em Proceedings of the 31st Annual
  IEEE Symposium on Foundations of Computer Science (FOCS 1990)}, volume~II,
  pages 746--757, Piscataway, 1990. IEEE.
\newblock \href {https://doi.org/10.1109/FSCS.1990.89597}
  {\path{doi:10.1109/FSCS.1990.89597}}.

\bibitem[Rab69]{rabin1969decidability}
Michael~O. Rabin.
\newblock Decidability of second-order theories and automata on infinite trees.
\newblock {\em Transactions of the American Mathematical Society}, 141:1--35,
  1969.
\newblock \href {https://doi.org/10.1090/S0002-9947-1969-0246760-1}
  {\path{doi:10.1090/S0002-9947-1969-0246760-1}}.

\bibitem[Rab72]{Rabin72}
Michael~Oser Rabin.
\newblock {\em Automata on Infinite Objects and {Church}'s Problem}, volume~13
  of {\em CBMS Regional Conference Series in Mathematics}.
\newblock American Mathematical Society, Providence, 1972.
\newblock \href {https://doi.org/10.1090/cbms/013}
  {\path{doi:10.1090/cbms/013}}.

\bibitem[Sis94]{Sistla94}
A.~Prasad Sistla.
\newblock Safety, liveness and fairness in temporal logic.
\newblock {\em Form. Asp. Comput.}, 6(5):495–511, September 1994.
\newblock \href {https://doi.org/10.1007/BF01211865}
  {\path{doi:10.1007/BF01211865}}.

\bibitem[ST87]{srikanth87}
T.~K. Srikanth and Sam Toueg.
\newblock Optimal clock synchronization.
\newblock {\em Journal of the ACM}, 34(3):626--645, 1987.
\newblock \href {https://doi.org/10.1145/28869.28876}
  {\path{doi:10.1145/28869.28876}}.

\bibitem[SW89]{SW89}
Nicola Santoro and Peter Widmayer.
\newblock Time is not a healer.
\newblock In Burkhard Monien and Robert Cori, editors, {\em Proceedings of the
  6th Annual Symposium on Theoretical Aspects of Computer Science (STACS
  1989)}, volume 349 of {\em Lecture Notes in Computer Science}, pages
  304--313, Heidelberg, 1989. Springer.
\newblock \href {https://doi.org/10.1007/bfb0028994}
  {\path{doi:10.1007/bfb0028994}}.

\bibitem[SWK09]{SWK09}
Ulrich Schmid, Bettina Weiss, and Idit Keidar.
\newblock Impossibility results and lower bounds for consensus under link
  failures.
\newblock {\em SIAM Journal on Computing}, 38(5):1912--1951, 2009.
\newblock \href {https://doi.org/10.1137/s009753970443999x}
  {\path{doi:10.1137/s009753970443999x}}.

\bibitem[Tho90]{thomas_automata_1990}
Wolfgang Thomas.
\newblock Automata on infinite objects.
\newblock In Jan van Leeuwen, editor, {\em Handbook of Theoretical Computer
  Science, Volume B: Formal Models and Semantics}, pages 133--191. Elsevier,
  Amsterdam, 1990.
\newblock \href {https://doi.org/10.1016/B978-0-444-88074-1.50009-3}
  {\path{doi:10.1016/B978-0-444-88074-1.50009-3}}.

\bibitem[Tho08]{Thomas08}
Wolfgang Thomas.
\newblock {Church}'s problem and a tour through automata theory.
\newblock In Arnon Avron, Nachum Dershowitz, and Alexander Rabinovich, editors,
  {\em Pillars of Computer Science}, volume 4800 of {\em Lecture Notes in
  Computer Science}, pages 635--655. Springer, Heidelberg, 2008.
\newblock \href {https://doi.org/10.1007/978-3-540-78127-1_35}
  {\path{doi:10.1007/978-3-540-78127-1_35}}.

\bibitem[vdMW05]{mw05}
Ron van~der Meyden and Thomas Wilke.
\newblock Synthesis of distributed systems from knowledge-based specifications.
\newblock In Mart{\'i}n Abadi and Luca de~Alfaro, editors, {\em Proceedings of
  the 16th International Conference on Concurrency Theory (CONCUR 2005)},
  volume 3653 of {\em Lecture Notes in Computer Science}, pages 562--576,
  Heidelberg, 2005. Springer.
\newblock \href {https://doi.org/10.1007/11539452_42}
  {\path{doi:10.1007/11539452_42}}.

\bibitem[Wil01]{wilke2001alternating}
Thomas Wilke.
\newblock Alternating tree automata, parity games, and modal $\mu$-calculus.
\newblock {\em Bulletin of the Belgian Mathematical Society -- Simon Stevin},
  8(2):359--391, 2001.
\newblock \href {https://doi.org/10.36045/bbms/1102714178}
  {\path{doi:10.36045/bbms/1102714178}}.

\bibitem[WSS19]{WSS19}
Kyrill Winkler, Manfred Schwarz, and Ulrich Schmid.
\newblock Consensus in rooted dynamic networks with short-lived stability.
\newblock {\em Distributed Computing}, 32(5):443--458, 2019.
\newblock \href {https://doi.org/10.1007/s00446-019-00348-0}
  {\path{doi:10.1007/s00446-019-00348-0}}.

\end{thebibliography}

\end{document}